\documentclass[10pt,oneside,reqno]{amsart}

\usepackage{cancel,bm, amssymb}
\usepackage{color}
\usepackage{color,caption}
\usepackage{graphicx,framed}
\usepackage[colorlinks=true, pdfstartview=FitV, linkcolor=blue, citecolor=blue, urlcolor=blue]{hyperref}
\definecolor{shadecolor}{rgb}{0.95, 0.95, 0.86}
\definecolor{darkgreen}{rgb}{0.2, 0.5,  0}

\def \gg{{\mathfrak g}}
\makeatletter
\@addtoreset{equation}{section}
\makeatother

\renewcommand{\d}{{\mathrm d}}

\newtheorem{theorem}{Theorem}[section]
\newtheorem{example}[theorem]{Example}
\newtheorem{exercise}[theorem]{Exercise}

\newtheorem{lemma}[theorem]{Lemma}
\newtheorem{remark}[theorem]{Remark}
\newtheorem{problem}[theorem]{Riemann-Hilbert Problem}

\newtheorem{proposition}[theorem]{Proposition} 
\newtheorem{corollary}[theorem]{Corollary} 
\newtheorem{conjecture}[theorem]{Conjecture}
\newtheorem{definition}[theorem]{Definition}
\def\le{\left}
\def\ri{\right}
\def\ds{\displaystyle}

\def\res{\mathop{\mathrm {res}}\limits}

\def\bt{\begin{theorem}}
\def\et{\end{theorem}}
\def\bc{\begin{corollary}}
\def\ec{\end{corollary}}
\def\bx{\begin{example}\small}
\def\ex{\end{example}}
\def\bxr{\begin{exercise}\small}
\def\exr{\end{exercise}}
\def\bl{\begin{lemma}}
\def\el{\end{lemma}}
\def\bd{\begin{definition}}
\def\ed{\end{definition}}
\def\bp{\begin{proposition}}
\def\ep{\end{proposition}}

\def\br{\begin{remark}}
\def\er{\end{remark}}

\def\be{\begin{equation}}
\def\ee{\end{equation}}
\def\&{&\hspace{-15pt}}
\def\mod{\, \mathrm{mod}\,\,}
\def\bea{\begin{eqnarray}}
\def\eea{\end{eqnarray}}
\def\beas{\begin{eqnarray*}}
\def\eeas{\end{eqnarray*}}

\def\C{{\mathbb C}}

\def\R{{\mathbb R}}
\def\N{{\mathbb N}}
\def\wh{\widehat}

\def\1{{\bf 1}}

\def\z{\zeta}

\def\eqref#1{(\ref{#1})}

\begin{document}

\title[Universality conjecture and results for a model of several coupled positive matrices]{Universality conjecture and results for a model of several coupled positive-definite matrices}

\author{Marco Bertola}
\address{Centre de recherches math\'ematiques,
Universit\'e de Montr\'eal, C.~P.~6128, succ. centre ville, Montr\'eal,
Qu\'ebec, Canada H3C 3J7 and,
Department of Mathematics and
Statistics, Concordia University, 1455 de Maisonneuve W., Montr\'eal, Qu\'ebec,
Canada H3G 1M8}
\email{Marco.Bertola@concordia.ca}

\author{Thomas Bothner}
\address{Centre de recherches math\'ematiques,
Universit\'e de Montr\'eal, C.~P.~6128, succ. centre ville, Montr\'eal,
Qu\'ebec, Canada H3C 3J7 and,
Department of Mathematics and
Statistics, Concordia University, 1455 de Maisonneuve W., Montr\'eal, Qu\'ebec,
Canada H3G 1M8}
\email{bothner@crm.umontreal.ca}



\thanks{The first author is supported
in part by the Natural Sciences and Engineering Research Council of Canada. The second author acknowledges support by Concordia University through a postdoctoral fellow
top-up award.}

\date{\today}

\begin{abstract}
The paper contains two main parts: in the first part, we analyze the general case of $p\geq 2$ matrices coupled in a chain subject to Cauchy interaction. Similarly to the Itzykson-Zuber interaction model, the eigenvalues of the Cauchy chain form a multi level determinantal point process. We first compute all correlations functions in terms of Cauchy biorthogonal polynomials and locate them as specific entries of a $(p+1)\times (p+1)$ matrix valued solution of a Riemann-Hilbert problem. 
In the second part, we fix  the external potentials as classical Laguerre weights. We then derive strong asymptotics for the Cauchy biorthogonal polynomials when the support of the equilibrium measures contains the origin. As a result, we obtain a new family of universality classes for multi-level random determinantal point fields which include the Bessel$_\nu$ universality for $1$-level and the Meijer-G universality for $2$-level. Our analysis uses the Deift-Zhou nonlinear steepest descent method and the explicit construction of a $(p+1)\times (p+1)$ origin parametrix in terms of Meijer G-functions. The solution of the full Riemann-Hilbert problem is derived rigorously only for $p=3$ but the general framework of the proof can be extended to the Cauchy chain of arbitrary length $p$.
\end{abstract}

\maketitle

\section{Introduction}
The general study of universal behaviors in random matrix models consists in identifying statistical properties of the fluctuations of eigenvalues near a point of the spectrum; 
for instance, the celebrated Tracy--Widom distribution was first derived \cite{TWII} in studying the fluctuations of the largest eigenvalue of a $n\times n$ GUE (Gaussian Unitary Ensemble) matrix around the edge of the limiting (macroscopic) density (which obeys the Wigner semicircle law). They connected the  probability (for the rescaled eigenvalues $x_i  = \sqrt{2} n^\frac 23 (\lambda_i- \sqrt{2})$) that $x_{max}<s$  to a special solution (Hastings-McLeod) of the second Painlev\'e\ equation, 
\begin{align*}
	\lim_{n\to\infty} \textnormal{Prob}\le(\lambda_{\textnormal{max}} \leq \sqrt{2} + \frac {\sqrt{2}s}{2n^{\frac 23}} \ri) &= \textnormal{Prob}\big(\hbox{no $x_i$'s in } [s,\infty)\big) = \exp\left[-\int_s^\infty (x-s) q(x)^2 \d x \right]\\
	q'' &= sq + 2q^3\ ,\ \ (')=\frac{\d}{\d s}; \hspace{0.7cm}  q(s) \sim \textnormal{Ai}(s)\ , \ \ s\to+\infty.  
\end{align*}

For the Laguerre Unitary Ensemble (LUE) of positive definite matrices, the analogous question deals with the fluctuations of the {\em smallest} eigenvalues; in this case the origin $z=0$ of the spectrum is a ``hard-edge'' because the matrices are conditioned to be positive definite. Tracy and Widom  also connected these fluctuations to a special solution of the Painlev\'e\ III equation \cite{TWIII} (see also \cite{GirottiBessel} for a different direct derivation).\smallskip

The universal character of these fluctuations is encoded in  the determinantal structure of the correlation functions; in both cases these distributions are obtained from the Fredholm determinant of a kernel. To prove these results (cf. \cite{KuijU} for a recent review on the subject) it is sufficient to show that the correlation kernels, in a suitable scaling, tend to a special form; for example the Airy kernel in the GUE case or the Bessel$_\nu$ kernel in the LUE case.

It is then a fundamental step to identify the possible types of kernels occurring in the scaling limit. A general question in the study of universality issues related to multi--matrix models (as opposed to single-matrix models) is whether they exhibit, in the suitable scaling limit, different types of statistical behaviors for their eigenvalues;  this can be addressed by investigating their limiting kernels. The literature on the subject is ever growing and we mention \cite{AB, ABK, ABKN, AIK, AKW, BZ, DK, DKMo}. The present work is precisely addressing the question of limiting kernels (thus leading to addressing fluctuations in a future publication) for a multi-matrix model that naturally generalizes the LUE; the model shall be termed ``Cauchy-chain matrix model". The Cauchy two-matrix model was  introduced in \cite{BGS1}, as a random matrix  model defined in terms of a probability measure on the space of pairs $M_1, M_2$ of $n\times n$ {\em positive definite} Hermitian matrices. We now consider an extension of the setting to an arbitrary number $p$ of positive definite 
Hermitian
 matrices $M_1,\dots, M_p$.  Their joint probability distribution function  
depends on the choice of $p$ scalar functions $U_j: \R_+ \to \R$, $j=1,\dots p$,  
called the {\em potentials}, and is defined as 
\begin{equation}\label{introchain}
	\d\mu(M_1,\dots, M_p) =c \frac{{\rm e}^{-\textnormal{tr} \sum_{j=1}^p U_j(M_j)}}{\prod_{j=1}^{p-1} \det (M_j + M_{j+1})^n}\,\d M_1\cdot\ldots\cdot\d M_p,
\qquad \d M=\prod_{j<k}\d\Re M_{jk}\,\d\Im M_{jk}\prod_{\ell}\d M_{\ell\ell}
\end{equation}

The model under study is an instance of a ``multi-matrix model''; a different one which is also actively studied  was introduced in \cite{EM}.
The difference consists in the choice of interaction between subsequent matrices in the chain:  instead of $\det(M_1+M_2)^{-n}$,  it was the exponential interaction ${\rm e}^{-\tau\, \textnormal{tr}(M_1M_2)}$ commonly known  as the ``Itzykson-Zuber'' (IZ) interaction.\smallskip

Following  \cite{EM} we shall show here that the eigenvalues of the $p$ matrices constitute what is known as a ``multi-level'' determinantal point field; the  correlation functions are computed in terms of determinants constructed from certain biorthogonal polynomials (see Section \ref{results}).\smallskip

The present paper has the following main goals: 
\begin{enumerate}
\item formulate the general properties of the model with $p$--matrices in  Cauchy interaction \eqref{introchain};
\item introduce the relevant biorthogonal polynomials (Definition \ref{bipoly}) and express them in terms of the solution of a Riemann-Hilbert problem (Theorem \ref{Chainthm});
\item express all kernels of the correlation functions in terms of the solution of the problem above (Theorem \ref{theo2});
\item for a simple choice of potentials, we study the correlation function in the scaling limit near the origin;  we complete the analysis for $p=3$ but indicate how it can be extended to $p=4,5,6$.
\item the limiting scaling fields can be expressed in terms of special functions, the Meijer-G functions. The method allows us to extend (at least conjecturally) the resulting formul\ae\ to the Cauchy-chain of arbitrary length $p$ (Definition \ref{defGjl}, Conjecture \ref{conjpchain} and Theorem \ref{thmmain1}).
\item we show how, in suitable limits, the limiting statistics at the origin of the $p$--chain decouples into two independent chains (Theorem \ref{chainsepn}). 
\end{enumerate}

The results above allow one to  express the joint fluctuation statistics of the smallest eigenvalues of the matrices in the chain in terms of a suitable Fredholm determinant with a matrix-valued kernel constructed from Definition \ref{defGjl}. 
In the next section we introduce the necessary notation to formulate the results in a precise form. The proofs of these results constitute the remainder of the paper.

\section{Statement of results}\label{sec1}
\label{results}
Consider the space $\mathcal{M}_+^p(n),p,n\in\mathbb{Z}_{\geq 2}$ consisting of $p$-tuples $(M_1,\ldots,M_p)$ of $n\times n$ positive-definite Hermitian matrices $M_j$.
Equipped with the probability measure \eqref{introchain}
the probability space  $(\mathcal{M}_+^p(n),\d\mu)$ is  referred to as the Cauchy chain-matrix model. Here, 
the external potentials $U_j:(0,\infty)\rightarrow\mathbb{R}$ are chosen so that
\begin{equation*}
\liminf_{x\to+ \infty}	\frac{U_j(x)}{\ln x}=+\infty,\hspace{0.75cm} 
\ \ \ -\limsup_{x\downarrow 0} \frac {U_j(x)}{\ln x} = a_j,
\end{equation*}
with parameters $a_j\in\mathbb{R}$ which satisfy
\begin{equation}\label{constra}
	a_{k\ell}\equiv \sum_{j=k}^{\ell}a_j>-1,\ \ \forall\,1\leq k\leq\ell\leq p.
\end{equation}
The reason for the  constraint \eqref{constra} is simply that the measure \eqref{introchain} be normalizable. Consider now the weight functions $\eta_p(x,y),\ p\geq 2$ on $\mathbb{R}^2_+$, given by
\begin{eqnarray*}
	\eta_2(x,y) &=& \frac{e^{-U_1(x)-U_2(y)}}{x+y},\\
	\eta_p(x,y) &=&\int_0^{\infty}\cdots\int_0^{\infty}\frac{e^{-U_1(x)}}{x+\xi_1}\left(\frac{e^{-\sum_{j=2}^{p-1}U_j(\xi_{j-1})}}{\prod_{j=1}^{p-3}(\xi_j+\xi_{j+1})}\right)\frac{e^{-U_p(y)}}{\xi_{p-2}+y}\,\d\xi_1\cdot\ldots\cdot\d\xi_{p-2},\ \ p\geq 3.
\end{eqnarray*}
The natural generalization of the biorthogonal polynomials introduced in \cite{BGS1} to general $p\geq 2$ is then given by:
\bd\label{bipoly} The monic (Cauchy) biorthogonal polynomials $\{\psi_n(x),\phi_n(x)\}_{n\geq 0}$ are defined by the requirements
\begin{eqnarray}
	\int_0^{\infty}\int_0^{\infty}&\&  \psi_n(x)\phi_m(y)\eta_p(x,y)\,\d x\d y=h_n\delta_{nm}\label{poly}\\
&\& \psi_n(x) = x^n+\mathcal{O}\left(x^{n-1}\right),\hspace{0.5cm}x\rightarrow\infty;\ \ \qquad 
	\phi_n(x) = x^n+\mathcal{O}\left(x^{n-1}\right),\hspace{0.5cm}x\rightarrow\infty.\nonumber
\end{eqnarray}
\ed
The pair $\{\psi_n(x),\phi_n(x)\},n\geq 1$ can always (see. e.g. \cite{MMN}) be constructed in terms of the {\it bimoment matrix} $I=[I_{j\ell}]_{j,\ell=0}^{n-1}$ with
\be\label{i:2} 
	I_{j\ell} = \int_0^{\infty}\int_0^{\infty}x^jy^{\ell}\eta_p(x,y)\,\d x\d y,\hspace{0.5cm}j,\ell\geq 0
\ee 
The convergence of the multiple integrals $I_{j\ell}$ also mandates condition \eqref{constra} and it is here simply a statement that allows the application of Fubini's theorem on the iterated integral in any order.  In terms of \eqref{i:2}, the biorthogonal polynomials can be written as 
\be
\label{bipolyformula}
\psi_n(x) = \frac {1}{\Delta_n}\det \big[{ I_{j\ell}\,|\, x^j}\big]_{j,\ell=0}^{n,n-1},\hspace{1cm}
\phi_n(y) = \frac {1}{\Delta_n}\det \le[{ I_{j\ell}\atop y^\ell}\ri]_{j,\ell=0}^{n-1,n};\hspace{0.5cm}\Delta_n=\det[I_{j\ell}]_{j,\ell=0}^{n-1}.
\ee
It is clear that the existence of the sequence of polynomials requires that all the principal minors of the bimoment matrix $I_{j\ell}$ be nonzero. More is true, in fact, as in the given case \eqref{introchain} of the Cauchy interaction they are known to be {\em positive}.
\bp 
\label{proppos}
All moment determinants $\Delta_n=\det[I_{j\ell}]_{j,\ell=0}^{n-1}$ are strictly positive, i.e. $\Delta_n>0$ for all $n\geq 1$.
\ep
\begin{proof} As observed in \cite{BGS1}, the Cauchy kernel $K(x,y)=\frac{1}{x+y}$ is totally positive on $\mathbb{R}^2_+$. But total positivity is stable under convolution \cite{KarlinBook}, thus  $\eta_p(x,y)$ is totally positive and therefore $\Delta_n>0$.
\end{proof}

\subsection{Part I: general structure}
We shall now describe all correlation functions in terms of the solution of a Riemann--Hilbert problem (RHP); this is conceptually parallel to the case of the unitary ensemble, see for example \cite{M1}. In the following we shall use $\chi_A$ for the indicator function of a set $A$.

\begin{problem}\label{ChainRHP} Let $W_{2j+1}(x)\equiv U_{2j+1}(x)$ for $x> 0$ and $W_{2j}(x)=U_{2j}(-x)$ for $x< 0$. Determine the piecewise analytic $(p+1)\times(p+1)$ matrix valued function $\Gamma(z)\equiv \Gamma(z;n)
=\big[\Gamma_{j\ell}(z;n)\big]_{j,\ell=1}^{p+1}
$ such that
\begin{itemize}
 \item $\Gamma(z)$ is analytic in $\C\backslash\R$
 \item $\Gamma(z)$ admits boundary values $\Gamma_{\pm}(z)$ for $z\in\mathbb{R}\backslash\{0\}$ which are related via
\be\label{Gammajumps}
\Gamma_+(z)=\Gamma_-(z)\le[\begin{array}{ccccc}
                        1 & w_1(z) & 0 & &  0 \\
			0 & 1 & w_2(z) &   & 0 \\
			  & 0 & 1 &\ddots  & 0\\
			  & &\ddots & \ddots  & w_p(z) \\
			0 & 0& & 0&  1   \\
                       \end{array}\ri],\hspace{0.5cm}z\in\mathbb{R}\backslash\{0\}.
\ee
Here,
\begin{equation*}
 	w_j(z) = e^{-W_j(z)}\chi_{(-1)^{j+1}\R_+}(z)
\end{equation*}
and the orientation of the jump contour is as shown in Figure \ref{fig1} below.
\begin{figure}[tbh]
\begin{center}
\includegraphics[width=0.4\textwidth]{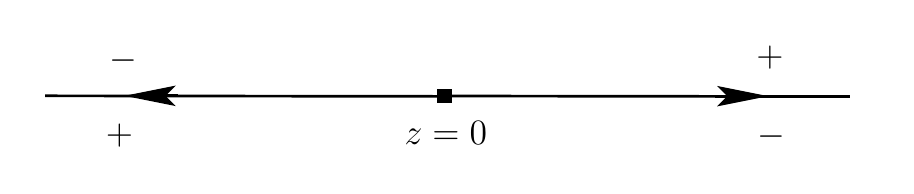}
\end{center}
\caption{The jump contour for $\Gamma(z)$ with fixed orientation: the half-ray $[0,\infty)$ is oriented towards $+\infty$ whereas $(-\infty,0]$ is oriented towards $-\infty$.}
\label{fig1}
\end{figure}
\item The columns of $\Gamma(z)$ have the following singular behavior near $z=0$;
\be\label{Gammasing}
	\Gamma_{\bullet,1}(z)=\mathcal{O}(1),\ \ z\rightarrow 0
\ee
and the precise behavior of the subsequent columns $\Gamma_{\bullet,\ell+1}(z)$ is the same as the behavior of the iterated Cauchy transforms 
\be
\label{zerobeh}
\mathcal{C}_{\ell+1} (z) = \int_0^1\cdots\int_0^1\left(\prod_{j=1}^{\ell-1}\frac{x_j^{a_j}}{x_j-x_{j+1}}\right)\frac{x_{\ell}^{a_{\ell}}}{x_{\ell}-z}\,\d x_1\cdot\ldots\cdot\d x_{\ell},\ \ 1\leq \ell\leq p
\ee
as $z\rightarrow 0$ (compare Remark \ref{zerobehrem} below for further clarification).
\item As $z$ tends to infinity we have the asymptotic behavior
  \be
\Gamma(z) =\big(I  + \mathcal O\left(z^{-1}\right)\big)\le[\begin{array}{ccccc}
z^n&&&&0\\
&  1& &&\\
&&\ddots &&\\
&&&1&\\
0&&&&z^{-n}
\end{array}\ri]
\label{Gammainfty}
\ee
\end{itemize}
\end{problem}
\br
\label{zerobehrem}
We preferred to state the behavior at the origin in a slightly cryptic form \eqref{zerobeh} rather than explicitly because it would entail too many case distinctions; in general, the behavior of iterated Cauchy transforms as in \eqref{zerobeh} near $z=0$ follows from Chapter 1, section 8.6 of \cite{Ga}. For example; 
\begin{enumerate}
\item 
if all $a_j$ are positive, then all columns are $\mathcal O(1)$;
\item if all $a_j=0$ then the $\ell$-th column behaves like $\mathcal O((\ln z)^{\ell-1})$;
\item  if all the $a_j$ are negative (but still with condition \eqref{constra} in place), then the $\ell$-th column has behavior $\mathcal O(|z|^{a_{1,\ell-1}})$.
\end{enumerate}
 The problem arises when trying to describe compactly all possible cases where the exponents can be positive, negative or zero.
\er

The solvability issue of the RHP \ref{ChainRHP} and the connection to the biorthgonal polynomials $\{\psi_n(x),\phi_n(x)\}_{n\geq 0}$ is addressed in the following Theorem, our first result.
\bt
\label{Chainthm}
The Riemann-Hilbert problem \ref{ChainRHP} for $\Gamma(z)=[\Gamma_{j\ell}(z;n)]_{j,\ell=1}^{p+1}$  has a unique solution if and only if $\Delta_{n}\neq 0$. If $\Gamma(z)$ is the solution of the problem, then 
\be\label{Gammaid}
\psi_n(z)= \Gamma_{11}(z;n)\ ,\qquad \phi_n(z)=(-1)^{n(p+1)}\Gamma_{p+1,p+1}^{-1}\big((-1)^{p+1}z;n\big).
\ee
\et
\br
The assumption $\Delta_n\neq 0$ of course applies in our case in view of Proposition \ref{proppos} {\em if the potentials $U_j$ are real}; however one may also want to consider more general settings in Theorem \ref{Chainthm} where the potentials are complex-valued (of course this would undermine any probabilistic application).
\er
We now turn our attention towards eigenvalue correlations. In \cite{EM}, Eynard and Mehta analyzed the Itzykson-Zuber chain of matrices, defined through the probability measure
\begin{equation*}
	\d\nu(M_1,\ldots,M_p)\propto \exp\left[-\textnormal{tr}\left(\sum_{j=1}^pU_j(M_j)-\sum_{j=1}^{p-1}\tau_jM_jM_{j+1}\right)\right]\d M_1\cdot\ldots\cdot\d M_p
\end{equation*}
on the real vector space of $n\times n$ Hermitian matrices with coupling constants $\tau_j\in\mathbb{R}$. They proved that a general correlation function for the Itzykson-Zuber chain can be written in closed determinantal form. But for this to work, the precise form of the interaction was not used at all.
 What is important for the determinantal reduction is the fact that in both models, Itzykson-Zuber and Cauchy, the underlying distribution functions are of the form
\begin{equation*}
	\d\lambda(M_1,\ldots,M_p)\propto e^{-\textnormal{tr}\sum_{j=1}^pU_j(M_j)}\prod_{j=1}^{p-1}I_j(M_j,M_{j+1})\,\d M_1\cdot\ldots\cdot\d M_p
\end{equation*}
with the interaction functions
\begin{equation*}
	I_j(A,B) = \begin{cases}
	e^{\tau_j\textnormal{tr}(AB)},&A,B\ \textnormal{Hermitian}\hspace{3cm} {\it Itzykson-Zuber}\\
	\det(A+B)^{-n},&A,B\ \textnormal{positive-definite Hermitian}\ \ \ \ {\it Cauchy},
	\end{cases}
\end{equation*}
which are invariant under unitary conjugations $I_j(A,B)=I_j(UA\overline{U}^T,UB\overline{U}^T)$. In either model we can then integrate out the angular variables with the help of a generalized Harish-Chandra formula: there exists a function $F(x,y)$ such that for any diagonal matrices $X=\textnormal{diag}[x_1,\ldots,x_n]$ and $Y=\textnormal{diag}[y_1,\ldots,y_n]$ we have 
\begin{equation*}
	\int_{\mathcal{U}(n)}I(X,UY\overline{U}^T)\,\d U \propto\frac{\det\big[F(x_j,y_k)\big]_{j,k=1}^n}{\Delta(X)\Delta(Y)},\ \ \ \Delta(X)=\prod_{i<j}(x_j-x_i).
\end{equation*}
This is the crucial step for the reduction to a biorthogonal polynomial ensemble and thus the result of \cite{EM} for the corresponding correlation function can serve as our guideline. To be more precise, consider the probability density for the eigenvalues of all $p$ matrices
\begin{equation}\label{corr1}
	\mathcal{P}\big(\{x_{1j}\}_{j=1}^n,\ldots,\{x_{pj}\}_{j=1}^n\big)=\frac{1}{\mathcal{Z}_n}\Delta(X_1)\Delta(X_p)e^{-\sum_{m=1}^p\sum_{j=1}^nU_m(x_{mj})}\prod_{\alpha=1}^{p-1}\det\big[K(x_{\alpha i},x_{\alpha+1,k})\big]_{i,k=1}^n
\end{equation}
with the Vandermonde determinants $\Delta(X_k) = \prod_{i<j}(x_{k j}-x_{k i})$, the Cauchy kernel $K(x,y)=\frac{1}{x+y}$ and the partition function
\begin{equation*}
	\mathcal{Z}_n=\int_{\R_+^n}\cdots\int_{\R_+^n}\Delta(X_1)\Delta(X_p)\exp\left[-\sum_{m=1}^p\sum_{j=1}^nU_m(x_{mj})\right]\prod_{\alpha=1}^{p-1}\det\big[K(x_{\alpha i},x_{\alpha+1,k})\big]_{i,k=1}^n
\prod_{j=1}^p\prod_{\ell=1}^n\d x_{j\ell}.
\end{equation*}
Identity \eqref{corr1} is a direct adjustment of formula $(1.5)$ of \cite{EM} to the given Cauchy matrix-chain, moreover the $(\ell_1,\ldots,\ell_p)$-point correlation function equals, see formula $(1.6)$ in loc.cit,
\begin{align}\label{corr2}
	\mathcal{R}^{(\ell_1,\ldots,\ell_p)}&\big(\{x_{1j}\}_{j=1}^{\ell_1},\ldots,\{x_{pj}\}_{j=1}^{\ell_p}\big)=\left[\prod_{j=1}^p\frac{n!}{(n-\ell_j)!}\right]\int_{\R_+^{n-\ell_1}}\cdots\int_{\R_+^{n-\ell_p}}\mathcal{P}\big(\{x_{1j}\}_{j=1}^n,\ldots,\{x_{pj}\}_{j=1}^n\big)\\
	&\times\prod_{j=1}^p\prod_{m_j=\ell_j+1}^n\d x_{jm_{j}}.\nonumber
\end{align}
Introduce the collection of functions $\{\Psi_{\ell n}(x),\Phi_{\ell m}(x)\}_{\ell=1}^p$ for $m,n\geq 0$ and $x>0$, given by
\beas
  \Psi_{1n}(x) =\,\psi_n(x)e^{-\frac{1}{2}U_1(x)},\hspace{0.5cm} \Psi_{\ell n}(x) &\&= \int_0^{\infty}\Psi_{\ell-1,n}(y)w_{\ell-1}(y,x)\,\d y,\hspace{0.5cm}
  \qquad 
  \ell=2,\ldots,p\\
  \Phi_{pm}(x) =\,\phi_m(x)e^{-\frac{1}{2}U_p(x)},\hspace{0.5cm} \Phi_{\ell m}(x) &\&= \int_0^{\infty}\Phi_{\ell+1,m}(y)w_{\ell}(x,y)\,\d y,\hspace{0.5cm}\qquad 
  \ell=1,\ldots,p-1
\eeas
where
\be 
  w_{\ell}(x,y) = \frac{e^{-\frac{1}{2}U_{\ell}(x)-\frac{1}{2}U_{\ell+1}(y)}}{x+y}.
\ee 
Although the functions $\Psi_{\ell n}(x),\Phi_{\ell m}(x)$ are in general non-polynomial, they are orthogonal by construction, namely with \eqref{poly} for $1\leq\ell\leq p$
\be 
	\int_0^{\infty}\Psi_{\ell n}(x)\Phi_{\ell m}(x)\,\d x = \int_0^{\infty}\int_0^{\infty}\psi_n(x)\phi_m(y)\eta_p(x,y)\,\d x\d y=h_n\delta_{nm}.
\ee 
\br 
If the potentials admit analytic continuation outside of $\R_+$ (as it will be the case) then 
 the functions $\{\Psi_{\ell n}(z),\Phi_{\ell m}(z)\}_{\ell=1}^p$ can be analytically extended as well.
\er
Introduce also the {\it kernel functions}, i.e. for $1\leq i,j\leq p$,
\bea
&\&  \mathbb{K}_{ij}(x,y) = H_{ij}(x,y)-E_{ij}(x,y),\hspace{0.5cm} H_{ij}(x,y) = \sum_{\ell=0}^{n-1}\Phi_{i\ell}(x)\Psi_{j\ell}(y)\frac{1}{h_{\ell}}
\\
&\&   E_{ij}(x,y) =\begin{cases}
               0, &\textnormal{for}\ \ \  i\geq j\\
	w_i(x,y),&\textnormal{for}\ \ \ i=j-1\\
	\ds\int_0^{\infty}\cdots\int_0^{\infty}w_i(x,\xi_1)w_{i+1}(\xi_1,\xi_2)\cdots w_{j-1}(\xi_{j-i-1},y)\d\xi_1\cdots\d\xi_{j-i-1},&\textnormal{for}\ \ \ i<j-1.
              \end{cases}\nonumber
\eea
The main result in \cite{EM} - tailored here to the Cauchy chain - shows that the correlation function \eqref{corr1} is equal to
\be\nonumber
	\mathcal{R}\equiv
	\mathcal{R}^{(\ell_1,\ldots,\ell_p)}\big(\{x_{1j}\}_{j=1}^{\ell_1},\ldots,\{x_{pj}\}_{j=1}^{\ell_p}\big)
	= \det\big[\mathbb{K}_{ij}(x_{ir},x_{js})\big]_{i,j=1;\,\substack{r=1,\ldots,\ell_i\\ s=1,\ldots,\ell_j}}^{p}.
\ee 
This identity involves a determinant of size $(\sum_1^p\ell_j)\times(\sum_1^p\ell_j)$, more precisely
\begin{equation}\label{eynardmeh}
	\mathcal{R}=\det\le[\begin{array}{cccc}
\ \ \begin{array}{|c|}\hline
\mathbb{K}_{11}(x_{1r},x_{1s})\\
 {\scriptstyle 1\leq r\leq \ell_1,1\leq s\leq \ell_1}\\ \hline
\end{array} & \begin{array}{|c|}\hline
\mathbb{K}_{12}(x_{1r},x_{2s})\\
{\scriptstyle 1\leq r\leq \ell_1,1\leq s\leq \ell_2}\\ \hline
\end{array} & \cdots & 
\begin{array}{|c|}\hline
\mathbb{K}_{1p}(x_{1r},x_{ps})\\
{\scriptstyle 1\leq r\leq \ell_1,1\leq s\leq \ell_p}\\ \hline
\end{array}\ \ \\
& & & \\
\ \ \begin{array}{|c|}\hline
\mathbb{K}_{21}(x_{2r},x_{1s})\\
{\scriptstyle 1\leq r\leq \ell_2,1\leq s\leq \ell_1}\\ \hline
\end{array}& 
\begin{array}{|c|}\hline
\mathbb{K}_{22}(x_{2r},x_{2s})\\
{\scriptstyle 1\leq r\leq \ell_2,1\leq s\leq \ell_2}\\ \hline
\end{array} &\cdots & 
\begin{array}{|c|}\hline
\mathbb{K}_{2p}(x_{2r},x_{ps})\\
{\scriptstyle 1\leq r\leq \ell_2,1\leq s\leq \ell_p}\\ \hline
\end{array}\ \ \\
\vdots & \vdots & \ddots & \vdots\\
\ \ \begin{array}{|c|}\hline
\mathbb{K}_{p1}(x_{pr},x_{1s})\\
{\scriptstyle 1\leq r\leq \ell_p,1\leq s\leq \ell_1}\\ \hline
\end{array} & 
\begin{array}{|c|}\hline
\mathbb{K}_{p2}(x_{pr},x_{2s})\\
{\scriptstyle 1\leq r\leq \ell_p,1\leq s\leq \ell_2}\\ \hline
\end{array} & \cdots &
\begin{array}{|c|}\hline
\mathbb{K}_{pp}(x_{pr},x_{ps})\\
{\scriptstyle 1\leq r\leq \ell_p,1\leq s\leq \ell_p}\\ \hline
\end{array}\ \ 
\end{array}\ri]_{(\sum\ell_i)\times(\sum\ell_i)}
\end{equation}
where each block $\mathbb{K}_{ij}(x_{ir},x_{js})$ is a matrix of size $\ell_i\times \ell_j$. If the eigenvalues $\{x_{jr}\}$ of a matrix $M_j$ are not observed, i.e. if $\ell_j=0$, then no row or column corresponding to them appears in \eqref{eynardmeh}. Identity \eqref{eynardmeh} shows how general correlation functions in the Cauchy chain model can be computed explicitly for finite $n$ in terms of (Cauchy) biorthogonal polynomials. However, in order to analyze the behavior of the eigenvalue correlations asymptotically as the sizes $n$ of matrices tend to infinity, it is preferable to express the kernel functions in terms of the solution of the RHP stated in Definition \ref{ChainRHP}. This connection constitutes our second main result: rewrite \eqref{eynardmeh} as 
\begin{equation}\label{red}
	\mathcal{R}=\left(\prod_{j=1}^p\prod_{\alpha_j=1}^{\ell_j}e^{-U_j(x_{j\alpha_j})}\right)\,
 \det\big[\mathbb{M}_{ij}(x_{ir},x_{js})\big]_{i,j=1;\,\substack{r=1,\ldots,\ell_i\\ s=1,\ldots,\ell_j}}^{p}.
%
\end{equation}
where $\mathbb K$ and $\mathbb M$ are related as follows
\begin{equation}
	 \mathbb{K}_{j\ell}(x,y)=e^{-\frac{1}{2}U_j(x)-\frac{1}{2}U_{\ell}(y)}\,\mathbb{M}_{j\ell}(x,y),\ \ x,y>0.
	 \label{MK}
\end{equation}
More explicitly and for future reference, we have
\bea
\label{lrow}
	\mathbb{M}_{p1}(x,y) &\&= \sum_{\ell=0}^{n-1}\phi_{\ell}(x)\psi_{\ell}(y)\frac{1}{h_{\ell}},\hspace{0.5cm}\mathbb{M}_{p,i+1}(x,y)=\int_0^{\infty}\mathbb{M}_{pi}(x,z)\frac{e^{-U_i(z)}}{z+y}\,\d z,\ \ \ i=1,\ldots,p-1
\\
\label{row}
	\mathbb{M}_{i,i+1}(x,y)&\& =\int_0^{\infty}\mathbb{M}_{i+1,i+1}(z,y)\frac{e^{-U_{i+1}(z)}}{x+z}\,\d z-\frac{1}{x+y},\hspace{0.5cm}i=1,\ldots,p-1
\\
\label{row1}
	\mathbb{M}_{ij}(x,y)&\& =\int_0^{\infty}\mathbb{M}_{i+1,j}(z,y)\frac{e^{-U_{i+1}(z)}}{x+z}\,\d z,
	\hspace{0.5cm}i=1,\ldots,p-1,\ \ \ j=1,\ldots,p, \ i+1\neq j.
\eea
In particular all kernels can be constructed from $\mathbb{M}_{p1}(x,y)$ by means of suitable transformations and we notice that $\mathbb{M}_{p1}(x,y)$ is a reproducing kernel, i.e.
\be 
	\int_0^{\infty}\int_0^{\infty}\mathbb{M}_{p1}(x,\xi_1)\mathbb{M}_{p1}(\xi_2,y)\eta_p(\xi_1,\xi_2)\,\d\xi_1\d\xi_2 = \mathbb{M}_{p1}(x,y),\hspace{0.5cm}\int_0^{\infty}\int_0^{\infty}\mathbb{M}_{p1}(x,y)\eta_p(y,x)\,\d x\d y=n.
\ee 
The connection to the solution of the RHP for $\Gamma=\Gamma(z;n)$ in Definition \ref{ChainRHP} is as follows
\bt\label{theo2} Let $x,y>0$. The correlation kernels \eqref{lrow},\eqref{row} and \eqref{row1} equal
\be\label{rep}
  \mathbb{M}_{j\ell}(x,y) = \frac{(-1)^{\ell-1}}{(-2\pi i)^{j-\ell+1}}\left[\frac{\Gamma^{-1}(w;n)\Gamma(z;n)}{w-z}\right]_{j+1,\ell}
  \Bigg|_{ {w=x(-1)^{j+1}\atop  z=y(-1)^{\ell-1}}},\hspace{0.5cm}1\leq j,\ell\leq p
\ee
where the choice of limiting values $(\pm)$ in the matrix entry upon evaluation at $w=x(-1)^{j+1},z=y(-1)^{\ell-1}$ is immaterial.
\et
\subsection{Part II: asymptotic eigenvalue distribution near the origin in the $p$-Laguerre case.}
After establishing the general results in Theorem \ref{Chainthm} and \ref{theo2} we intend to analyze the correlation kernels asymptotically as $n\rightarrow\infty$ for the specific choice of Laguerre
--type
 weights, i.e. for the choice of external potentials
\begin{equation}\label{potch}
	U_j(x)=NV_j(x)-a_j\ln x,\ \ a_j>-1:\ \ a_{k\ell}=\sum_{j=k}^{\ell}a_j>-1;\ \ \ \lim_{x\rightarrow+\infty}\frac{V_j(x)}{\ln x}=+\infty
\end{equation}
with $V_j(x)$ real-analytic on $[0,\infty)$ and $N$ independent. The parameter $N>0$ is a scaling parameter: in the study of the large-size limit $n\to \infty$ it is chosen in such a way that $\frac n N \to T\in \R_+$. In the asymptotic study here we shall simply choose $n=N$ and therefore $T=1$.\smallskip

We derive an asymptotic solution of the RHP for $\Gamma=\Gamma(z;n)$ as $n\rightarrow\infty$ through the nonlinear steepest descent method of Deift and Zhou, cf. \cite{DZ,DKM,DKMVZ}. As opposed to the Riemann-Hilbert analysis carried out in \cite{BGS2}, the choice of potential \eqref{potch} allows for an overlap of the supports of the equilibrium measures (compare section \ref{part2} below). 
Hence we face the necessity to carry out a local analysis near the overlap point and we consider the construction of the new parametrix the main technical contribution of the paper to the nonlinear steepest descent literature. The relevant parametrix is constructed for the general $(p+1)\times (p+1)$ RHP using Meijer G-functions. These special functions have appeared recently in a variety of problems \cite{BZ,AIK,AKW,AB,ABKN,ABK} analyzing the statistics of singular values of products of Ginibre random matrices. In particular, they also appeared in the context of the Cauchy-Laguerre two-matrix model, i.e. with $p=2$ in \eqref{introchain} and $
	U_j(x)=Nx-a_j\ln x,\ \ a_1,a_2>-1,\ a_1+a_2>-1.$
In fact, it was shown in \cite{BGS3} that the biorthogonal polynomials in Definition \ref{bipoly} can be written explicitly as Meijer G-functions. Thus for the Cauchy-Laguerre two chain one can analyze the correlation kernels asymptotically without any Riemann-Hilbert analysis. However this feature does not seem to carry over to general $p\geq 2$, which motivates our current initiative based on nonlinear steepest descent techniques.
In order to state our results for the scaling analysis, we first pose the following Definition:
\bd[Meijer-G random point field for $p$-chain] \label{defGjl}
Let $\{a_j\}_{j=1}^p\subset\mathbb{R}$ satisfy the condition \eqref{constra} 
with $a_{10}\equiv 0$
and define the polynomial $K(u)$
\be
 K(u)  = (-1)^p \prod_{s=0}^{p} (u - a_{1s})\ . \label{Kdef}
\ee
 The Meijer-G random point field consists of the (multi-level) determinantal random point field of $p$ point fields in $\R_+$  with correlation functions 
 \be
 \label{Gcorr}
	\mathcal{G}^{(\ell_1,\ldots,\ell_p)}\left(\xi_{11},\ldots,\xi_{1\ell_1};\ldots;\xi_{p1},\ldots,\xi_{p\ell_p}\right)= \det\big[\mathcal {G}^{(p)}_{ij}(\xi_{ir},\xi_{js})\big]_{i,j=1;\,\substack{r=1,\ldots,\ell_i\\ s=1,\ldots,\ell_j}}^{p}.
\ee 
with the determinant above analogous to \eqref{eynardmeh}.
 The kernels appearing above are defined as follows:
\begin{align}
	\mathcal G^{(p)}_{j\ell}(\xi,\eta; &\{a_1,\dots, a_p\}) =\frac{1} {(-1)^{\ell} \eta - (-1)^{j} \xi}\label{scalelimit3}\\
	&\times\,\frac{1}{(2\pi i)^2}
 \int_L\int_{\wh L}
 \frac{\prod_{s=0}^{\ell-1}\Gamma(u-a_{1s})}{\prod_{s=\ell}^p\Gamma(1+a_{1s}-u)}
\frac{\prod_{s=j}^p\Gamma(a_{1s}-v)}{\prod_{s=0}^{j-1}\Gamma(1-a_{1s}+v)}
\frac{K(u)-K(v)}{u-v}\,\xi^{v}\eta^{-u}\,\d v\,\d u.\nonumber
\end{align}
Here, the integration contours for $u\in L, v\in \wh L$ are chosen so as to leave all the poles of the integrand in $u,v$ to the left, right and to extend to $\infty$ in the left, right half plane.
Alternatively, and equivalently, we have the formula 
\begin{align}
	\mathcal G^{(p)}_{j\ell}(\xi,\eta;&\{a_1,\dots, a_p\})=\frac{1}{(2\pi i)^2}\int_L\int_{\wh L}\frac{\prod_{s=0}^{\ell-1}\Gamma(u-a_{1s})}{\prod_{s=\ell}^p\Gamma(1+a_{1s}-u)}
	\frac{\prod_{s=j}^p\Gamma(a_{1s}-v)}{\prod_{s=0}^{j-1}\Gamma(1-a_{1s}+v)}\frac{\xi^{v}\eta^{-u}}{1-u+v}\,\d v\,\d u\nonumber\\
&+\sum_{s\in\mathcal{P}\cup\{0\}}
\res_{v=s}\frac{\prod_{s=0}^{\ell-1}\Gamma(1+v-a_{1s})}{\prod_{s=\ell}^p\Gamma(a_{1s}-v)}
	\frac{\prod_{s=j}^p\Gamma(a_{1s}-v)}{\prod_{s=0}^{j-1}\Gamma(1+v-a_{1s})}\frac{\xi^v \eta^{-v}}{ {(-1)^{j} \xi-(-1)^{\ell} \eta}}
\label{Gpsecond}
\end{align}
where now the contours are meant to be small circles around the poles of the integrands, with the circles in the $v$ variable smaller than those in the $u$ variable, 
and where $\mathcal P = \{a_{1\ell},\, 1\leq\ell\leq p\}$.
\ed
We now state our second result,  in the form of a conjecture which is then proven for $p=3$ (and we indicate how to prove it also for $p=4,5,6$ in Remark \ref{remp456}).
\begin{conjecture}[Universality]\label{conjpchain}
For any $p\in\mathbb{Z}_{\geq 2}$, there exists $c_0=c_0(p)>0$ and $\{\varpi_j\}_{j=1}^p$ which depend  on the parameters $\{a_j\}_{j=1}^p$ introduced in \eqref{potch} such that
\begin{equation}\label{conj}
	\lim_{n\rightarrow\infty}\frac{c_0}{n^{p+1}} n^{\varpi_{\ell}-\varpi_j}\mathbb{K}_{j\ell}\left(\frac{c_0}{n^{p+1}}\xi,\frac{c_0}{n^{p+1}}\eta\right)=  
	c_0^{\frac{\varpi_{\ell}-\varpi_j}{p+1}}\xi^{\frac{1}{2}a_j}\eta^{\frac{1}{2}a_{\ell}}
 {\xi^{-a_{1j}} \eta^{a_{1\ell-1}}}
 \mathcal G^{(p)}_{j\ell}(\xi,\eta;\{a_1,\dots, a_p\})
\end{equation}
with $\mathcal G_{j\ell}^{(p)}$ as in Definition \ref{defGjl}.
The limit holds uniformly for $\xi,\eta$ chosen from compact subsets of $(0,\infty)$.
\end{conjecture}
\br
The correlation functions of the kernels on the right side of \eqref{conj} are the same as those of the kernels $\mathcal G^{(p)}_{j\ell}$ \eqref{Gcorr} because the corresponding matrices in the determinants \eqref{eynardmeh} are conjugate of each other by a diagonal matrix.
\er
Conjecture \ref{conjpchain} expresses our belief that the Meijer-G random point field \eqref{scalelimit3} is universal in the scaling limit $z\mapsto z\,c_0n^{-(p+1)}$ within the Cauchy $p$-chain \eqref{introchain} for the choice \eqref{potch}. This expectation is based on a rigorous proof of the following Theorem
\bt
\label{thmmain1}
Conjecture \ref{conjpchain} holds for $p=2,3$ and potentials as in \eqref{CLaguerre}.
\et
The case $p=2$ for the Cauchy-Laguerre chain was addressed completely in \cite{BGS3} without the necessity of a complicated asymptotic analysis because of a lucky occurrence by which the biorthogonal polynomials for any $n$ can be expressed {\em exactly} in terms of Meijer G-functions, and therefore the asymptotic analysis follows from relatively simple estimates on their integral representations. Clearly, we have verified that our conjecture matches the existing result, see Section \ref{comparep2}. 
In addition, in Section \ref{comparep3}, we show that the limiting kernel of Kuijlaars and Zhang (\cite{BZ}, Theorem $5.3.$) which appears in the analysis of the singular values of products of Ginibre random matrices, is exactly one of the kernels in the family \eqref{Gpsecond}.
\bigskip

We have stated the Conjecture \ref{conjpchain}  based on our rigorous analysis of the Cauchy-Laguerre $p=3$ chain with the choice of external potentials
\begin{equation}\label{CLaguerre}
	U_j(x)=Nx-a_j\ln x,\ \ a_j>-1:\ \ a_{k\ell}=\sum_{j=k}^{\ell}a_j>-1,\ \ \ \forall\,1\leq k\leq\ell\leq p.
\end{equation}
Indeed, we will solve the relevant $4\times 4$ Riemann-Hilbert problem asymptotically and prove \eqref{conj} with explicit values for $c_0$ and $\varpi_j$. 
The reader with some experience in the Deift-Zhou steepest-descent analysis will know that the method relies on two main hinges:
\begin{itemize}
\item the construction of appropriate {\em equilibrium measures} representing the asymptotic densities of eigenvalues of the matrices of the chain (replacing the Wigner semicircle law for GUE or the Mar$\check c$henko--Pastur law);
\item the construction of local parametrices near the points where the equilibrium densities vanish or diverge.
\end{itemize}
For the first point it is known that the equilibrium measures minimize a certain functional \cite{BB} and that their Stieltjes transforms then solve a certain algebraic equation that can be viewed as a Riemann surface (algebraic plane curve). The logic can be turned on its head in special cases: one can (and often does), based on a body of experience and heuristic expectations, {\em postulate} an appropriate Riemann-surface-Ansatz  and subsequently verify that the Ansatz leads to the appropriate equilibrium measures by verifying a certain set of equalities and inequalities that characterize the equilibrium measures.
We have followed this second route and postulated the Ansatz of the algebraic equation \eqref{spec}, and then verified the appropriate necessary and sufficient properties in Proposition \ref{behave}. Although not completely satisfactory from a general point of view, the approach is quite effective in these special cases. Given that this is not the main focus of the paper, it would be however too long and possibly even too vague to try and formulate a clear set of guiding principles that lead to an effective Ansatz. We did, nonetheless, follow the same principles to postulate the algebraic curves for the cases $p=4,5,6$ in Remark \ref{remp456}; in these cases we did not provide the corresponding analog of Proposition \ref{behave} because we are not using those results in the sequel. We believe that the reader, if interested, can easily adapt the idea of Proposition \ref{behave} since it amounts to a straightforward exercise in calculus.\smallskip

For the second point the crux of the matter is the construction of a {\em local parametrix}, $ \mathbb G(\z) $, that solves a suitable local model RHP near the origin. We shall detail this construction for general $p\geq 2$ in Section \ref{originparm} in terms of  Meijer G-functions. The connection to the "physical", i.e. spectral variable $z$ of the RHP is carried out  only   for $p=3$ with the specific choice \eqref{CLaguerre}. The main reason for this lies in the use of a (vector) $\mathfrak{g}$-function transformation, which we achieve through the spectral curve method rather than via the analysis of the underlying equilibrium problem. However, as universality theorems have been established in many areas of random matrix theory, we expect the specific choice of the potentials $V_j(z)$ in \eqref{potch} not to violate the scaling behavior near the origin, thus our conjecture \eqref{conj}.\smallskip

The key ingredient for the explicit construction of the vector-equilibrium solution for $p=4,5,6$ is given (without proof) in Remark \ref{remp456}. The reason we cannot fully claim to have proven \eqref{conj} also for $p=4,5,6$ is simply because we are not  providing the necessary error analysis of the final approximation in the Riemann-Hilbert problem. On the other hand we believe that it should be clear to the experienced reader that such a proof can be obtained by simply repeating the steps we are taking now for $p=3$.  

\subsection{Chain separation  in the $p$-chain Meijer-G case}
Consider the $p$--chain Meijer-G random point field of Definition \ref{defGjl}. We refer to the random point fields of the eigenvalues of the three chain as the $(j)$--fields, $j=1,2,3$.
The $(2)$-field interacts with both the $(1)$-field and $(3)$-field. For a longer $p$--chain the $(j)$-field for $2\leq j \leq p-1$, interacts with both the $(j-1)$ and $(j+1)$ fields.\smallskip

In the general chain, the exponent $a_q, 1\leq q \leq p$ measures the strength of the {\em repulsion} of the $(q)$-field from the origin: the larger $a_q$ is, the more suppressed is the empirical statistics of the $(q)$-field at the origin. This simply follows from the observation that the probability measure  $\d\mu $ in \eqref{introchain} is proportional to $\det(M_q)^{a_q}$. 
For the scaling field at the origin, therefore, the $(q)$-field becomes statistically irrelevant as $a_q\to\infty$: thus it is expectable that if $a_1$ or $a_p$ tend to infinity, the corresponding field will disappear and the remaining ones obey the same limiting statistics as the chain of one unit shorter.  
If one of the $a_q$, corresponding to a field  in middle of the chain, tends to infinity, then we should observe that the remaining fields obey the statistics of two independent chains of length $q-1$ and $p-q$, respectively: i.e. the $p$--chain is broken into two independent subchains.\smallskip

The formalization of the above discussion is contained in the following Theorem \ref{chainsepn}; for the case $p=3$ we have either $q=1,3$ or $q=2$; in the former case Theorem \ref{chainsepn} states that the remaining parts of the field obey the same statistics as the $2$--level Meijer-G field obtained in \cite{BGS3}. 
In the latter case, $p=2$, the chain is split into two ``one-chains'' of equal length. In this case we show in Section \ref{onechain} that the $p=1$--chain is nothing but the Bessel field appearing in the scaling limit of the Laguerre Unitary Ensemble.
\bt[Chain separation]
\label{chainsepn}
Let $1\leq q \leq p$ and consider the kernels $\mathcal G^{(p)}_{j\ell} (\z,\eta;\{a_1,\dots, a_q\})$. 
In the limit as $\Lambda=a_q\to\infty$ we have the following behaviors; 
\begin{equation*}
	\Lambda^{p-q+1} \Big[\mathcal G^{(p)}_{j\ell} (\Lambda^{p-q+1}  \z,\Lambda^{p-q+1}  \eta;\{a_1,\dots, a_q\})\Big]_{j,\ell=1}^p  = \le[
	\begin{array}{cc}
	\begin{array}{|c|}\hline
		\ds \mathcal G^{(q-1)}_{j\ell} (\xi, \eta;\{a_1,\dots,a_{q-1}\})\\
		{\scriptstyle 1\leq j,\ell \leq q-1}\\ \hline
	\end{array} & \mathcal{O}(1)\\ 
	&\\
	\mathcal{O}\left(\Lambda^{-1}\right) & \mathcal{O}\left(\lambda^{-1}\right)
\end{array}\ri],
\end{equation*}	
\begin{equation*}
	\Lambda^{q} \Big[\mathcal G^{(p)}_{j\ell} (\Lambda^{q}  \z,\Lambda^{q}  \eta;\{a_1,\dots, a_q\})\Big]_{j,\ell=1}^p = \le[
	\begin{array}{cc}
	\mathcal O(\Lambda^{-1}) & \mathcal O(1)\\ 
	&\\
\mathcal O(\Lambda^{-1}) & \begin{array}{|c|}\hline
	\big(\frac \xi \eta\big)^{a_{1q}}  \mathcal G^{(p-q)}_{j\ell} (\xi, \eta;\{a_{q+1},\dots,a_{p}\}) \\
	{\scriptstyle 1\leq j,\ell \leq p-q}\\ \hline
	\end{array}\end{array}
\ri].\vspace{0.35cm}
\end{equation*}
That is, the $p$-chain random point field split into two independent multi-level random point fields corresponding to two subchains of lengths $q-1,\,p-q$ with scaling at the indicated rates.
In the case that $p-q= q-1$ (i.e. $p$ is odd and $p = 2q-1$) so that the two subchains scale at the same rate, we have 
\begin{equation*}
\Lambda^{q}\Big[\mathcal G^{(p)}_{j\ell} (\Lambda^{q}  \z,\Lambda^{q}  \eta;\{a_1,\dots, a_q\})\Big]_{j,\ell=1}^p  =
\le[
\begin{array}{ccc}
	 \begin{array}{|c|}\hline
	\scriptstyle{\mathcal G^{(q-1)}_{j\ell} (\xi, \eta;\{a_k\}_{k=1}^{q-1})}\\
	 \scriptstyle{1\leq j,\ell \leq q-1}\\ \hline
	\end{array} & \mathcal O(1)& \mathcal O(1)\\
	& &\\
\mathcal O(\Lambda^{-1}) &\mathcal O(\Lambda^{-1}) &\mathcal O(1)\\ 
&&\\
\mathcal O(\Lambda^{-1}) &\mathcal O(\Lambda^{-1}) & \begin{array}{|c|}\hline
\scriptstyle{(\frac \xi \eta)^{a_{1q}}  \mathcal G^{(p-q)}_{j\ell} (\xi, \eta;\{a_k\}_{k=q+1}^p\})}\\
 \scriptstyle{1\leq j,\ell \leq p-q}\\ \hline
	\end{array}
\end{array}\ri],
\end{equation*}
and hence they still  are independent subchains because the correlation functions factorize to leading order.
\et
\br
We would like to offer an explanation regarding the scalings in Theorem \ref{chainsepn}; this is based on the heuristics (see Conjecture \ref{conjpchain}) that for a  chain of length $p$ the scaling of the eigenvalues at the origin is $n^{-p-1}$. The chain separation occurs when one of the exponents $a_q$ in the potentials \eqref{potch} scales as $a_q = n\beta$. 
Then the chain separates into two independent chains of lengths $p-q$ and $q-1$. The $q-1$ subchain should be now scaled by $n^{-q}$; but since the variables $\zeta, \eta$ had been previously scaled as $n^{p+1}$ then the effective scaling in $a_q\propto n$ is $n^{p-q+1}$, exactly as in the latter Theorem. A similar argument explains the scaling of the other subchain.
\er
\br
For the single-matrix case and $a_1$ scaled with $n$ in the Mar$\check c$enko--Pastur density, one also observes that the spectrum gets ``detached'' from the origin. This detachment is the underlying mechanism of the chain separation.
\er
We conclude this introduction with a short outline for the remainder of the article. In section \ref{part1} we prove Theorems \ref{Chainthm} and \ref{theo2}. After that section \ref{part2} contains the most technical part of the paper, the rigorous asymptotical analysis of the Cauchy-Laguerre three matrix chain \eqref{CLaguerre}: this includes in particular the construction of the vector $\mathfrak{g}$-function, a series of explicit transformations (including the construction of the origin parametrix) and a, somewhat tedious, error analysis at the end. After that we are ready to prove Theorem \ref{theorig} which forms an intermediate step on the way to Theorem \ref{thmmain1}. Followed by that, we complete the proof of Theorem \ref{thmmain1} by deriving double contour integral representations for the entries under scrutiny in \eqref{scalelimit}. This step is again carried out for the general $p\geq 2$ chain and it allows us to derive Theorem \ref{chainsepn}.

\section{Part I. Correlation kernels for finite $N$: proof of Theorems \ref{Chainthm} and \ref{theo2}}\label{part1}
%
\bl
\label{detlemma}
The determinant of $\Gamma(z)$ is constant and equal to $1$.
\el
\begin{proof}
The usual argument is that $\det \Gamma(z)$ has no jumps in $\C\backslash\{0\}$ with a possible isolated singularity at the origin. 
Then one estimates the possible growth near $z=0$; if $\det \Gamma(z) = o(z^{-1})$, the possible singularity at $z=0$ has to be removable. Thus $\det\Gamma(z)$ is an entire function  that tends to $1$ at infinity (compare \eqref{Gammainfty}) and hence identically equal to $1$ by Liouville's theorem.\smallskip

However for negative $a_j$'s, we have $\det\Gamma(z) = \mathcal O(z^{\sum_{\ell=1}^pa_{1\ell}}),\,z\rightarrow 0$ but from \eqref{constra} it only follows that $\sum_{\ell=1}^pa_{1\ell}>-p$, hence the above argument fails. To cover also these cases we use a different argument:  if $-p< \sum_{\ell=1}^pa_{1\ell}\leq -q,\  q\in \N$  we can only argue that $\det \Gamma(z) = Q(z)/z^q$ with $Q(z)$ a monic polynomial of degree $q$ (so that $\det \Gamma(z) \to 1$ as $z\rightarrow\infty$). Suppose $q\geq 1$ and let $z_0\in\mathbb{C}$ be a root of $Q(z)$; then there is a linear combination of the rows $\Gamma_{1,\bullet}(z),\dots,\Gamma_{p+1,\bullet}(z)$ of $\Gamma(z)$ such that $r(z)= \sum_{j}r_j\Gamma_{j,\bullet}(z)$ vanishes at $z=z_0$  but is otherwise not identically zero (if $z_0\in \R$, since we have assumed the potential real-analytic, a simple argument  shows that both boundary values of $r(z)$ vanish at $z=z_0$). Then $r(z)/(z-z_0)$ is a bounded row-solution of the jump condition \eqref{Gammajumps} which at infinity has the behavior $(\mathcal O(z^{n-1}), \mathcal O(z^{-1}), \dots,\mathcal O(z^{-1}), \mathcal O(z^{-n-1}))$. But this implies that we could add any multiple of $r(z)$ to the first row, therefore altering its entries. But as we shall see in a few moments (without using the unique solvability of the RHP \ref{ChainRHP}) the first row $\Gamma_{1,\bullet}(z)$ contains the polynomial $\psi_n(x)$, which is uniquely determined, compare Proposition \ref{proppos}. Hence we must have $q=0$ and unimodularity of $\Gamma(z)$ follows. 
\end{proof}
\begin{proof}[Proof of Theorem \ref{Chainthm}]
Uniqueness of the solution follows in the standard way. By Lemma \ref{detlemma},  $\det\Gamma(z)$ is an entire function and by \eqref{Gammainfty} with Liouville's theorem, $\det\Gamma(z)\equiv 1$. This shows that the ratio of two solutions, $\Gamma_1(z)$ and $\Gamma_2(z)$, is first well-defined and secondly from \eqref{Gammajumps}, $\Gamma_1(z)\Gamma_2^{-1}(z)$ is analytic in $\C\backslash\{0\}$ with a removable singularity at the origin. Hence by another application
of Liouville's theorem, we have $\Gamma_1(z)\equiv\Gamma_2(z)$.\smallskip

For existence, the jump condition \eqref{Gammajumps} and behavior \eqref{Gammasing}, \eqref{zerobeh} imply that the first column of $\Gamma(z)=\Gamma(z;n)$ must consist of entire functions; on the other hand from the 
asymptotic behavior at infinity, the first column $\Gamma_{\bullet,1}(z)$ of $\Gamma(z)$ consists of polynomials, more precisely
\be
\Gamma_{\bullet,1}(z) = \big(\pi_n(z) ,\psi_{n-1}^{(1)}(z) ,\dots, \psi_{n-1}^{(p)}(z)\big)^T
\ee
where $\pi_n(z)$ is a monic polynomial of exact degree $n$ and 
\be\label{poly1}
\psi_{n-1}^{(j)}(z) = \sum_{m=0}^{n-1}\wh {\psi}_{m}^{(j)}z^{m},\hspace{0.5cm} j=1,\ldots,p
\ee
are polynomials of degree $\leq n-1$ whose coefficients will be determined uniquely later on. The jump condition \eqref{Gammajumps} and 
asymptotics \eqref{Gammainfty} imply the following formul\ae\ for the remaining columns
\bea
 \Gamma_{\bullet,\,\ell+1}(z) &=& {\bf e}_{\ell+1} +\frac{1}{2\pi i}\int_0^{\infty}\Gamma_{\bullet,\ell-}\left((-1)^{\ell+1}w\right)e^{-U_{\ell}(w)}
\frac{\d w}{w+z(-1)^{\ell}},\hspace{0.5cm}1\leq\ell\leq p-1\label{e:1}\\
\Gamma_{\bullet,\,p+1}(z) &=& \frac{1}{2\pi i}\int_0^{\infty}\Gamma_{\bullet,\,p-}\left((-1)^{p+1}w\right)e^{-U_p(w)}
\frac{\d w}{w+z(-1)^p}.\nonumber
\eea
where ${\bf e}_j$ denotes the $j$-th cartesian unit (column) vector
. Here and in the following, all integrals are ordinary Lebesgue integrals, not oriented line integrals. The asymptotic behavior \eqref{Gammainfty} for the $(p+1)^{\textnormal{st}}$ column poses certain conditions on the polynomials $\pi_n(z),\psi_{n-1}^{(j)}(z)$ which we now read off:
\be\label{Gammasys}
\int_0^{\infty}\cdots\int_0^{\infty}\Gamma_{\bullet,1}(w_1)w_p^{\ell}
\frac{e^{-\sum_{j=1}^pU_j(w_j)}}{\prod_{j=1}^{p-1}(w_j+w_{j+1})}\,\d w_1\cdot\ldots\cdot\d w_p=\le[\begin{array}{c}
0\\
-(2\pi i)J_{\ell,2}\\
\vdots \\
-(2\pi i)^{p-1}J_{\ell,p}\\
(-2\pi i)^p(-1)^{(p+1)\ell}\delta_{\ell,n-1}
\end{array}\ri],
\ee
valid for $0\leq\ell\leq n-1$ and where
\be\nonumber
J_{\ell,m} = \int_0^{\infty}\cdots\int_0^{\infty}w_p^{\ell}\frac{e^{-\sum_{j=m}^pU_j(w_j)}}{\prod_{j=m}^{p-1}(w_j+w_{j+1})}\,\d w_m\cdot\ldots\cdot\d w_p,
\hspace{0.5cm}m=1,\ldots,p.
\ee
Let us consider the first row in \eqref{Gammasys}, it reads as
\be
0=\int_0^{\infty}\cdots\int_0^{\infty}\pi_n(w_1)w_p^{\ell}
\frac{e^{-\sum_{j=1}^pU_j(w_j)}}{\prod_{j=1}^{p-1}(w_j+w_{j+1})}\d w_1\cdot\ldots\cdot\d w_p=\int_0^{\infty}\int_0^{\infty}\pi_n(x)y^{\ell}\eta_p(x,y)\,\d x\d y
\ee
and has to hold for any $\ell\in\{0,\ldots,n-1\}$, i.e. $\pi_n(x)$, which is a monic polynomial of exact degree $n$, must be the $n^{\textnormal{th}}$ monic orthogonal polynomial $\psi_n(x)$ subject to \eqref{poly}.
The next $(p-1)$ rows in \eqref{Gammasys} can be written as
\be\label{Gammasys2}
\sum_{m=0}^{n-1}\wh {\psi}_m^{(j-1)}I_{m\ell} = -(2\pi i)^{j-1}J_{\ell,j},\hspace{0.5cm} j=2,\ldots,p
\ee
and these equations have to hold for any $\ell\in\{0,\ldots,n-1\}$. A similar equation also follows from the last row in \eqref{Gammasys}, it differs from the
latter only by a replacement of the right hand side in \eqref{Gammasys2}. Fixing $j$ in \eqref{Gammasys2}, we can rewrite the corresponding equation as an $n\times n$ linear 
system of equations on the unknown coefficients $\wh {\psi}_0^{(j-1)},\ldots,\wh {\psi}_{n-1}^{(j-1)}$. In this system however the coefficient matrix is given by the moment matrix
$[I_{m\ell}]_{m,\ell=0}^{n-1}$. Hence assuming $\Delta_n\neq 0$ ensures solvability of \eqref{Gammasys2}, which in turn guarantees existence of the polynomials in \eqref{poly1}
and therefore the solution of the RHP \ref{ChainRHP}. Conversely assuming solvability of the RHP for $\Gamma(z)$ we have already seen that this solution has to be
unique. Hence following our previous logic, all resulting systems from \eqref{Gammasys2} have to be uniquely solvable, i.e. $\Delta_n\neq 0$.\smallskip

As for the remaining identity \eqref{Gammaid}, we know from the previous part that $\psi_n(z)=\Gamma_{11}(z;n)$. In order to find $\phi_n(z)$, we let 
$\widehat{\Gamma}(z)=\Gamma^{-1}(z),z\in\C\backslash\R$. This leads to the following jump relation for $\widehat{\Gamma}(z)$
\be
\widehat{\Gamma}_+(z)=\le[\begin{array}{ccccc}
                        1 & -w_1(z) & 0 & &  0 \\
			0 & 1 & -w_2(z) &   & 0 \\
			  & 0 & 1 &\ddots  & 0\\
			  & &\ddots & \ddots  & -w_p(z) \\
			0 & 0& & 0&  1   \\
                       \end{array}\ri]\widehat{\Gamma}_-(z),\hspace{0.5cm}z\in\mathbb{R}\backslash\{0\},
\ee
which follows from \eqref{Gammajumps}, and adjusted behavior at infinity
\be\nonumber
\widehat{\Gamma}(z) = \le[\begin{array}{ccccc}
z^{-n}&&&&0\\
&  1& &&\\
&&\ddots &&\\
&&&1&\\
0&&&&z^{n}
\end{array}\ri]\le(I  + \mathcal O\left(z^{-1}\right) \ri),\hspace{0.5cm}z\rightarrow\infty.
\ee
Solving this problem recursively as we did it before for $\Gamma(z)$ (here row by row, instead of  column by column), we first see that 
\be\nonumber
\widehat{\Gamma}_{p+1,\bullet}(z)=\big(\widehat{\psi}_{n-1}^{\,(1)}(z),\ldots,\widehat{\psi}_{n-1}^{\,(p)}(z),\widehat{\pi}_n(z)\big)
\ee
where $\widehat{\pi}_n(z)=\widehat{\Gamma}_{p+1,p+1}(z;n)$ is a monic polynomial
of exact degree $n$ and $\widehat{\psi}_n^{\,(j)}(z)$ (uniquely determined) polynomials of degree $\leq n-1$. Next
\bea
  \widehat{\Gamma}_{\ell,\bullet}(z) &=& {\bf e}_{\ell}^T-\frac{1}{2\pi i}\int_0^{\infty}\widehat{\Gamma}_{\ell+1,\bullet-}\left((-1)^{\ell+1}w\right)
e^{-U_{\ell}(w)}\frac{\d w}{w+z(-1)^{\ell}},\hspace{0.5cm}\ell=2,\ldots,p\label{e:2}\\
  \widehat{\Gamma}_{1,\bullet}(z)&=&-\frac{1}{2\pi i}\int_0^{\infty}\widehat{\Gamma}_{2,\bullet-}(w)e^{-U_1(w)}\frac{\d w}{w-z}\nonumber
\eea
and recalling the behavior at infinity in the $\widehat{\Gamma}$-RHP therefore
\be
\int_0^{\infty}\int_0^{\infty}x^{\ell}\,\widehat{\pi}_n\left((-1)^{p+1}y\right)\eta_p(x,y)\,\d x\d y=0,\hspace{0.5cm}\ell\in\{0,\ldots,n-1\},
\ee
thus $\Gamma_{p+1,p+1}^{-1}(z;n)=\widehat{\pi}_n(z) = (-1)^{n(p+1)}\phi_n\left((-1)^{p+1}z\right)$ which completes the proof.
\end{proof}
We state several corollaries to the latter Theorem which are used later on.
\bc
The entry $(p+1,1)$ of the solution $\Gamma(z)=\Gamma(z;n)$ of the RHP \ref{ChainRHP} is given by 
\be\nonumber
\Gamma_{p+1,1}(z) = (-2\pi i)^{p}(-1)^{(n-1)(p+1)}\frac {\Delta_{n-1}}{\Delta_n} \psi_{n-1}(z) 
\ee
and the ``norms'' $h_n$ in \eqref{poly} equal 
\be
h_n = \frac {\Delta_{n+1}}{\Delta_n}\label{hn}.
\ee
\ec
\begin{proof}
From \eqref{Gammasys} we see that the entry under scrutiny must be proportional to $\psi_{n-1}(z)$, on the other hand the representation \eqref{bipolyformula}
gives us
\be\nonumber
\int_0^{\infty}\int_0^{\infty}\psi_n(x)y^m\eta_p(x,y)\,\d x\,\d y = \delta_{nm}\frac{\Delta_{n+1}}{\Delta_n},\hspace{0.5cm}m\leq n
\ee
and therefore the claim follows from \eqref{Gammasys}.
\end{proof}
\bc
\label{corSchles}
The solution of the RHP \ref{ChainRHP} is such that 
\be
\Gamma(z;n) = \le(I + \frac{Y_{1n}}z + \frac{Y_{2n}}{z^2} + \mathcal O\left(z^{-3}\right)\ri)z^{n(E_{11} - E_{p+1,p+1})},\hspace{0.5cm}E_{j\ell} = \big[\delta_{j\alpha}\delta_{\beta\ell}\big]_{\alpha,\beta=1}^{p+1}
\ee
where 
\be\label{norms}
[Y_{1n}]_{1,p+1} = \frac {(-1)^{n(p+1)}}{(-2\pi i)^p} \frac {\Delta_{n+1}}{\Delta_{n}} =  \frac {(-1)^{n(p+1)}h_n}{(-2\pi i)^p}
\ee
\ec
\begin{proof}
The matrix entry $[Y_{1n}]_{1,p+1}$ is the coefficient in $z^{-n-1}$ of the asymptotic expansion of $\Gamma_{1,1}(z;n)$
in the proof of Theorem \ref{Chainthm}, namely
\begin{align*}
\frac{1}{(-2\pi i)^p}\int_0^\infty\cdots\int_0^{\infty}\psi_n(w_1)&(-1)^{n(p+1)}w_p^n\frac{e^{-\sum_{j=1}^pU_j(w_j)}}{\prod_{j=1}^{p-1}(w_j+w_{j+1})}\,
\d w_1\cdot\ldots\cdot\d w_p\\
&=\frac{(-1)^{n(p+1)}}{(-2\pi i)^p}\int_0^{\infty}\int_0^{\infty}\psi_n(x)y^n\eta_p(x,y)\,\d x\d y = \frac{(-1)^{n(p+1)}h_n}{(-2\pi i)^p}.
\end{align*}
\end{proof}

We will prove Theorem \ref{theo2} by induction on $n\in\mathbb{Z}_{\geq 0}$ and for that we need to analyze the action of the shift $n\mapsto n+1$ on $\Gamma(z;n)$. In the Riemann-Hilbert problem, this shift corresponds to an elementary Schlesinger transformation in the sense of \cite{JM} which takes on the following form. We first observe that $\Gamma(z;n+1)\Gamma^{-1}(z;n)$ is a linear affine function, more precisely 
\begin{equation}\label{schles}
	\Gamma(z;n+1)\Gamma^{-1}(z;n)= zA_n+B_n \equiv R_n(z),\hspace{0.5cm}z\in\mathbb{C}.
\end{equation}
Indeed, the expression on the right side of \eqref{schles} is immediately seen to have no jumps on the real axis, and an isolated singularity at the origin. However, due to \eqref{constra} one finds that this singularity is $o(z^{-1})$ and thus concludes that the expression is analytic at $z=0$. The asymptotic behavior at $z=\infty$ implies that the expression grows at most linear and by Liouville's theorem we conclude that it must be an affine function in $z$. 
The coefficients $A_n$ and $B_n$ are determined from the asymptotics \eqref{Gammainfty}, we have (see \cite{JM}, formula (A.$1$))
\be\nonumber
 A_n = E_{11},\hspace{0.5cm}B_n = 
\le[\begin{array}{ccccc}
B_{11}  & B_{12} & \cdots & B_{1p}& B_{1,p+1}\\
B_{21} & 1 & \dots & 0 & 0\\
\vdots &&\ddots && \vdots\\
B_{p1} & 0 & &1 &  0\\
B_{p+1,1} & 0& \dots &0&  0
\end{array}\ri] 
,\ \ B_{1\ell} = -[Y_{1n}]_{1\ell},\ \ \ 2\leq\ell\leq p+1
\ee
and
\be\nonumber
B_{11}=\frac{\sum_{j=2}^{p+1}[Y_{1n}]_{1j}[Y_{1n}]_{j,p+1}-[Y_{2n}]_{1,p+1}}{[Y_{1n}]_{1,p+1}},\  B_{p+1,1}=\frac{1}{[Y_{1n}]_{1,p+1}},\ 
B_{\ell,1} = -\frac{[Y_{1n}]_{\ell,p+1}}{[Y_{1n}]_{1,p+1}},\ 2\leq\ell\leq p
\ee
where we recall from \eqref{norms} that $[Y_{1n}]_{1,p+1}\neq0$. By similar reasoning as above, one also finds that
\be 
	R_n^{-1}(z) = zE_{p+1,p+1}+C_n
\ee 
with
\be\nonumber
 C_n = 
\le[\begin{array}{ccccc}
0  & 0 & \cdots & 0& C_{1,p+1}\\
0 & 1 & \dots & 0 & C_{2,p+1}\\
\vdots &&\ddots && \vdots\\
0 & 0 & &1 &  C_{p,p+1}\\
C_{p+1,1} & C_{p+1,2}& \dots & C_{p+1,p}&  C_{p+1,p+1}
\end{array}\ri] 
\ee
where
\be\nonumber
C_{\ell,p+1} = [Y_{1n}]_{\ell,p+1},\ \ell\in\{1,\ldots,p\}; \hspace{0.75cm} C_{p+1,\ell}=-\frac{[Y_{1n}]_{1\ell}}{[Y_{1n}]_{1,p+1}},\ \ \ \ell\in\{2,\ldots,p\}
\ee
and
\be\nonumber
C_{p+1,1} = -\frac{1}{[Y_{1n}]_{1,p+1}},\hspace{0.75cm} C_{p+1,p+1}= [Y_{1n}]_{p+1,p+1}-\frac{[Y_{2n}]_{1,p+1}}{[Y_{1n}]_{1,p+1}}.
\ee
Using the previous identities, we derive the following Proposition, which will be important in the proof of Theorem \ref{theo2}
\bp For any $n\in\mathbb{Z}_{\geq 0}$
\begin{equation}\label{crucial}
	R_n^{-1}(w)R_n(z) = I-(z-w)\frac{E_{p+1,1}}{[Y_{1n}]_{1,p+1}},\hspace{0.5cm}z,w\in\mathbb{C}.
\end{equation}
\ep
At this point we are ready to derive Theorem \ref{theo2}.
\begin{proof}[Proof of Theorem \ref{theo2}] We use induction on $n\in\mathbb{Z}_{\geq  0}$ and apply \eqref{crucial}. During this we employ the  notation
\begin{equation*} 
	\big[\Gamma^{-1}(x(-1)^{j+1})\Gamma(y(-1)^{\ell-1})\big]_{j+1,\ell}\equiv \big[\Gamma^{-1}_{\pm}(w)\Gamma_{\pm}(z)\big]_{j+1,\ell}\bigg|_{w=x(-1)^{j+1},\ z=y(-1)^{\ell-1}}
\end{equation*}
and $\mathbb{M}_{j\ell}(x,y)\equiv \mathbb{M}_{j\ell}(x,y;n)$ to indicate the $n$-dependency.
\smallskip

{\it First case: $1\leq\ell\leq j\leq p$}. In the base case, use that both, $\Gamma(z;0)$ and $\Gamma^{-1}(z;0)$ are upper triangular, thus
\begin{equation*} 
	\Big[\Gamma^{-1}\big(x(-1)^{j+1};0\big)\Gamma\big(y(-1)^{\ell-1};0\big)\Big]_{j+1,\ell}=0
\end{equation*}
which matches the left hand side in \eqref{rep}, since by \eqref{lrow} and \eqref{row1} the corresponding kernels $\mathbb{M}_{j\ell}(x,y)$ always contain an empty sum. For the induction step, apply \eqref{schles}, thus
\begin{align}
	\Big[\Gamma^{-1}&\big(x(-1)^{j+1};n+1\big)\Gamma\big(y(-1)^{\ell-1};n+1\big)\Big]_{j+1,\ell} = \Big[\Gamma^{-1}\big(x(-1)^{j+1};n\big)\Gamma\big(y(-1)^{\ell-1};n\big)\Big]_{j+1,\ell}\nonumber\\
	&-\big(y(-1)^{\ell-1}-x(-1)^{j+1}\big)\Big[\Gamma^{-1}\big(x(-1)^{j+1};n\big)E_{p+1,1}\Gamma\big(y(-1)^{\ell-1};n\big)\Big]_{j+1,\ell}\frac{1}{[Y_{1n}]_{1,p+1}}\nonumber\\
	&=\big(x(-1)^{j+1}-y(-1)^{\ell-1}\big)\bigg\{\mathbb{M}_{j\ell}(x,y;n)(-2\pi i)^{j-\ell+1}(-1)^{\ell-1}\nonumber\\
	&+\Gamma^{-1}_{j+1,p+1}\big(x(-1)^{j+1};n)\Gamma_{1\ell}\big(y(-1)^{\ell-1};n\big)(-2\pi i)^p(-1)^{n(p+1)}\frac{1}{h_n}\bigg\}\label{int:1}
\end{align}
where we used the induction hypothesis as well as \eqref{norms} in the last equality. For $j=p$ and $1\leq \ell\leq p$, (compare \eqref{e:1},\eqref{e:2})
\begin{eqnarray*}
	\Gamma^{-1}_{p+1,p+1}(z;n) &=& (-1)^{n(p+1)}\phi_n\big((-1)^{p+1}z\big),\hspace{0.5cm}\Gamma_{11}(z;n)=\Psi_n(z)\\
	\Gamma_{1\ell}(z;n) &=&\frac{1}{2\pi i}\int_0^{\infty}\Big(\Gamma_{1,\ell-1}\big((-1)^{\ell}w;n\big)\Big)_-\frac{e^{-U_{\ell-1}(w)}}{w+z(-1)^{\ell-1}}\d w,\ \ 2\leq \ell\leq p
\end{eqnarray*}
and therefore with \eqref{lrow} back in \eqref{int:1}
\beas
	\Big[\Gamma^{-1}\big(x(-1)^{p+1};n+1\big)\Gamma\big(y(-1)^{\ell-1};n+1\big)\Big]_{p+1,\ell} =&\,\,(-2\pi i)^{p-\ell+1}(-1)^{\ell-1}\mathbb{M}_{p\ell}(x,y;n+1)\\
	&\times \big(x(-1)^{p+1}-y(-1)^{\ell-1}\big)
\eeas
in accordance with \eqref{rep}. Similarly, for $1\leq j\leq p-1$ and $1\leq\ell\leq j$, we use in addition
\begin{equation*} 
	\Gamma_{j+1,p+1}^{-1}(z;n) = -\frac{1}{2\pi i}\int_0^{\infty}\Big(\Gamma_{j+2,p+1}^{-1}\big((-1)^{j+2}w;n\big)\Big)_-\frac{e^{-U_{j+1}(w)}}{w+z(-1)^{j+1}}\d w
\end{equation*}
and obtain from \eqref{lrow} and \eqref{row1} back in \eqref{int:1}
\beas
	\Big[\Gamma^{-1}\big(x(-1)^{j+1};n+1\big)\Gamma\big(y(-1)^{\ell-1};n+1\big)\Big]_{j+1,\ell} =&\,\,(-2\pi i)^{j-\ell+1}(-1)^{\ell-1}\mathbb{M}_{j\ell}(x,y;n+1)\\
	&\times \big(x(-1)^{j+1}-y(-1)^{\ell-1}\big).
\eeas
This completes the induction for $1\leq\ell\leq j\leq p$.\bigskip

{\it Second case: $\ell=j+1$}. In the base case, we have to take into account that 
\begin{equation*} 
		\Big[\Gamma^{-1}\big(x(-1)^j;0\big)\Gamma\big(y(-1)^j;0\big)\Big]_{j+1,j+1} = 1.
\end{equation*}
	But from \eqref{row}, we get
\begin{equation*} 
		\mathbb{M}_{j,j+1}(x,y;0) = -\frac{1}{x+y}=(-1)^j\frac{1}{x(-1)^{j+1}-y(-1)^j},
\end{equation*}
i.e. the base case is completed. The induction step is as before:
\begin{align*}
	\Big[\Gamma^{-1}\big(x(-1)^{j+1}&;n+1\big)\Gamma\big(y(-1)^j;n+1\big)\Big]_{j+1,j+1} = \big(x(-1)^{j+1}-y(-1)^j\big)\bigg\{\mathbb{M}_{j,j+1}(x,y;n)(-1)^j\\
	&+\Gamma^{-1}_{j+1,p+1}\big((-1)^{j+1}x;n\big)\Gamma_{1,j+1}\big((-1)^jy;n\big)(-2\pi i)^p(-1)^{n(p+1)}\frac{1}{h_n}\bigg\}\\
	&=(-1)^j\mathbb{M}_{j,j+1}(x,y;n)\big(x(-1)^{j+1}-y(-1)^j\big)
\end{align*}
where all three identities \eqref{row}, \eqref{lrow} and \eqref{row1} are used in the last equality. This completes the induction in case $\ell=j+1$.\bigskip

{\it Third case: $\ell>j+1$}. We need to use that
\be\nonumber
	\Gamma(z;0) = \le[\begin{array}{ccccc}
                 1 & W_{11} & W_{12}&\cdots & W_{1p}\\
		 0 & 1 & W_{22} & &W_{2p}\\
		 \vdots  & 0& \ddots &\ddots &\vdots\\
		  & & & &W_{pp}\\
		0 & 0& \cdots & & 1
                \end{array}\ri]
\ee 
with
\be\nonumber
  W_{j\ell}(z) = \frac{1}{(2\pi i)^{\ell-j+1}}\int_0^{\infty}\dots\int_0^{\infty}\frac{e^{-\sum_{m=j}^{\ell}U_m(w_m)}}{\prod_{m=j}^{\ell-1}(w_m+w_{m+1})}
  \frac{\d w_j\cdots\d w_{\ell}}{w_{\ell}+z(-1)^{\ell}},\hspace{0.5cm} 1\leq j\leq \ell\leq p,
\ee
and also
\be\nonumber
  \Gamma^{-1}(z;0) = \le[\begin{array}{ccccc}
                 1 & \widehat{W}_{11} & \widehat{W}_{12}&\cdots & \widehat{W}_{1p}\\
		 0 & 1 & \widehat{W}_{22} & &\widehat{W}_{2p}\\
		 \vdots  & 0& \ddots &\ddots &\vdots\\
		  & & & &\widehat{W}_{pp}\\
		0 & 0& \cdots & & 1
                \end{array}\ri]
\ee
where
\be\nonumber	
  \widehat{W}_{j\ell}(z) = \frac{1}{(-2\pi i)^{\ell-j+1}}\int_0^{\infty}\cdots \int_0^{\infty}\frac{e^{-\sum_{m=j}^{\ell}U_m(w_m)}}{\prod_{m=j}^{\ell-1}(w_m+w_{m+1})}
  \frac{\d w_j\cdots\d w_{\ell}}{w_{j}+z(-1)^j},\hspace{0.5cm} 1\leq j\leq \ell\leq p.
\ee
Hence certain combinations of $\widehat{W}_{j\ell}(w)$ and $W_{j\ell}(z)$ will appear in the base case. On the other hand \eqref{row} gives additional terms inside the integrals and using partial fraction decomposition, we can verify the base case. The induction step is again a direct application of \eqref{crucial} combined with \eqref{lrow},\eqref{row} and \eqref{row1}.
\end{proof}
\section{Part II: asymptotic for the $p$--Laguerre chain}
\label{part2}
In the rest of the paper we specialize the potentials to the choice \eqref{CLaguerre};
due to the form of the potentials, we shall refer this chain model as the Cauchy-Laguerre $p$-chain. In the interest of concreteness, we also choose $p=3$, that is the first case which is not analyzed already in the literature. This choice is dictated mostly by convenience, as the overall logic can be carried out along similar lines for arbitrary $p$. The only step where a general theorem would be needed is in the construction of the so--called $\mathfrak{g}$--function. One of the key features (which is verified here) would be that the macroscopic densities  $\rho_j(x)$ of the eigenvalues of the matrices $M_j$ should have the following local behavior near the origin 
\be
\rho_j (x)  \sim C |x|^{-\frac {p}{p+1}}.
\label{beh0}
\ee
For $p=1$ (i.e. the ordinary Laguerre unitary ensemble) the density is the arcsine law and has precisely the behavior \eqref{beh0}. For $p=2$ this is verified in \cite{BGS3}; for $p=3$ it is verified in the present paper and for $p=4,5,6$ see Remark \ref{remp456}. For general $p$ (and general potential) a proof of this can only follow from potential theoretic methods. 
On a different note, the same singular behavior \eqref{beh0} has also been found in the analysis of products of random matrices, cf. \cite{BJLNS,Z}.

\subsection{Riemann-Hilbert analysis for the Cauchy-Laguerre three-chain}
We shall now address  the asymptotic analysis of  
Problem  \ref{ChainRHP}  for $p=3$ and choice of potentials  \eqref{CLaguerre}
to be analyzed in the limit $n=N\to\infty$.

Following the well established nonlinear steepest descent method of Deift and Zhou \cite{DZ,DKMVZ,DKM} a sequence of explicit and invertible transformations is carried out to simplify the initial problem for $\Gamma=\Gamma(z;n)$ and to derive an iterative solution valid as $n\rightarrow\infty$. The overall logic for this is well-known in the literature and we shall begin with a normalization transformation, the introduction of the (vector) $\mathfrak{g}$-functions.

\subsubsection{$\mathfrak{g}$-function transformation}
We transform the initial problem
\bea
\label{g:0}
	Y(z)&\& =\mathcal{L}\,\Gamma(z)\mathcal{G}(z)\mathcal{L}^{-1},\ \ z\in\mathbb{C}\backslash\mathbb{R}
\\
	\mathcal{L} =  &\&\textnormal{diag}\,\left[e^{-\frac{n}{4}\mathfrak{l}_1},e^{-\frac{n}{4}\mathfrak{l}_2},e^{-\frac{n}{4}\mathfrak{l}_3},e^{-\frac{n}{4}\mathfrak{l}_4}\right],\ \ \ \mathcal{G}(z)=\textnormal{diag}\,\left[e^{-n\mathfrak{g}^{(1)}(z)},e^{-n\mathfrak{g}^{(2)}(z)},e^{-n\mathfrak{g}^{(3)}(z)},e^{-n\mathfrak{g}^{(4)}(z)}\right].\nonumber
\eea
The diagonal matrices $\mathcal{G}(z)$ and $\mathcal{L}$ contain functions and normalization parameters which are constructed as follows. Start from the algebraic equation
\begin{equation}
		y^4-\frac{z-2}{2z}\,y^2+\frac{(3z+4)(3z-8)^2}{432 z^3}=0.\label{spec}
\end{equation}
The algebraic equation \eqref{spec} will be used to construct the $\gg$ function and all the required equalities and inequalities will be rigorously verified in Proposition \ref{behave}. The equation itself was the result of an Ansatz based on heuristic guidelines and then subsequent rigorous verification of its suitability.
The  equation \eqref{spec}
defines a Riemann surface $X=\left\{(y,z):\ \textnormal{satisfy}\, \eqref{spec}\right\}$ which consists of four sheets $X_j,j=1,\ldots,4$ glued together in the usual crosswise manner along $[a,0]$ and $[0,b]$ where
\begin{equation}
	a=-\frac{4}{3},\hspace{0.5cm} b=\frac{64}{27}
\end{equation}
are zeros of the discriminant of \eqref{spec}. We denote with $y:X\rightarrow \mathbb{CP}^1$ the bijective mapping such that $y_j=y|_{X_j},j=1,2,3,4$ are the four roots of \eqref{spec}. Since we usually identify the sheets $X_j$ with copies of the complex plane, $y_j=y_j(z)$ are defined on $\mathbb{C}$ with appropriate cuts. In more detail, we have
\beas
	y_1(z)&=\frac{1}{2z}\left(z^2-2\left(z(z-b)\right)^{\frac{1}{2}}-2z\right)^{\frac{1}{2}},\ \ \ \ \ \,y_4(z)=-y_1(z),\\
	y_2(z)&=-\frac{1}{2z}\left(z^2+2\left(z(z-b)\right)^{\frac{1}{2}}-2z\right)^{\frac{1}{2}},\ \ \ y_3(z)=-y_2(z)
\eeas
with principal branches for all fractional exponents, in particular $(z(z-b))^{\frac{1}{2}}$ is defined and analytic for $z\in\mathbb{C}\backslash(0,b)$ such that $(z(z-b))^{\frac{1}{2}}\sim z$ as $z\rightarrow+\infty,\ \textnormal{arg}\,z=0$ and
\begin{equation}\label{g:1}
	y_1(z)=\frac{1}{2}-\frac{1}{z}-\frac{11}{27z^2}+\mathcal{O}\left(z^{-3}\right),\ \ \ y_2(z)=-\frac{1}{2}+\frac{16}{27z^2}+\mathcal{O}\left(z^{-3}\right),\ \ \ z\rightarrow\infty.
\end{equation}
Notice that $y_1(z)$ is analytic for $z\in\mathbb{C}\backslash(0,b)$ whereas $y_2(z)$ is analytic for $z\in\mathbb{C}\backslash(a,b)$. In particular, 
\be\nonumber 
	y_{1+}(z)=y_{2-}(z),\ \ \ y_{1-}(z)=y_{2+}(z),\ z\in(0,b);\hspace{0.5cm}y_{4+}(z)=y_{3-}(z),\ \ \ y_{4-}(z)=y_{3+}(z),\ z\in(0,b)
\ee 
and
\be\nonumber 
	y_{2+}(z)=y_{3-}(z),\ \ \ y_{2-}(z)=y_{3+}(z),\ \ z\in(a,0).
\ee 
We can visualize this behavior as shown in Figure \ref{sheets}.

Moreover, the Riemann surface $X$ is of genus $g=0$ with a rational uniformization given by
\begin{eqnarray}
	z=-\frac{1}{210}\frac{t^4}{(t-1)(t-\frac{8}{7})(t-\frac{8}{5})(t-2)}\label{unizt},\qquad
	y=-\frac{99}{2}+\frac{210}{t}-\frac{288}{t^2}+\frac{128}{t^3},\ \ \ t\in\mathbb{CP}^1\nonumber
\end{eqnarray}
which defines a bijective map $\mathbb{T}:\mathbb{CP}^1\rightarrow X,\ t\mapsto(z(t),y(t))$ with branch points $\{t_j^{\ast}\}_{j=1}^4$ where
\be 
	\underbrace{0}_{=t_1^{\ast}}<1<\underbrace{\frac{96}{67}-\frac{8}{67}\sqrt{10}}_{=t_2^{\ast}}<\frac{8}{7}<\underbrace{\frac{4}{3}}_{=t_3^{\ast}}<\frac{8}{5}<\underbrace{\frac{96}{67}+\frac{8}{67}\sqrt{10}}_{=t_4^{\ast}}<2.
\ee 

\noindent\begin{minipage}{0.5\textwidth}
\includegraphics[width=0.9\textwidth]{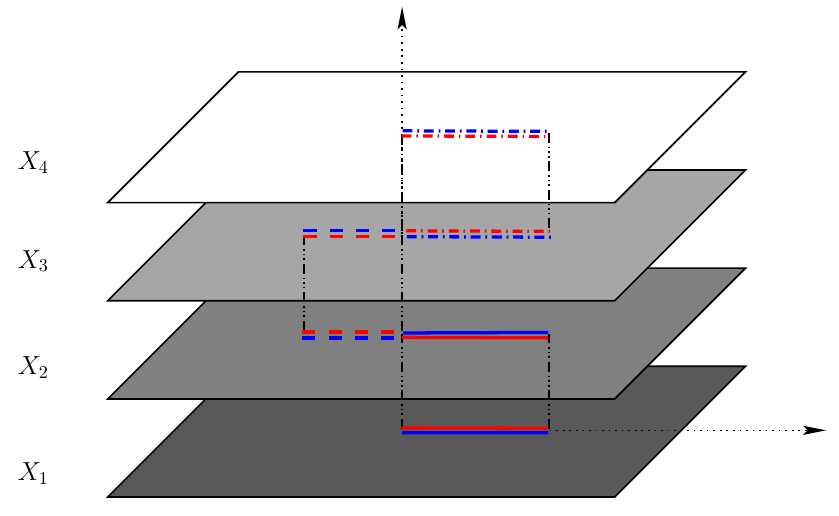}
\captionof{figure}{The four sheeted Riemann surface $X$. The endpoint of the cuts are $z=a$ on the left and $z=b$ on the right.}
\label{sheets}
\end{minipage}
\hspace{-0.1\textwidth}
\begin{minipage}{0.65\textwidth}
\begin{center}
\includegraphics[width=0.6\textwidth]{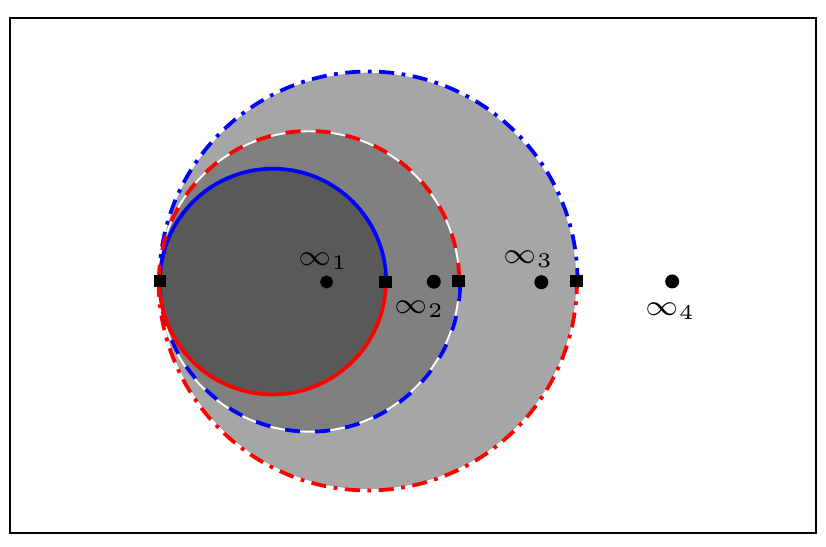} 
\captionof{figure}{Schematics of the mapping of the sheets $X_j$ to the complex $t$ plane. All sheets meet at the branch point $t_1^{\ast}$ shown as black box on the very left. The other branch points $t_j^{\ast}$ correspond to the other boxes. We give the boundary pieces $\mathcal{C}_j^{\pm}=\mathcal{C}_j\cap\{\Im\,t,\gtrless 0\},\,j=1,2,3$ the same orientation as the branch cuts shown in Figure \ref{sheets}, i.e they are oriented from $t_1^{\ast}$ to $t_j^{\ast},j\neq 1$. The labeling of $\mathcal{C}_j^{\pm}$ is according to the labeling of sheets $X_j$.}
\label{uni}
\end{center}
\end{minipage}

In particular, under the map $\mathbb{T}=\mathbb{T}(t)$, we have the following correspondences:
\be 
	1\mapsto\le\{\begin{array}{cc}
	z=\infty_1\\
	y=\frac{1}{2};
	\end{array}\ri.  \hspace{0.5cm}\frac{8}{7}\mapsto\le\{\begin{array}{cc}
	z=\infty_2\\
	y=-\frac{1}{2};
	\end{array}\ri.  \hspace{0.5cm}\frac{8}{5}\mapsto\le\{\begin{array}{cc}
	z=\infty_3\\
	y=\frac{1}{2};
	\end{array}\ri.  \hspace{0.5cm}2\mapsto\le\{\begin{array}{cc}
	z=\infty_4\\
	y=-\frac{1}{2},
	\end{array}\ri.  
\ee 
and we depict the partitioning of $\C \mathbb P^1\ni t$ into the four sheets under the uniformization map $\mathbb{T}^{-1}:X\rightarrow\mathbb{CP}^1$ in Figure \ref{uni}. With the jump behavior of the $y_j$'s  in mind, we introduce the  functions
\bea\nonumber 
	\mathfrak{g}^{(1)}(z)=\frac{\mathfrak{l}_1}{4}+\frac{z}{2}-\int_0^zy_1(\lambda)\,\d\lambda,\ \ \ \mathfrak{g}^{(4)}(z)=\frac{\mathfrak{l}_4}{4}-\frac{z}{2}-\int_0^{z}y_4(\lambda)\,\d\lambda,\ \ \ \ z\in\mathbb{C}\backslash(0,b),
\\
\nonumber 
	\mathfrak{g}^{(2)}(z)=\frac{\mathfrak{l}_2}{4}-\frac{z}{2}-\int_0^zy_2(\lambda)\,\d\lambda,\ \ \ \mathfrak{g}^{(3)}(z)=\frac{\mathfrak{l}_3}{4}+\frac{z}{2}-\int_0^zy_3(\lambda)\,\d\lambda,\ \ \ \ z\in\mathbb{C}\backslash(a,b).
\eea
The integration contours are chosen in the upper half plane and avoid crossing the branch cuts $(a,0)\cup(0,b)$. Furthermore, the constants $\mathfrak{l}_j,j=1,\ldots,4$ are chosen in such a way as to ensure the normalization
\be\label{gnorm} 
	\mathfrak{g}^{(1)}(z)=\ln z+\mathcal{O}\left(z^{-1}\right),\ \ \mathfrak{g}^{(4)}(z)=-\ln z+\mathcal{O}\left(z^{-1}\right),\ \ \mathfrak{g}^{(j)}(z)=\mathcal{O}\left(z^{-1}\right),\ j=2,3\ \ \ \ z\rightarrow\infty.
\ee 
As can be seen from \eqref{g:1}, this is achieved by 
\begin{eqnarray*}
	\frac{\mathfrak{l}_1}{4}&=&\ln b-\frac{b}{2}+\int_0^by_{1+}(\lambda)\,\d\lambda+\int_b^{\infty}\left(y_1(\lambda)-\frac{1}{2}+\frac{1}{\lambda}\right)\,\d\lambda,\hspace{0.5cm}\mathfrak{l}_4=-\mathfrak{l}_1\\
	\frac{\mathfrak{l}_2}{4}&=&\frac{b}{2}+\int_0^by_{2+}(\lambda)\,\d\lambda+\int_b^{\infty}\left(y_2(\lambda)+\frac{1}{2}\right)\,\d\lambda,\hspace{0.5cm}\mathfrak{l}_3=-\mathfrak{l}_2.
\end{eqnarray*}
We summarize certain analytical properties of the $\mathfrak{g}$-functions which are consequences of the jumps of $y_j(z)$ in the following Proposition. 
\bp\label{behave} Let 
\be 
	\omega_{j,j+1}(z)=\mathfrak{g}^{(j)}_-(z)-\mathfrak{g}^{(j+1)}_+(z)-(-1)^{j+1}z-\frac{\mathfrak{l}_j}{4}+\frac{\mathfrak{l}_{j+1}}{4},\ \ z\in\mathbb{R},\ \  j=1,2,3.
\ee 
Then
\be 
	\omega_{12}(z)=\omega_{34}(z)=0,\ \ z\in(0,b);\hspace{0.5cm} \omega_{23}(z)=0,\ \ z\in(a,0)
\ee 
and
\begin{eqnarray*}
	\omega_{12}(z)&=&-\frac{1}{2}\int_b^z\left(\sqrt{\lambda^2+2\sqrt{\lambda(\lambda-b)}-2\lambda}+\sqrt{\lambda^2-2\sqrt{\lambda(\lambda-b)}-2\lambda}\,\right)\frac{\d\lambda}{\lambda}<0,\ \ z>b,\\
	\omega_{23}(z)&=&-\int_z^a\sqrt{\lambda^2-2\sqrt{|\lambda(\lambda-b)|}-2\lambda}\,\frac{\d\lambda}{|\lambda|}<0,\ \ z<a,\\
	\omega_{34}(z)&=&-\frac{1}{2}\int_b^z\left(\sqrt{\lambda^2+2\sqrt{\lambda(\lambda-b)}-2\lambda}+\sqrt{\lambda^2-2\sqrt{\lambda(\lambda-b)}-2\lambda}\,\right)\frac{\d\lambda}{\lambda}<0,\ \ z>b.
\end{eqnarray*}
\ep
In order to perform subsequent steps in the Riemann-Hilbert analysis, we also require
\bd 
\label{effpotsdef}
We introduce the effective potentials
\begin{eqnarray*}
	\varphi_1(z)&=&z+\frac{\mathfrak{l}_1}{4}-\frac{\mathfrak{l}_2}{4}-\mathfrak{g}^{(1)}(z)+\mathfrak{g}^{(2)}(z)=\int_0^z\big(y_1(\lambda)-y_2(\lambda)\big)\,\d\lambda\\
	\varphi_2(z)&=&-z+\frac{\mathfrak{l}_2}{4}-\frac{\mathfrak{l}_3}{4}-\mathfrak{g}^{(2)}(z)+\mathfrak{g}^{(3)}(z)=\int_0^z\big(y_2(\lambda)-y_3(\lambda)\big)\,\d\lambda\\
	\varphi_3(z)&=&z+\frac{\mathfrak{l}_3}{4}-\frac{\mathfrak{l}_4}{4}-\mathfrak{g}^{(3)}(z)+\mathfrak{g}^{(4)}(z)=\int_0^z\big(y_3(\lambda)-y_4(\lambda)\big)\,\d\lambda=\varphi_1(z)
\end{eqnarray*}
\ed
\bl\label{behave2} There is a neighborhood of $(0,b)$ for which $\Re\left(\varphi_1(z)\right)<0,\Re\left(\varphi_3(z)\right)<0$ away from the interval $(0,b)$. Similarly there is a neighborhood of $(a,0)$ for which $\Re\left(\varphi_2(z)\right)<0$ away from the interval $(a,0)$.
\el
\begin{proof}
Let $\pi_j(z)=\mathfrak{g}_+^{(j)}(z)-\mathfrak{g}_-^{(j)}(z),z\in\mathbb{R}$ and notice that
\be 
	\pi_1(z)=-\varphi_{1+}(z)=\varphi_{1-}(z),\ \ \ \pi_4(z)=\varphi_{3+}(z)=-\varphi_{3-}(z),\ \ \ \  z\in(0,b),
\ee 
\be 
	\pi_2(z)=-\varphi_{2+}(z)=\varphi_{2-}(z),\ \ \ \pi_3(z)=\varphi_{2+}(z)=-\varphi_{2-}(z),\ \ \ \ z\in(a,0).
\ee 
Thus the continuations of $\varphi_j(z)$ into the upper and lower half plane are ensured and since $\Im(\pi_1(z)),z\in(0,b)$ and $\Im(\pi_2(z)),z\in(a,0)$ are both 
strictly decreasing on $(0,b)$, resp. on $(a,0)$, the sign conditions on $\Re\big(\varphi_1(z)\big)$ and $\Re\big(\varphi_2(z)\big)$ follow from the Cauchy-Riemann equations.
\end{proof}
\br
\label{remp456}
We state here, without proof, the spectral curves to use for the analysis of the longer chains $p=4,5,6$. They have been obtained by an educated guess starting from a uniformization of the Riemann sphere of degree $p+1$ and subsequent verification that they define positive equilibrium measures.
A general {\em existence} proof for arbitrary $p$ (and in general arbitrary potentials) requires a vector-potential theoretic approach. This framework is partly contained in \cite{BB}; however, the potentials that are of interest here do not satisfy all the properties in loc. cit.: in particular those requirements which were invoked to guarantee that the supports of the equilibrium measures have a finite distance from the origin. In all cases the behavior of the various branches of the  solutions $y(z)$ near $z=0$ is $y(z) \sim c z^{-\frac 1{p+1}} $.
The spectral curves below and their corresponding vector-equilibrium measures could be used as a starting point for a steepest descent analysis in the  corresponding $p=4,5,6$ cases.
\begin{eqnarray}
E_4  &=& {y}^{5}-\frac 3 5\,{y}^{3}+{\frac { \left( 2\,{z}^{2}-25 \right) {y}^{
2}}{ 25 {z}^{2}}}+{\frac { \left( 12\,{z}^{2}-25
 \right) y}{ 125 {z}^{2}}}-{\frac {288\,{z}^{4}-3000\,{
z}^{2}+3125}{12500{z}^{4}}}\\
E_5 &=& 
{y}^{6}+ {\frac { \left( -3\,z+4 \right) {y}^{4}}{4z}} + {\frac { \left( 75\,{z}^{3}-200\,{z}^{2}+256 \right) {y}^{2}}{
 400 {z}^{3}}}+{\frac { \left( 4-5\,z \right)  \left( 
25\,{z}^{2}-40\,z-64 \right) ^{2}}{  200000{z}^{5}}}\\
E_6 &=&
{y}^{7}-\frac 6 7 {y}^{5}+{\frac { \left( 2\,z-7 \right)  \left( 2\,z
+7 \right) {y}^{4}}{ 49   {z}^{2}}}+  {\frac { \left( 87\,{z
}^{2}-98 \right) {y}^{3}}{ 343{z}^{2}}} -{\frac {
 \left( 2916\,{z}^{4}-30429\,{z}^{2}+19208 \right) {y}^{2}}{ 64827 {z}^{4}}}\nonumber\\
\bigskip
&&   -
{\frac { 8 \left( 54\,{z}^{2}-49 \right)  \left( 27
\,{z}^{2}-49 \right) y}{ 453789{z}^{4}}}
  +{\frac {16 \le( 
236196\,{z}^{6}-2250423\,{z}^{4}+2722734\,{z}^{2}-823543 \ri) }{600362847 {z}^{6}}}
\end{eqnarray}
\er

\br
\label{remchainsep}
For $p=3$ we can consider the following more general case where the exponent $a_2$ is allowed to scale with $n$ according to $a_2 = n \beta,\ \beta>0$. In this case the spectral curve is the following one
\begin{align}
	y^4 - \frac {z^2 - 2z + \beta^2}{2z^2} y^2  + \frac {Q_0(z)}{16 z^4} =0 \label{csep}\qquad 
	 Q_0 (z)= z^4 - 4z^3 - 2z^2 \beta (\beta+4) +4z\le((1+\beta)^2 q - \beta^2\ri)+ \beta^4
\end{align}
where $q = q(\beta)$ is the unique positive root of the following polynomial (in $q$)
\be
	27 (1 + \beta)^2 q^3 -16 (9 \beta^2 + 9 \beta  + 4) q^2 + 16 \beta^2(\beta^2 + \beta+8) q - 64 \beta^4\label{qeq}.
\ee
The existence of $q(\beta)>0$ follows from the following reasoning:
the discriminant of \eqref{qeq} equals $\Delta =-4096\, \left( 1+\beta \right)  \left( 1+2\,\beta \right) ^{2} \left(3\,{\beta}^{2}+3\,\beta+32 \right) ^{3}{\beta}^{5}
$ and hence it is negative for $\beta>0$. Therefore there must be at least one pair of complex roots. Since the degree of \eqref{qeq} is three there is only one (positive) real root.
The condition \eqref{qeq} guarantees that the spectral curve \eqref{csep} is of genus $0$ (with one nodal point). The solutions of \eqref{csep} are the four sheets
\be
y_{1,2,3,4}(z) = \pm \frac 1 2 ,{\frac {\sqrt {{z}^{2}-2\,z+{\beta}^{2}\pm 2\, \left( 1+\beta
 \right) \sqrt {z \left( z-q \right) }}}{z}}
\ee
and thus $z=0, q$ are  branchpoints connecting two pairs of sheets; the other two branchpoints are the zeros of the radicand of the outer root, which turn out to be the roots of $Q_0(z)$ in \eqref{csep}; the equation \eqref{qeq} is simply the vanishing of the discriminant w.r.t. $z$ of $Q_0(z)$, which guarantees that one root of $Q_0(z)$ is double.
A full inspection (
left to the reader
) reveals in turn that the roots of $Q_0(z)$ are all real: the simple ones are negative, and the positive one is double and greater than $q(\beta)$. These observations 
 can be used to obtain a complete proof
  that \eqref{csep} is the correct spectral curve for the construction of the relevant $\mathfrak{g}$--functions; 
  since we do not need them for our paper, the proof is also omitted for general $\beta>0$. Only the case $\beta=0$ is needed and proved in Proposition \ref{behave}
.\smallskip

\begin{figure}
\includegraphics[width=0.19\textwidth]{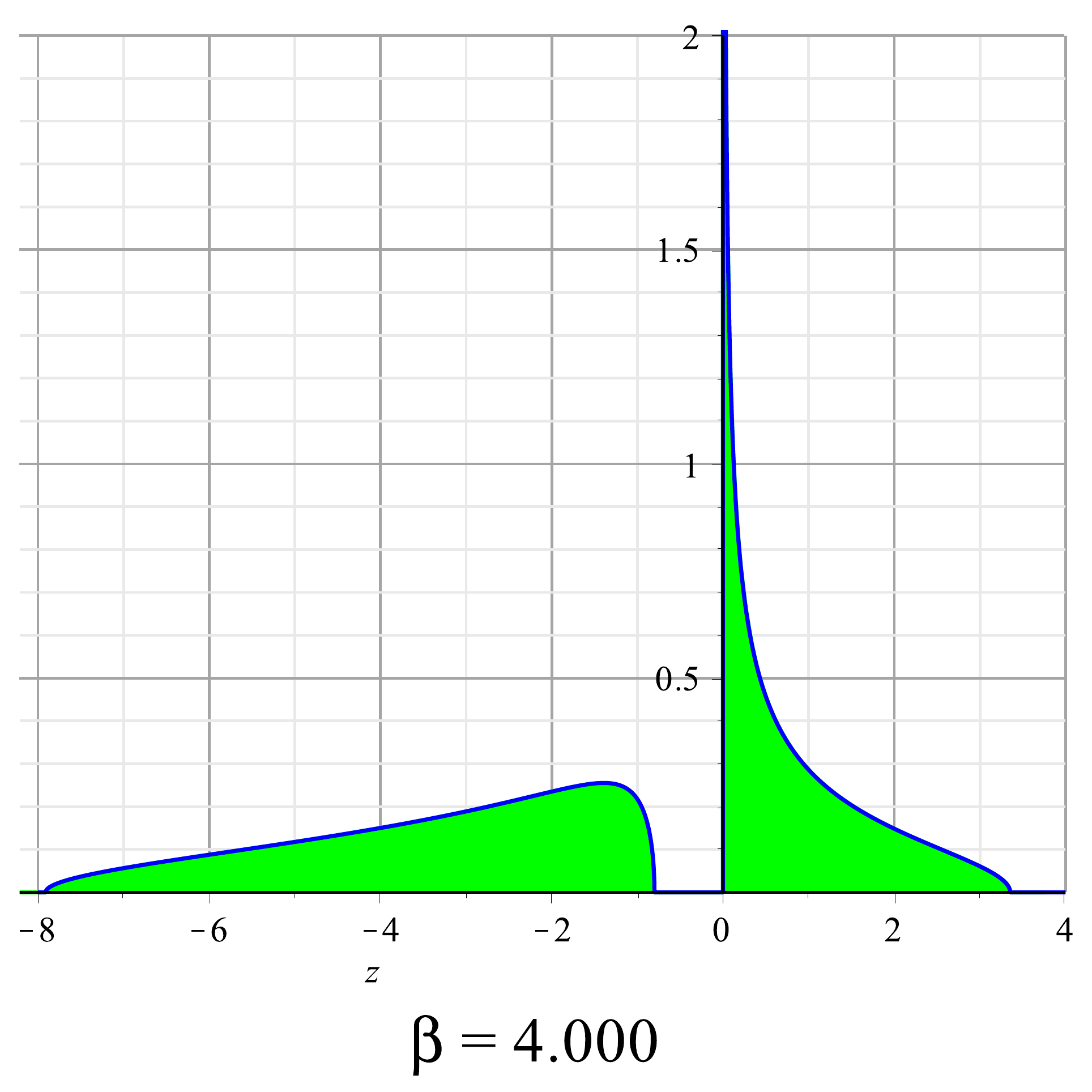}
\includegraphics[width=0.19\textwidth]{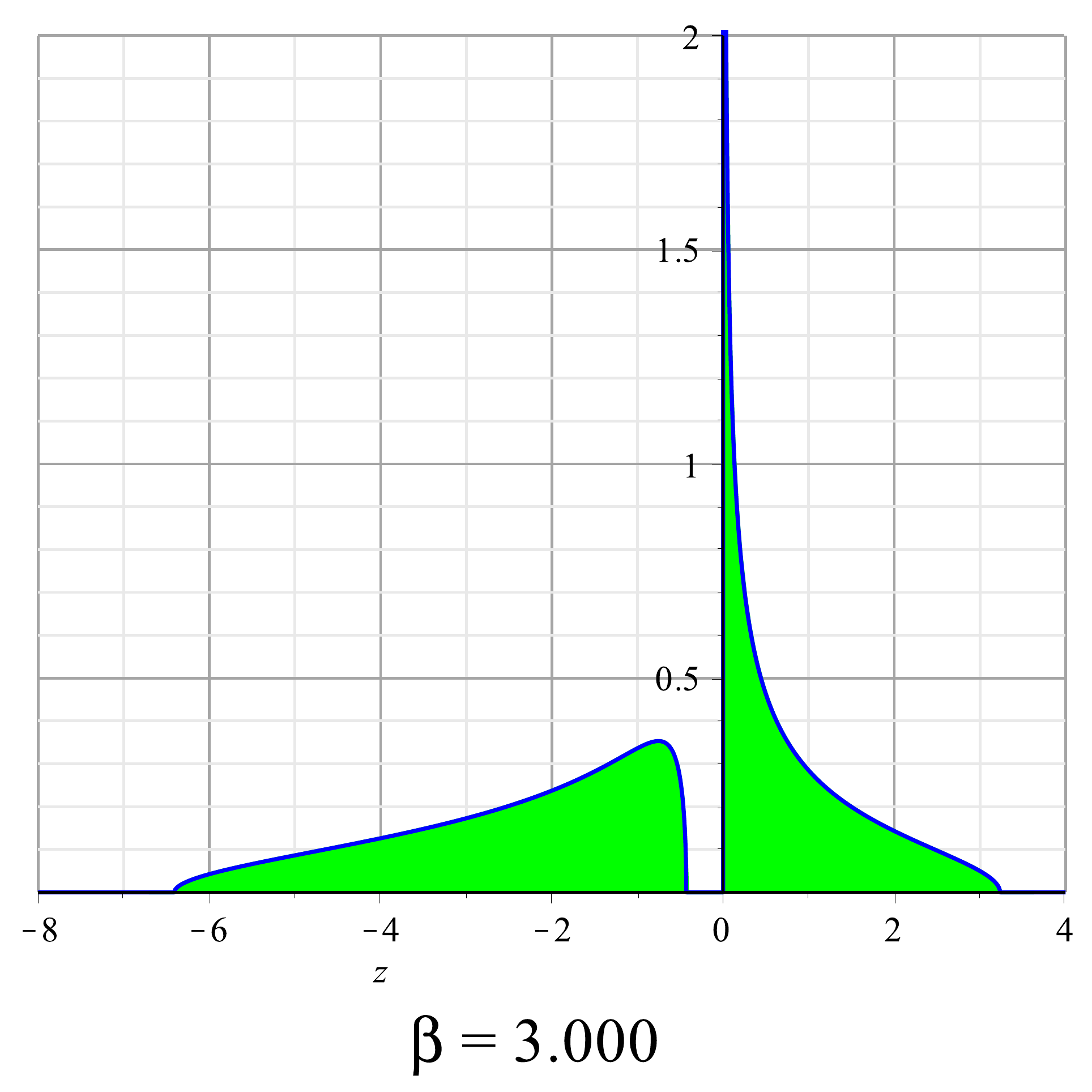}
\includegraphics[width=0.19\textwidth]{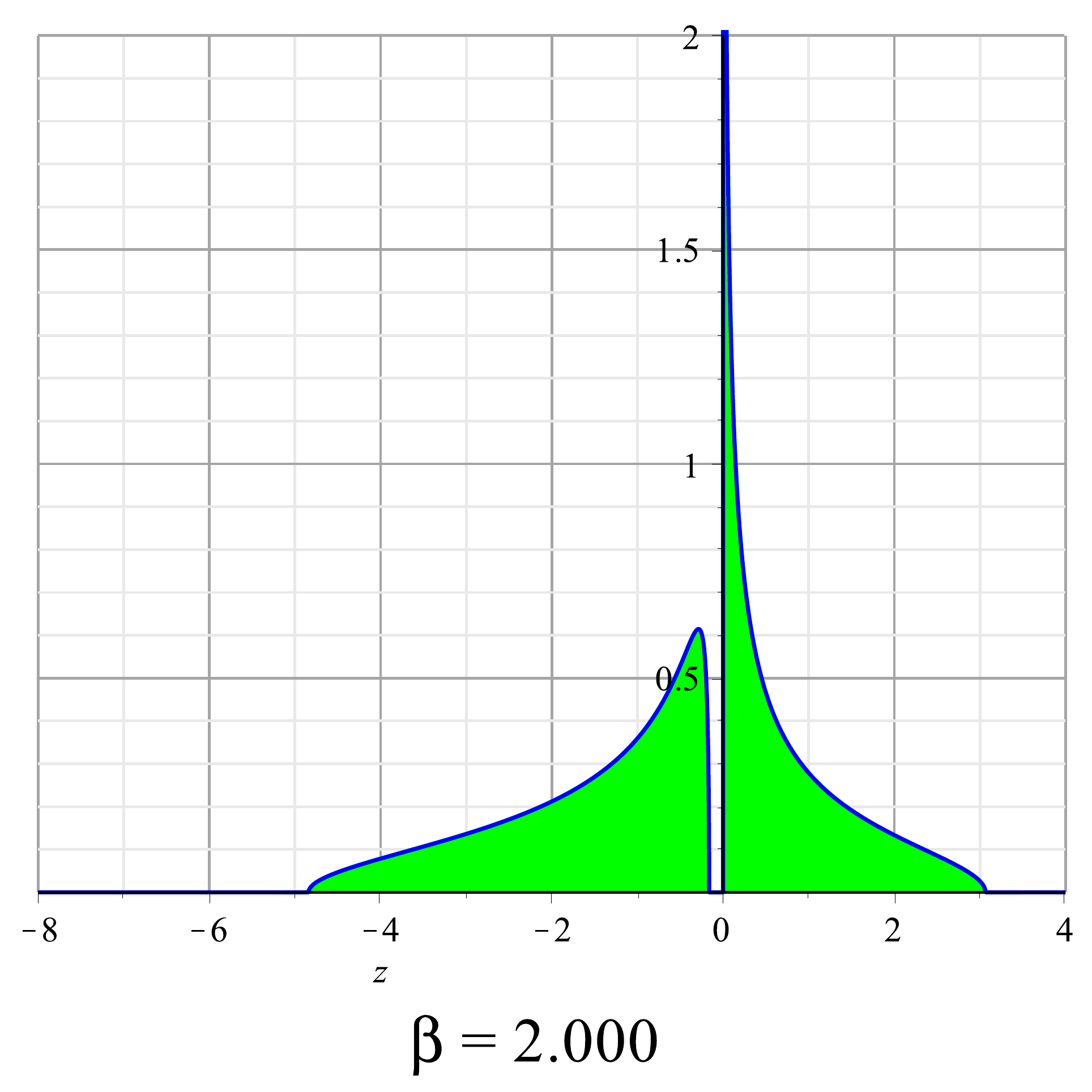}
\includegraphics[width=0.19\textwidth]{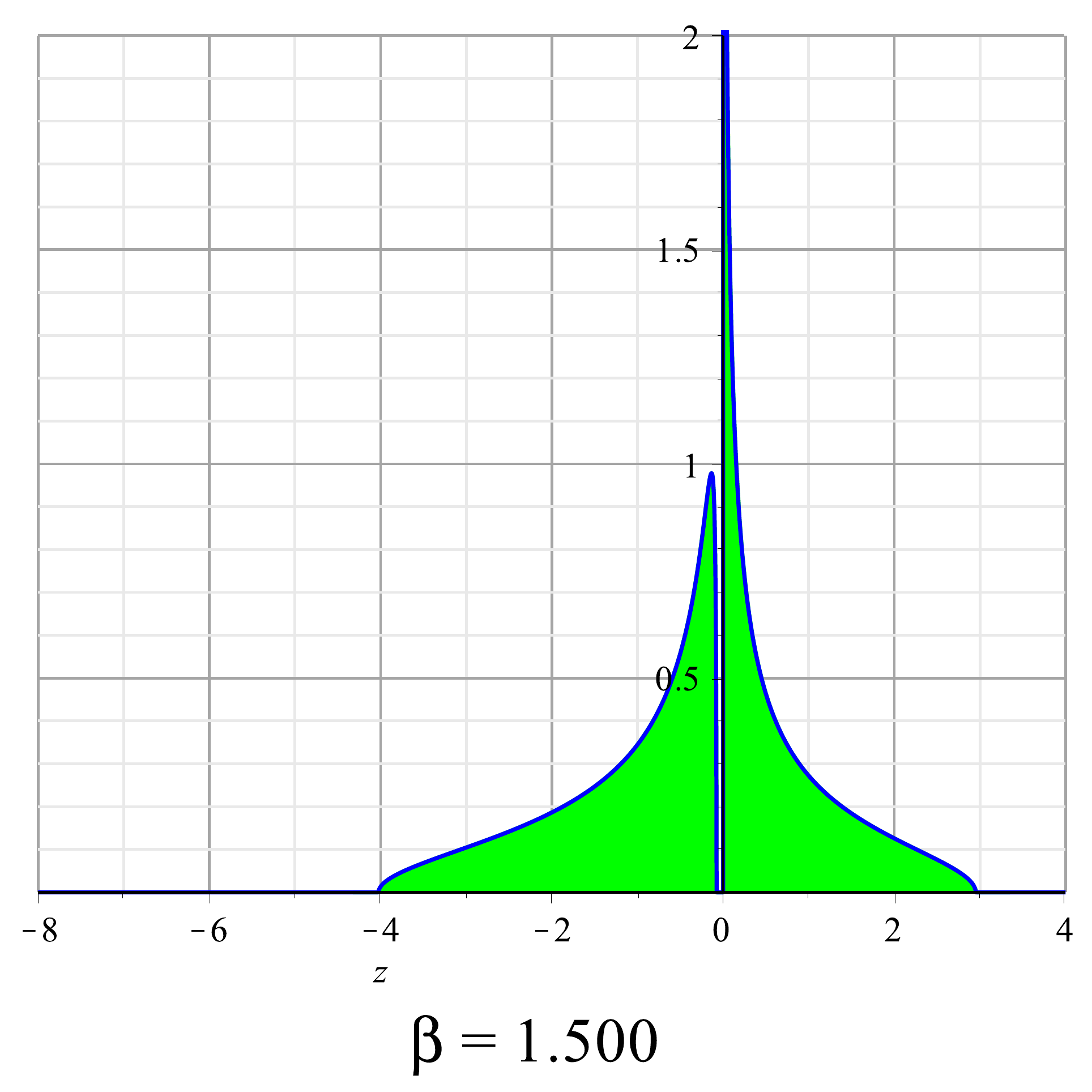}
\includegraphics[width=0.19\textwidth]{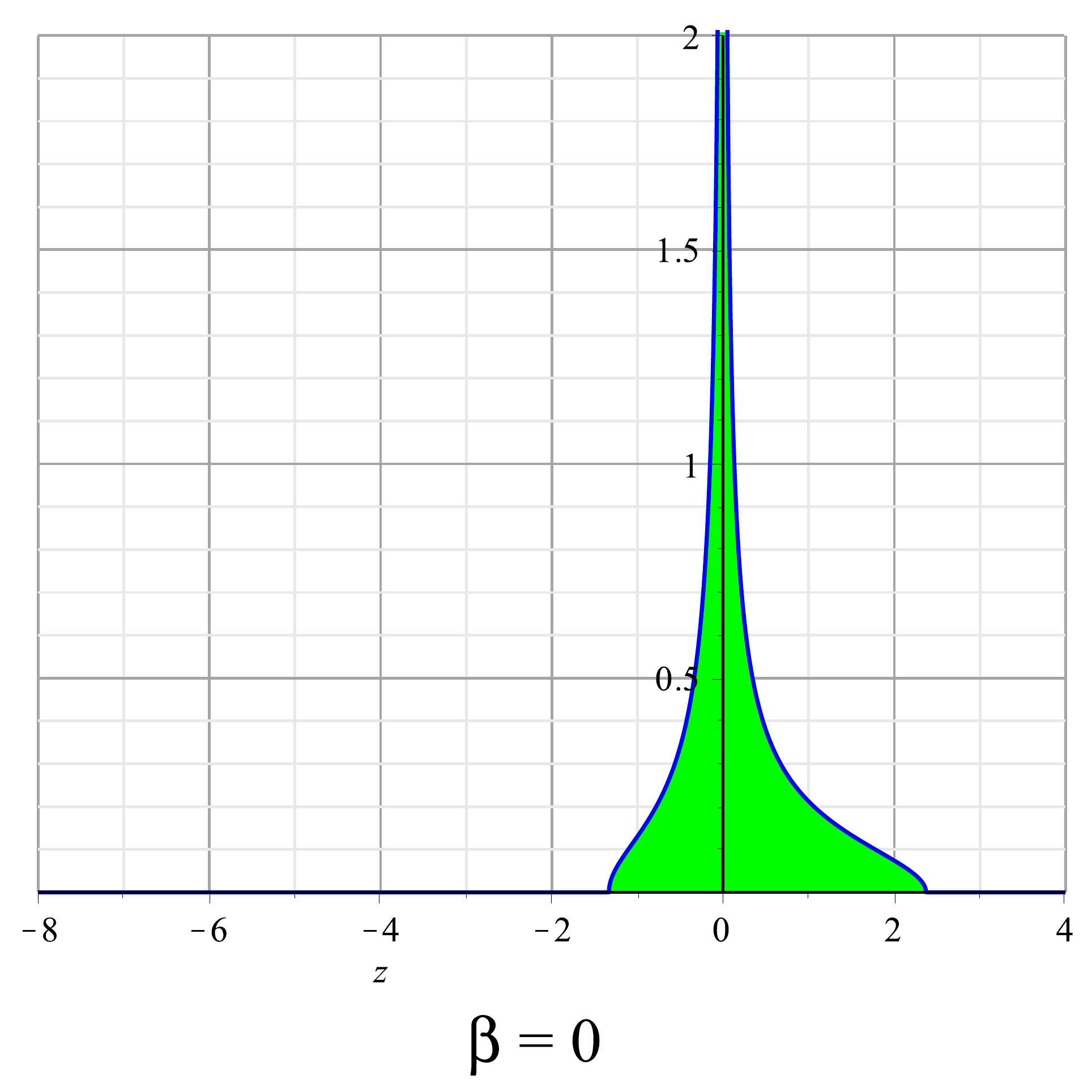}
\caption{The limiting densities of eigenvalues for the $p=3$ separated chain for  $a_2 = n\beta$ and different values of $\beta$ 
and $a_1$ is independent of $n$.
 Note that the support of the density of the matrix $M_2$ is separated from the origin. By comparison we also show the densities for $\beta =0$ (connected chain). 
The profile of the density on the negative axis is the asymptotic macroscopic density of the eigenvalues of  $M_2$ reflected about the origin, while on the positive axis the profile corresponds to the densities of $M_1, M_3$ (they are identical).
 }
\label{csepfig}
\end{figure}

Moreover, for $\beta=0$ the curve reduces to  \eqref{spec} (with $q = \frac {64}{27}$). As $\beta\to +\infty$ we have  $q\to 4$.
The plots of the relevant densities are shown in Figure \ref{csepfig}; the density on the negative axis is the density of the spectrum of $M_2$ while the densities of $M_1, M_2$ are equal to each other and equal to the density on the positive axis.
\er

Returning now  to \eqref{g:0}, we obtain a transformed $Y$-RHP with jump matrices 
\begin{eqnarray*}
	G_Y(z)&=&\begin{bmatrix}
	e^{-n\pi_1(z)} & z^{a_1}e^{n\omega_{12}(z)} \\
	0 & e^{-n\pi_2(z)}
	\end{bmatrix}\oplus\begin{bmatrix}
	e^{-n\pi_3(z)} & z^{a_3}e^{n\omega_{34}(z)}\\
	 0 & e^{-n\pi_4(z)}\\
	\end{bmatrix},\ \ z>0\\
	G_Y(z)&=&
	e^{-n\pi_1(z)}\oplus\begin{bmatrix}
	e^{-n\pi_2(z)} & (-z)^{a_2}e^{n\omega_{23}(z)}\\
	0 & e^{-n\pi_3(z)}
	\end{bmatrix}\oplus e^{-n\pi_4(z)},\ \ z<0
\end{eqnarray*}
which can be simplified using  Proposition \ref{behave} and Lemma \ref{behave2},
\begin{eqnarray*}
	G_Y(z)&=&\begin{bmatrix}
	e^{-n\pi_1(z)} & z^{a_1}\\
	0 & e^{n\pi_1(z)}\\
	\end{bmatrix}\oplus\begin{bmatrix}
	e^{-n\pi_3(z)} & z^{a_3}\\
	0 & e^{n\pi_3(z)}\\
	\end{bmatrix},\ \ z\in(0,b),\\
	G_Y(z)&=&\begin{bmatrix}
	1 & z^{a_1}e^{n\omega_{12}(z)}\\
	0 & 1\\
	\end{bmatrix}\oplus\begin{bmatrix}
	1 & z^{a_3}e^{n\omega_{34}(z)}\\
	0 & 1\\
	\end{bmatrix},\ \ z>b,
\end{eqnarray*}
as well as
\be\nonumber 
	G_Y(z)=1\oplus\begin{bmatrix}
	e^{-n\pi_2(z)} & (-z)^{a_2}\\
	0 & e^{n\pi_2(z)}\\
	\end{bmatrix}\oplus 1,\ \ z\in(a,0);\hspace{0.5cm}G_Y(z)=1\oplus\begin{bmatrix}
	1 & (-z)^{a_2}e^{n\omega_{23}(z)}\\
	0 & 1\\
	\end{bmatrix}\oplus 1,\ \ z<-a.
\ee 
In the latter, we also used that (compare earlier)
\be\nonumber 
	\pi_1(z)=-\pi_2(z),\ \ \ \pi_3(z)=-\pi_4(z),\ \ z\in(0,b);\hspace{0.75cm}\pi_2(z)=-\pi_3(z),\ \ z\in(a,0)
\ee 
and we emphasize the normalization
$ 
	Y(z)=I+\mathcal{O}\left(z^{-1}\right),\ \ z\rightarrow\infty,
$
following from \eqref{gnorm} and \eqref{g:0}.
\br From now on the  notation $ A\oplus B\oplus C\dots$ with $A,B,C,\dots$ square matrices (each of different sizes in general), stands for a block diagonal matrix with $A,B,C,\dots,$ along the diagonal.
\er
Next, we factorize the jump matrices on the segments $(a,0)\cup(0,b)$. For the corresponding $2\times 2$ blocks this means (recall Lemma \ref{behave2})
\begin{eqnarray*}
	\begin{bmatrix}
	e^{-n\pi_1(z)} & z^{a_1}\\
	0 & e^{n\pi_1(z)}\\
	\end{bmatrix}&=&\begin{bmatrix}
	1 & 0\\
	z^{-a_1}e^{n(\varphi_1(z))_-} & 1\\
	\end{bmatrix}\begin{bmatrix}
	0 & z^{a_1}\\
	-z^{-a_1} & 0\\
	\end{bmatrix}\begin{bmatrix}
	1 & 0\\
	z^{-a_1}e^{n(\varphi_1(z))_+} & 1\\
	\end{bmatrix},\ \ z\in(0,b),\\
	\begin{bmatrix}
	e^{n\pi_4(z)} & z^{a_3}\\
	0 & e^{-n\pi_4(z)}\\
	\end{bmatrix}&=&\begin{bmatrix}
	1 & 0\\
	z^{-a_3}e^{n(\varphi_3(z))_-} & 1\\
	\end{bmatrix}\begin{bmatrix}
	0 & z^{a_3}\\
	-z^{-a_3} & 0\\
	\end{bmatrix}\begin{bmatrix}
	1 & 0\\
	z^{-a_3}e^{n(\varphi_3(z))_+} & 1\\
	\end{bmatrix},\ \ z\in(0,b),\\
	\begin{bmatrix}
	e^{-n\pi_2(z)} & (-z)^{a_2}\\
	0 & e^{n\pi_2(z)}\\
	\end{bmatrix}&=&\begin{bmatrix}
	1 & 0\\
	z_-^{-a_2}e^{i\pi a_2}e^{n(\varphi_2(z))_-} & 1\\
	\end{bmatrix}\begin{bmatrix}
	0 & |z|^{a_2}\\
	-|z|^{-a_2} & 0\\
	\end{bmatrix}\begin{bmatrix}
	1 & 0\\
	z_+^{-a_2}e^{-i\pi a_2}e^{n(\varphi_2(z))_+} & 1\\
	\end{bmatrix},\ \ z\in(a,0).
\end{eqnarray*}

\subsubsection{Opening of lenses}
\begin{figure}[tbh]
\begin{center}
\includegraphics[width=0.6\textwidth]{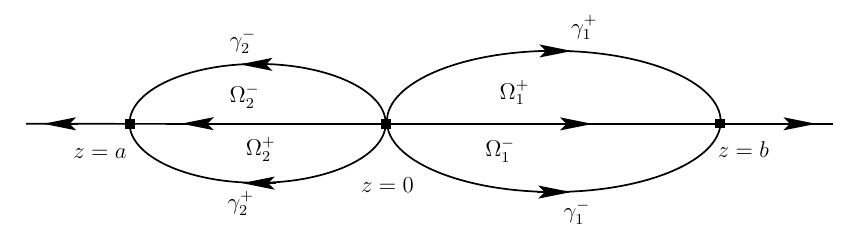}
\end{center}
\caption{Opening of lenses and the resulting jump contours in the $S$-RHP}
\label{openup}
\end{figure}

If we let
\be\nonumber 
		S_{L_1}^{(\pm)}(z) = \bigoplus_{j=1,3}\begin{bmatrix}
		1 & 0\\
		z^{-a_j}e^{n(\varphi_j(z))_{\pm}} & 1\\
		\end{bmatrix},\hspace{0.5cm} S_{L_2}^{(\pm)}(z) = 1\oplus\begin{bmatrix}
		1 & 0\\
		z_{\pm}^{-a_2}e^{\mp i\pi a_2}e^{n(\varphi_2(z))_{\pm}} & 1\\
		\end{bmatrix}\oplus 1,
\ee 
Lemma \ref{behave2} allows us to perform ``opening of lenses", i.e. we consider the transformation (compare Figure \ref{openup})
\be\label{openlens} 
	S(z)=\begin{cases}
		Y(z)\big(S_{L_j}^{(+)}(z)\big)^{-1},& z\in\Omega_j^{(+)}\\
		Y(z)\big(S_{L_j}^{(-)}(z)\big),& z\in\Omega_j^{(-)}\\
		Y(z),& \textnormal{else}
		\end{cases},\hspace{0.5cm}   j=1,2
\ee 
which leads to the following RHP
\begin{problem}\label{SRHP} Determine the $4\times 4$ piecewise analytic function $S(z)$ such that
\begin{itemize}
	\item $S(z)$ is analytic for $z\in\mathbb{C}\backslash(\mathbb{R}\cup\gamma_1^+\cup\gamma_1^-\cup\gamma_2^+\cup\gamma_2^-)$
	\item The jump conditions are as follows
	\begin{eqnarray*}
		S_+(z) &=&S_-(z)\bigoplus_{j=1,3}\begin{bmatrix}
		0 & z^{a_j} \\
		-z^{-a_j} & 0 \\
		\end{bmatrix},\ \ z\in(0,b)\\
		S_+(z)&=&S_-(z)\left(1\oplus\begin{bmatrix}
		0 & (-z)^{a_2}\\
		-(-z)^{-a_2}&0 \\
		\end{bmatrix}\oplus 1\right),\ \ z\in(a,0)\\
		S_+(z)&=&S_-(z)S_{L_j}^{(\pm)}(z),\ \ z\in\gamma_j^{\pm},\ \ j=1,2\\
		S_+(z)&=&S_-(z)\bigoplus_{j=1,3}\begin{bmatrix}
		1 & z^{a_j}e^{n\omega_{j,j+1}(z)}\\
		0 & 1\\
		\end{bmatrix},\ \ z>b\\
		S_+(z)&=&S_-(z)\left(1\oplus\begin{bmatrix}
		1 & (-z)^{a_2}e^{n\omega_{23}(z)}\\
		0 & 1 \\
		\end{bmatrix}\oplus 1\right),\ \ z<a
	\end{eqnarray*}
	\item The behavior at the origin is dictated as in \eqref{Gammasing} and \eqref{zerobeh} as long as we approach $z=0$ from the exterior of the lenses $\Omega_j^{(\pm)}$. From within the behavior is slightly changed due to the effect of $S_{L_j}^{(\pm)}$, compare \eqref{openlens}
	\item For $z\rightarrow\infty$, we have $S(z)\rightarrow I$
\end{itemize}
\end{problem}
As $\omega_{j,j+1}(z)<0$ for $z\in\mathbb{R}\backslash[a-\delta,b+\delta]$ with any fixed $\delta>0$ and $S_{L_j}^{(\pm)}(z)\rightarrow I$ as $n\rightarrow\infty$ exponentially fast away from the real line, we are naturally lead to the construction of the following model functions.
\subsubsection{Outer parametrix} We consider the following auxiliary RHP. Find $M:\mathbb{C}\backslash[a,b]\rightarrow\mathbb{C}^{4\times 4}$ such that
\begin{itemize}
	\item $M(z)$ is analytic for $z\in\mathbb{C}\backslash[a,b]$
	\item We have jumps
	\begin{eqnarray}
		M_+(z)&=&M_-(z)\bigoplus_{j=1,3}\begin{bmatrix}
		0 & z^{a_j}\\
		-z^{-a_j} & 0 \\
		\end{bmatrix},\ \ \ z\in(0,b),\label{outj1}\\
		M_+(z)&=&M_-(z)\left(1\oplus\begin{bmatrix}
		0 & (-z)^{a_2}\\
		 -(-z)^{-a_2}&0\\
		\end{bmatrix}\oplus 1\right),\ \ \ z\in(a,0)\label{outj2}
	\end{eqnarray}
	\item As $z\rightarrow\infty$,
	\be 
		M(z)=I+\mathcal{O}\left(z^{-1}\right)
	\ee 
\end{itemize}
Jump conditions in the form of \eqref{outj1}, \eqref{outj2} have appeared in the literature before, we shall use ideas similar to \cite{KAW} in the proof of the following Proposition.
\bp
\label{propM}
 Put
\be\nonumber 
	M(z)= \le[M_j\left (\mathbb T^{-1} \left(z,y_k(z)\right) \right)\ri]_{j,k=1}^4
\ee 
where $\mathbb{T}=\mathbb{T}(t)$ denotes the map $\mathbb{T}:\mathbb{CP}^1\rightarrow X$ introduced in \eqref{unizt} and 
\be\nonumber 
	M_j(t)=m_j\frac{\prod_{k=1,k\neq j}^4(t-t_k)}{(t^3(t-t_2^{\ast})(t-t_3^{\ast})(t-t_4^{\ast}))^{\frac{1}{2}}}\,\mathcal{D}(t),\ \ 
\ee 
with
\be\nonumber 
	m_1=\frac{35}{3}\frac{i}{\sqrt{67}}\big(\mathcal{D}(t_1)\big)^{-1},\ \ m_2=-\frac{40}{3}\sqrt{\frac{2}{67}}\,\big(\mathcal{D}(t_2)\big)^{-1},\ \ m_3=-\frac{56}{3}\sqrt{\frac{2}{67}}\,i\big(\mathcal{D}(t_3)\big)^{-1},\ \ m_4=\frac{70}{3}\frac{1}{\sqrt{67}}\,\big(\mathcal{D}(t_4)\big)^{-1}.
\ee 
Here $\{t_j\}_{j=1}^4=\{t_1=1,t_2=\frac{8}{7},t_3=\frac{8}{5},t_4=2\}$ and the square root function $(\prod_{j=2}^4t(t-t_j^{\ast}))^{\frac{1}{2}}$ is defined and analytic for $t\in\mathbb{C}\backslash\cup_1^3\mathcal{C}_j^-$ such that $(\prod_{j=2}^4t(t-t_j^{\ast}))^{\frac{1}{2}}\sim t^3$ as $t\rightarrow+\infty$. Moreover the scalar Szeg\"o function $\mathcal{D}(t)$ is given by
\begin{equation}\label{szego}
	\mathcal{D}(t)=\le\{\begin{array}{cc}
		\big(\frac{t-t_2}{\beta t}\big)^{a_1}\big(\frac{t-t_3}{\beta t}\big)^{a_{12}}\big(\frac{t-t_4}{\beta t}\big)^{a_{13}},&t\in \mathbb U^{-1}(X_1)\smallskip\\
		\big(\frac{\beta t}{t-t_1}\big)^{a_1}\big(\frac{t-t_3}{\beta t}\big)^{a_2}\big(\frac{t-t_4}{\beta t}\big)^{a_{23}},&t\in \mathbb U^{-1}(X_2)\smallskip\\
		\big(\frac{\beta t}{t-t_1}\big)^{a_{12}}\big(\frac{\beta t}{t-t_2}\big)^{a_2}\big(\frac{t-t_4}{\beta t}\big)^{a_3},&t\in \mathbb U^{-1}(X_3)\smallskip\\
		\big(\frac{\beta t}{t-t_1}\big)^{a_{13}}\big(\frac{\beta t}{t-t_2}\big)^{a_{23}}\big(\frac{\beta t}{t-t_3}\big)^{a_3},&t\in \mathbb U^{-1}(X_4).\smallskip
		\end{array}\ri.  
\end{equation}
which involves the normalization factor $\beta=\sqrt[4]{-\frac{1}{210}}$. Then, $M(z)$ has jumps as in \eqref{outj1},\eqref{outj2}, and we have the behavior
\begin{equation}\label{mp:1}
	M(z)=\mathcal{O}\left(z^{-\frac{3}{8}}z^{\frac{A}{4}}\right), \ \ z\rightarrow 0,\hspace{0.5cm}M(z)=I+\mathcal{O}\left(z^{-1}\right),\ \ z\rightarrow\infty,
\end{equation}
where
\begin{equation}\label{Adef}
	A=\textnormal{diag}\left[-(3a_1+2a_2+a_3),a_1-2a_2-a_3,a_1+2a_2-a_3,a_1+2a_2+3a_3\right]=\big[A_j\delta_{jk}\big]_{j,k=1}^4.
\end{equation}
\ep
\begin{proof}
The stated jump conditions \eqref{outj1} and \eqref{outj2} imply for the first row entries of $M(z)$,
\be\nonumber 
	\begin{cases}
		M_{11+}(z)=-z^{-a_1}M_{12-}(z),&z\in(0,b)\\
		M_{12+}(z)=z^{a_1}M_{11-}(z),&z\in(0,b)\\
		M_{13+}(z)=-z^{-a_3}M_{14-}(z),&z\in(0,b)\\
		M_{14+}(z)=z^{a_3}M_{13-}(z),&z\in(0,b)
	\end{cases}  \hspace{0.5cm}\begin{cases}
	M_{11+}(z)=M_{11-}(z),&z\in(a,0)\\
	M_{12+}(z)=-(-z)^{-a_2}M_{13-}(z),&z\in(a,0)\\
	M_{13+}(z)=(-z)^{a_2}M_{12-}(z),&z\in(a,0)\\
	M_{14+}(z)=M_{14-}(z),&z\in(a,0)
	\end{cases} 
\ee 
We lift the problem to the Riemann surface $X$ and treat $M_{11}(z)=M_{11}(z,y_1(z))$ as defined on the first sheet $X_1$, similarly $M_{12}$ on $X_2$, $M_{13}$ on $X_3$ and $M_{14}$ on $X_4$. Using the uniformization map $\mathbb T^{-1}:X\rightarrow\mathbb{CP}^1$, define
\begin{equation}\label{m1def}
	M_1(t)=\le\{\begin{array}{cc}
		M_{11}(z(t),y(t)),&t\in \mathbb T^{-1}(X_1)\\
		M_{12}(z(t),y(t)),&t\in \mathbb T^{-1}(X_2)\\
		M_{13}(z(t),y(t)),&t\in \mathbb T^{-1}(X_3)\\
		M_{14}(z(t),y(t)),&t\in \mathbb T^{-1}(X_4).
		\end{array}\ri.  
\end{equation}
With this the jumps for $M_{1j},j=1,2,3,4$ are translated into the $t$-plane (compare Figure \ref{uni}) as follows
\be\nonumber 
	M_{1+}(t)=\pm z^{\pm a_1}M_{1-}(t),\ t\in\mathcal{C}_1^{\pm};\ \ M_{1+}(t)=\pm (-z)^{\pm a_2}M_{1-}(t),\ t\in\mathcal{C}_2^{\pm};\ \ M_{1+}(t)=\pm z^{\pm a_3}M_{1-}(t),\ t\in\mathcal{C}_3^{\pm}
\ee 
where $z=z(t)$ as in \eqref{unizt}. We also enforce the normalization $M_{11}(z)\rightarrow 1, M_{1\ell}(z)\rightarrow 0,\,\ell=2,3,4$ as $z\rightarrow\infty$. In terms of $t$, this means that
\be\nonumber 
	M_1(1)=1,\ \ M_1\left(\frac{8}{7}\right)=0,\ \ M_1\left(\frac{8}{5}\right)=0,\ \ M_1(2)=0.
\ee 
We will seek $M_1(t)$ in the form
\be\nonumber 
	M_1(t)=c_1\frac{(t-\frac{8}{7})(t-\frac{8}{5})(t-2)}{(t^3(t-t_2^{\ast})(t-t_3^{\ast})(t-t_4^{\ast}))^{\frac{1}{2}}}\,\mathcal{D}(t),\ \ t\in\mathbb{C}\backslash\cup_1^3\mathcal{C}_j^-
\ee 
with a cut along $\mathcal{C}_1^-\cup\mathcal{C}_2^-\cup\mathcal{C}_3^-$. But this means that $\mathcal{D}(t)$ should be analytic in $\mathbb{C}\backslash\cup_1^3\mathcal{C}_j^-$ with jumps
\be\nonumber 
	\mathcal{D}_+(t)=z^{\pm a_1}\mathcal{D}_-(t),\ t\in\mathcal{C}_1^{\pm};\ \ \ \ \mathcal{D}_+(t)=(-z)^{\pm a_2}\mathcal{D}_-(t),\ t\in\mathcal{C}_2^{\pm};\ \ \ \ \mathcal{D}_+(t)=z^{\pm a_3}\mathcal{D}_-(t),\ t\in\mathcal{C}_3^{\pm}
\ee 
where $z=z(t)$. By straightforward computation, we check that $\mathcal{D}(t)$ as given in \eqref{szego} indeed satisfies the latter jumps and in order to ensure the correct normalization for $M_1(t)$ we must have
\be\nonumber 
	1=c_1\left(-\frac{3}{35}\right)\sqrt{-67}\,\mathcal{D}(1)\ \ \Leftrightarrow\ \ c_1=\frac{35}{3}\frac{i}{\sqrt{67}}\big(\mathcal{D}(1)\big)^{-1}.
\ee 
To get back from \eqref{m1def} to $M_{11}(z),M_{12}(z),M_{13}(z)$ and $M_{14}(z)$ we use
\be\nonumber 
	M_{1\ell}(z)=M_1\big(\mathbb T^{-1}(z,y_{\ell}(z))\big),\ \ \ell=1,2,3,4.
\ee 
The strategy for the remaining second, third and fourth row is identical to the previous, we obtain jumps for $M_2(t),M_3(t)$ and $M_4(t)$ as before, but we enforce slightly different normalizations, namely $M_j(t_k)=\delta_{jk}$. The remaining behavior at the origin follows from the observation that $M_j(t)(\mathcal{D}(t))^{-1}=\mathcal{O}(t^{-\frac{3}{2}})$ as $t\rightarrow 0$ and this combined with \eqref{unizt} gives \eqref{mp:1}.
\end{proof}
\br A somewhat more detailed representation for $M(z)$ near $z=0$ than \eqref{mp:1} is given by the following identity
\begin{equation}\label{outereasy}
	M(z)=\wh {M}(z)z^{-\frac{1}{8}\lambda_4}U(z)z^{\frac{A}{4}},\ \ |z|<r,\ \ z\notin\mathbb{R}
\end{equation}
where we choose principal branches for fractional exponents.
Then, $\wh {M}(z)$ is analytic at  $z=0$ and  we have
\begin{equation}\label{Umatrix}
	\lambda_4 = \textnormal{diag}\left[3,1,-1,-3\right];\hspace{0.5cm}
	U(z)=\begin{cases}
	U^+\left(e^{-i\frac{\pi}{4}A_{1}\sigma_3}\oplus e^{i\frac{\pi}{4}A_{4}\sigma_3}\right)	,& \textnormal{arg}\,z\in(0,\pi)\\
	U^-\left(e^{i\frac{\pi}{4}A_{1}\sigma_3}\oplus e^{-i\frac{\pi}{4}A_{4}\sigma_3}\right)	,& \textnormal{arg}\,z\in(-\pi,0)
	\end{cases} 
\end{equation}
with
\be 
	U^+=\begin{bmatrix}
	-\omega^{-3} & \omega^3 & -\omega^{-1} & \omega\\
	\omega^{-1} & -\omega & \omega^{-\frac{1}{3}} & -\omega^{\frac{1}{3}}\\
	-\omega & \omega^{-1} & -\omega^{\frac{1}{3}} & \omega^{-\frac{1}{3}}\\
	\omega^3 &-\omega^{-3} & \omega & -\omega^{-1}\\
	\end{bmatrix},\ \ \ \omega=e^{i\frac{3\pi}{8}},\hspace{0.5cm}U^-=U^+\begin{bmatrix}
	0 & -1 & 0& 0\\
	1 & 0 & 0&0 \\
	0& 0& 0 & -1\\
	0&0 & 1 & 0\\
	\end{bmatrix}.
\ee 
\er

\subsubsection{Local RHP at the origin $z=0$}
\begin{figure}[tbh]
\begin{center}
\includegraphics[width=0.6\textwidth]{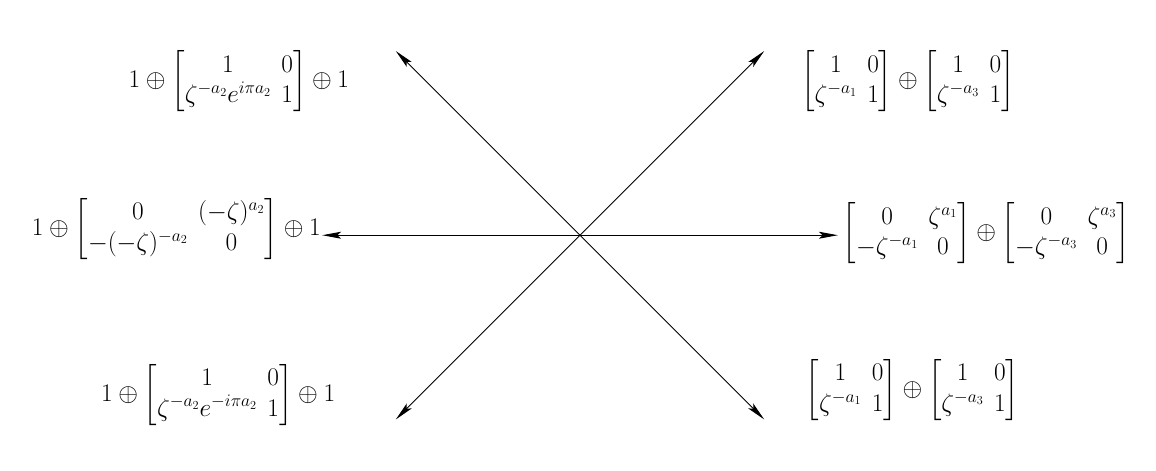}
\end{center}
\caption{A jump behavior near $\z=0$ which can be constructed explicitly using Meijer $G$-functions.}
\label{fig9}
\end{figure}

Near the origin we are looking for $4\times 4$ matrix valued function $Q(z)$ defined inside the disk $D(0,r)= \{ z\in \C:\ \ |z|< r\}$ with $0<r<\frac{4}{3}$ sufficiently small
 such that
 \\
 $\bullet$\  $Q(z)$ is analytic for $z\in D(0,r)\backslash ((-r,r)\cup\gamma_j^{\pm})$ \\
$\bullet$\  It satisfies the boundary relations (see Figure \ref{openup} for the orientations; all roots are principal)
\begin{align*}
	Q_+(z)&=Q_-(z)\bigoplus_{j=1,3}\begin{bmatrix}
		1 & 0\\
		z^{-a_j}e^{n\varphi_j(z)} & 1\\
		\end{bmatrix},\,z\in\gamma_1^{\pm};&\!\!\!\!\!\!\!\!\!\!\!\!\!\!\!\!\!\!\!\!\!\!\!\!\!\!\!\!\!\!\!\!\!Q_+(z)&=Q_-(z)\bigoplus_{j=1,3}\begin{bmatrix}
		0 & z^{a_j}\\
		-z^{-a_j} & 0\\
		\end{bmatrix},\,z\in(0,r);\\
	Q_+(z)&=Q_-(z)\left(1\oplus\begin{bmatrix}
		1 & 0\\
		z^{-a_2}e^{\mp i\pi a_2}e^{n\varphi_2(z)} & 1\\
		\end{bmatrix}\oplus 1\right),\,z\in\gamma_2^{\pm};\\
		Q_+(z)&=Q_-(z)\left(1\oplus\begin{bmatrix}
		0 & (-z)^{a_2}\\
		-(-z)^{-a_2} & 0\\
		\end{bmatrix}\oplus 1\right),\,z\in(-r,0)
\end{align*}
$\bullet$\  Near the origin it has the singular behaviour as in the RHP \ref{SRHP} for $S(z)$\\
$\bullet$\ As $n\rightarrow\infty$, we have uniformly for $|z|=r$,
	\begin{equation}\label{desire}
		Q(z)=\big(I+o(1)\big)M(z).
	\end{equation}
Our first step consists in modeling the jump behavior shown in Figure \ref{fig9} near the origin - we construct a bare parametrix $G^{(3)}(\z)$.
This construction makes use of the Meijer G-function, cf. \cite{N}, which can be defined through the Mellin-Barnes integral formula
\begin{equation}\label{MG:1}
	G_{p,q}^{\,m,n} \!\left( \le.{ a_1, \dots, a_p \atop  b_1, \dots, b_q } \; \right| \, \z \right) = \frac{1}{2\pi i}\int_L\frac{\prod_{\ell=1}^m\Gamma(b_{\ell}+s)}{\prod_{\ell=m}^{q-1}\Gamma(1-b_{\ell+1}-s)}\frac{\prod_{\ell=1}^n\Gamma(1-a_{\ell}-s)}{\prod_{\ell=n}^{p-1}\Gamma(a_{\ell+1}+s)}\,\z^{-s}\,\d s
\end{equation}
where $a_j,b_j\in\mathbb{C}$, we have $0\leq m\leq q,\,0\leq n\leq p$ and the integration contour $L$ is chosen in such a way that it separates the poles of the factors $\Gamma(b_{\ell}+s)$ from those of the factors $\Gamma(1-a_{\ell}-s)$.
The general construction for $G^{(p)}(\z)$ with $p\in\mathbb{Z}_{\geq 2}$ is accomplished in Section \ref{originparm}, Theorem \ref{originparmsol}. We avoid repeating the construction for the special case $p=3$, compare Theorem \ref{originparmsol}, and only list the relevant analytical properties of $G^{(3)}(\z)$ at this point.
\bc\label{bareprop} Let
\begin{equation*}
	\mathbb{G}^{(\pm)}(\z)=\big[(\Delta_{\z}-a_{1,k-1})^{j-1}g_k^{(\pm)}(\z)\big]_{j,k=1}^4,\ \ \ \z\in\mathbb{C}\backslash(-\infty,0];\hspace{0.5cm}\Delta_{\z}=\z\frac{\d}{\d\z}
\end{equation*}
with
\begin{equation*}
	g_m^{(\pm)}(\z)=\frac{c_m}{2\pi i}\int_L\frac{\prod_{\ell=1}^m\Gamma(s+a_{\ell,j-1})}{\prod_{\ell=m}^p\Gamma(1+a_{j\ell}-s)}e^{\pm i\pi s\sigma_m}\z^{-s}\,\d s,\ \ \z\in\mathbb{C}\backslash(-\infty,0],\hspace{0.75cm}1\leq m\leq 4.
\end{equation*}
Here, $\sigma_m\equiv (m+1)\mod 2$ and $c_m=2(2\pi i)^{4-m}(2\pi)^{-\frac{3}{2}}$. With
\begin{equation}\label{GRHdef0}
	\mathbb{G}(\z)=\begin{cases}
	\mathbb{G}^{(+)}(\z),&\,\,\,\,\,0<\textnormal{arg}\,\z<\pi\\
	\mathbb{G}^{(-)}(\z),&-\pi<\textnormal{arg}\,\z<0
	\end{cases},
\end{equation}
the bare parametrix
\begin{equation}\label{GRHdef}
	G^{(3)}(\z)=\begin{cases}
	\mathbb{G}(\z),&\textnormal{arg}\,\z\in(-\frac{3\pi}{4},-\frac{\pi}{4})\cup(\frac{\pi}{4},\frac{3\pi}{4})\\
	\mathbb{G}(\z)\left(1\oplus\begin{bmatrix}
	1 & 0\\
	\z^{-a_2}e^{i\pi a_2} & 1\\
	\end{bmatrix}\oplus 1\right),&\textnormal{arg}\,\z\in(\frac{3\pi}{4},\pi)\smallskip\\
	\mathbb{G}(\z)\left(\begin{bmatrix}
	1 & 0\\
	-\z^{-a_1} & 1\\
	\end{bmatrix}\oplus\begin{bmatrix}
	1 & 0\\
	-\z^{-a_3} & 1\\
	\end{bmatrix}\right),&\textnormal{arg}\,\z\in(0,\frac{\pi}{4})\smallskip\\
	\mathbb{G}(\z)\left(1\oplus\begin{bmatrix}
	1 & 0\\
	-\z^{-a_2}e^{-i\pi a_2} & 1\\
	\end{bmatrix}\oplus 1\right),&\textnormal{arg}\,\z\in(-\pi,-\frac{3\pi}{4})\smallskip\\
	\mathbb{G}(\z)\left(\begin{bmatrix}
	1 & 0\\
	\z^{-a_1} & 1\\
	\end{bmatrix}\oplus\begin{bmatrix}
	1 & 0\\
	\z^{-a_3} & 1\\
	\end{bmatrix}\right),&\textnormal{arg}\,\z\in(-\frac{\pi}{4},0)
	\end{cases}
\end{equation}
has jumps on the six rays $\textnormal{arg}\,\z=0,\pi,\pm\frac{\pi}{4},\pm\frac{3\pi}{4}$ as shown in Figure \ref{fig9}. It has the same singular behavior at $\z=0$ as the one stated in the RHP \ref{SRHP} (we are allowed to locally deform the lens boundaries $\gamma_j^{\pm}$ as to match the aforementioned six rays). Moreover, as $\z\rightarrow\infty$ with $\epsilon>0$ fixed,
\begin{equation}\label{bareexpinf3}
	G^{(3)}(\z)=\z^{-\frac{1}{8}\lambda_4}U(\z)\left(I+\mathcal{O}\left(\z^{-\frac{1}{4}}\right)\right)\z^{\frac{A}{4}}\begin{cases}
	e^{-4\z^{\frac{1}{4}}\Omega},&\epsilon\leq \textnormal{arg}\,\z\leq\pi-\epsilon\\
	e^{-4\z^{\frac{1}{4}}\tilde{\Omega}},&-\pi+\epsilon\leq\textnormal{arg}\,\z\leq-\epsilon
	\end{cases}
\end{equation}
where $\lambda_4,U(\z)$ and $A$ have appeared in \eqref{outereasy} and
\begin{equation*}
	\Omega=\textnormal{diag}\,\left[e^{i\frac{3\pi}{4}},e^{-i\frac{3\pi}{4}},e^{i\frac{\pi}{4}},e^{-i\frac{\pi}{4}}\right],\hspace{0.8cm}\tilde{\Omega}=\textnormal{diag}\,\left[e^{-i\frac{3\pi}{4}},e^{i\frac{3\pi}{4}},e^{-i\frac{\pi}{4}},e^{i\frac{\pi}{4}}\right].
\end{equation*}
\ec
\br The functions $g_m^{(\pm)}(\z),m=1,\ldots,4$ involved in the latter construction are all Meijer G-functions, in fact
\begin{equation*}
	g_4^{(\pm)}(\z) = \frac{2}{(2\pi)^{\frac{3}{2}}}\;G_{0,4}^{\,4,0}\!\left(\le.{--\atop 0,a_3,a_{23},a_{13}}\; \right| \,e^{\mp i\pi}\z \right),\hspace{0.5cm} g_3^{(\pm)}(\z)=\frac{2i}{\sqrt{2\pi}}\;G_{0,4}^{\,3,0}\!\left(\le.{--\atop 0,a_2,a_{12},-a_3}\; \right| \,\z \right),
\end{equation*}
and
\begin{equation*}
	g_2^{(\pm)}(\z)=-2\sqrt{2\pi}\;G_{0,4}^{\,2,0}\!\left(\le.{--\atop 0,a_1,-a_2,-a_{23}}\; \right| \,e^{\mp i\pi}\z \right),\hspace{0.5cm}
	 g_1^{(\pm)}(\z)=-2i(2\pi)^{\frac{3}{2}}\;G_{0,4}^{\,1,0}\!\left(\le.{--\atop 0,-a_1,-a_{12},-a_{13}}\; \right| \,\z \right).
\end{equation*}
\er
We now connect the $\z$-plane to the $z$-plane. The effective potentials in Definition \eqref{effpotsdef} satisfy
\begin{eqnarray*}
	\varphi_1(z)=\varphi_3(z)&=&4b^{\frac{1}{4}}e^{\pm i\frac{\pi}{2}}\left[z^{\frac{1}{4}}e_1(z)-\frac{\sqrt{3}}{16}z^{\frac{3}{4}}e_2(z)\right],\ \ \ z\in\gamma_1^{\pm}\cap D(0,r)\\
	\varphi_2(z)&=&4\sqrt{2}\,b^{\frac{1}{4}}e^{\pm i\frac{\pi}{2}}\left[\left(e^{\pm i\pi}z\right)^{\frac{1}{4}}e_1(z) + \frac{\sqrt{3}}{16}\left(e^{\pm i\pi}z\right)^{\frac{3}{4}}e_2(z)\right],\ \ \ z\in\gamma_2^{\pm}\cap D(0,r)
\end{eqnarray*}
for $0<r<\frac{4}{3}$ sufficiently small. We have chosen principal branches for $z^{\frac{1}{4}}$ and both functions $e_1(z)$ and $e_2(z)$ are analytic at $z=0$; in fact
\be\nonumber 
	e_1(z) = 1-\frac{z}{40b}+\mathcal{O}\left(z^2\right),\ \ \ e_2(z)=1+\frac{3}{14}\left(\frac{1}{2b}-1\right)z+\mathcal{O}\left(z^2\right),\ \ \ \ z\rightarrow 0.
\ee 
The expansions for $\varphi_j(z)$ motivate the use of the locally conformal change of variables
\be\nonumber 
	\z=\z(z)=\frac{16}{27}n^4z\big(e_1(z)\big)^4,\ \ -\pi<\textnormal{arg}\,\z\leq\pi,\ \ z\in D(0,r)\ \ \ \Leftrightarrow\ \ \z^{\frac{1}{4}}(z)=\frac{2n}{3^{\frac{3}{4}}}z^{\frac{1}{4}}e_1(z),\ \ -\pi<\textnormal{arg}\,z\leq\pi
\ee 
as well as the definition of the origin parametrix
\begin{equation}\label{3matori}
	Q(z)=B_0(z)G^{(3)}\big(\z(z)\big)\left(\frac{2n}{3^{\frac{3}{4}}}e_1(z)\right)^{-A}\le\{\begin{array}{cc}
	e^{4\Omega\z^{\frac{1}{4}}(z)+\frac{n}{2}3^{-\frac{1}{4}}z^{\frac{3}{4}}e_2(z)\tilde{\Omega}},& 0<\textnormal{arg}\,z<\pi,\\
	e^{4\tilde{\Omega}\z^{\frac{1}{4}}(z)+\frac{n}{2}3^{-\frac{1}{4}}z^{\frac{3}{4}}e_2(z)\Omega},& -\pi<\textnormal{arg}\,z<0.
	\end{array}\ri.  
\end{equation}
with $G^{(3)}(\z)$ as in Corollary \ref{bareprop}. In \eqref{3matori} we have chosen 
\begin{eqnarray}
	B_0(z) &=& M(z)z^{-\frac{A}{4}}U^{-1}\big(\z(z)\big)\big(\z(z)\big)^{\frac{1}{8}\lambda_4},\ \ \ |z|<r \label{multi1}\\
	&=& \wh {M}(z)z^{-\frac{1}{8}\lambda_4}\big(\z(z)\big)^{\frac{1}{8}\lambda_4},\ \ z\rightarrow 0\nonumber
\end{eqnarray}
to be analytic at the origin, compare \eqref{outereasy}.
\br
In order to achieve a control over the matching condition \eqref{desire} on the boundary of the disk $D(0,r)$ it will be necessary to re-define the multiplier $B_0(z)$ in \eqref{multi1}. This shall be accomplished in \eqref{multi4}. See Proposition \ref{matprop}.
\er

By the jump properties $G^{(3)}(\z)$, compare Corollary \ref{bareprop}, the function $Q(z)$ has the following jumps near the origin (we match the jump contours in the $S$-RHP near the origin with those in the definition of the bare parametrix by a local contour deformation)
\begin{eqnarray*}
	Q_+(z)&=&Q_-(z)\left(\begin{bmatrix}
	1 & 0\\
	z^{-a_1}e^{n\varphi_1(z)} & 1\\
	\end{bmatrix}\oplus\begin{bmatrix}
	1 & 0\\
	z^{-a_3}e^{n\varphi_3(z)} & 1\\
	\end{bmatrix}\right),\ \ \ \ z\in\gamma_1^{\pm}\cap D(0,r)\\
	Q_+(z)&=&Q_-(z)\left(\begin{bmatrix}
	0 & z^{a_1}\\
	-z^{-a_1} & 0\\
	\end{bmatrix}\oplus\begin{bmatrix}
	0 & z^{a_3}\\
	-z^{-a_3}& 0\\
	\end{bmatrix}\right),\ \ \ \ z\in(0,r)\\
	Q_+(z)&=&Q_-(z)\left(1\oplus\begin{bmatrix}
	1 & 0\\
	z^{-a_2}e^{\mp i\pi a_2}e^{n\varphi_2(z)} & 1\\
	\end{bmatrix}\oplus 1\right),\ \ \ \ z\in\gamma_2^{\pm}\cap D(0,r)\\
	Q_+(z)&=&Q_-(z)\left(1\oplus\begin{bmatrix}
	0 & (-\z)^{a_2}\\
	-(-\z)^{-a_2} & 0\\
	\end{bmatrix}\oplus 1\right),\ \ \ \ z\in(-r,0).
\end{eqnarray*} 
This matches exactly the jumps of $S(z)$ in the RHP \ref{SRHP} near the origin. Also, as another consequence of Theorem  \ref{originparmsol}, $Q(z)$ and $S(z)$ have the same singular behavior at the origin. Thus, by construction, the function $Q(z)$ is related with the exact solution $S(z)$ of the  RHP \ref{SRHP} by a left analytic multiplier $N(z)$,
\begin{equation}\label{parae}
	S(z)=N(z)Q(z),\ \ \ |z|<r.\smallskip
\end{equation}

Let us now turn towards the matching between the local model functions $Q(z)$ and $M(z)$. From \eqref{bareexpinf3}, as $n\rightarrow\infty$ (hence $|\z|\rightarrow\infty$) for $0<|z|<r$ with $r$ sufficiently small,
\begin{equation}\label{mat1}
	Q(z)\big(M(z)\big)^{-1}\sim\wh {M}(z)z^{-\frac{1}{8}\lambda_4}\Big[I+\sum_{j=1}^{\infty}K_j\,\z^{-\frac{j}{4}}\Big]H(z)z^{\frac{1}{8}\lambda_4}\big(\wh {M}(z)\big)^{-1}
\end{equation}
where we introduced the function $H(z),z\in\mathbb{C}\backslash\mathbb{R}$ given by
\begin{equation}\label{Hdef}
	H(z)U(z)=U(z)\begin{cases}
		 e^{\frac{n}{2}3^{-\frac{1}{4}}z^{\frac{3}{4}}e_2(z)\tilde{\Omega}},&\,\,\,\,\,0<\textnormal{arg}\,z<\pi\\
		e^{\frac{n}{2}3^{-\frac{1}{4}}z^{\frac{3}{4}}e_2(z)\Omega},&-\pi<\textnormal{arg}\,z<0
		\end{cases} 
\end{equation}
with $U(z)$ as in \eqref{outereasy} and the $4\times 4$ matrix valued coefficients $K_j$ depend polynomial on $\{a_k\}_{k=1}^3$ but are independent of $\z$ and $z$. 
We could, in principle, compute all coefficients $K_j$ explicitly, however our analysis requires only a certain structural information which is stated after the next Proposition.

\bp\label{Hprop} Let $z^{\gamma}$ be defined for $-\pi<\textnormal{arg}\,z\leq\pi$ such that $z^{\gamma}>0$ for $z>0$. Then $z^{-\frac{1}{8}\lambda_4}H(z)z^{\frac{1}{8}\lambda_4}$ is an entire function with
\begin{equation}\label{Hzero}
	z^{-\frac{1}{8}\lambda_4}H(z)z^{\frac{1}{8}\lambda_4}=I+h_n(0)E_{14}-\frac{z}{2}h^2_n(0)(E_{13}+E_{24})-\frac{z^3}{120}h^5_n(0)E_{14}+\mathcal{E}_n(z),\ \ z\rightarrow 0,
\end{equation}
where
\be\nonumber 
	\big|\mathcal{E}_n(z)\big|\leq cn^3|z|^2,\ \ c>0,\ \ |z|<r;\hspace{0.5cm} h_n(z)=\frac{n}{2}\,3^{-\frac{1}{4}}e_2(z).
\ee  
\ep
\begin{proof} Notice that $z^{-\frac{1}{8}\lambda_4}H(z)z^{\frac{1}{8}\lambda_4}$ has no jump on the real line, since
\begin{equation}\label{Hjump}
	U^+\tilde{\Omega}^k\big(U^+\big)^{-1} = U^-\Omega^k\big(U^-\big)^{-1} = \begin{cases}
	E_{14}-E_{21}-E_{32}-E_{43},& k=1\\
	-E_{13}-E_{24}+E_{31}+E_{42},& k=2\\
	E_{12}+E_{23}+E_{34}-E_{41},& k=3\\
	-I,&k=4,
	\end{cases} 
\end{equation}
where $E_{jk}$ are again matrix units, i.e. $E_{jk}=[\delta_{j\ell}\delta_{\ell k}]_{\ell=1}^4$, and also
\be\nonumber 
	e^{-i\frac{\pi}{4}\lambda_4}U^+\wh {\Omega}^k\big(U^+\big)^{-1}e^{i\frac{\pi}{4}\lambda_4} = U^-\Omega^k\big(U^-\big)^{-1},\ \ \ \wh {\Omega}=\textnormal{diag}\,\left[e^{i\frac{3\pi}{4}},e^{-i\frac{\pi}{4}},e^{-i\frac{3\pi}{4}},e^{-i\frac{\pi}{4}}\right].
\ee 
This means $z^{-\frac{1}{8}\lambda_4}H(z)z^{\frac{1}{8}\lambda_4}$ could only have an isolated singularity at the origin $z=0$, but with the help of \eqref{Hjump} we can compute its expansion at $z=0$, in fact
\be 
	z^{-\frac{1}{8}\lambda_4}H(z)z^{\frac{1}{8}\lambda_4}=\sum_{m=0}^{\infty}A_m(-z)^m
\ee 
with coefficients 
\be 
	A_m=\begin{cases}
	\frac{h^{4m}(z)}{(4m)!}I+\frac{h^{4m+1}(z)}{(4m+1)!}E_{14}+B_m(z),&m\equiv 0\mod 3\smallskip\\
	\frac{h^{4m+1}(z)}{(4m+1)!}(-E_{21}-E_{32}-E_{43})+\frac{h^{4m+2}(z)}{(4m+2)!}(-E_{13}-E_{24}),&m\equiv 1\mod 3\smallskip\\
	\frac{h^{4m+2}(z)}{(4m+2)!}(E_{31}+E_{42})+\frac{h^{4m+3}(z)}{(4m+3)!}(E_{12}+E_{23}+E_{34}),&m\equiv 2\mod 3,
	\end{cases}  
\ee 
where 
\be\nonumber 
	h(z)=h_n(z)=\frac{n}{2}\,3^{-\frac{1}{4}}e_2(z),\ \ \ \ B_m(z)=\begin{cases}
	0,&m=0\\
	\frac{h^{4m-1}(z)}{(4m-1)!}E_{41},&m\equiv 0\mod 3,\ \ m\geq 3.
	\end{cases}
\ee 
Thus $z^{-\frac{1}{8}\lambda_4}H(z)z^{\frac{1}{8}\lambda_4}$ is analytic at $z=0$ and we obtain the first terms written in \eqref{Hzero}.
\end{proof} 
\br\label{struc1} Subsequently we will make use of the following structure of the error term $\mathcal{E}_n(z)$,\footnote{If an error estimate $\mathcal{O}$ is not multiplied by a matrix from the right, we interpret the error estimate entry wise on the full $4\times 4$ matrix.}
\begin{eqnarray*}
	\mathcal{E}_n(z)&=&\frac{z^2}{6}h^3_n(0)(E_{12}+E_{23}+E_{34})+\big\{zh_n'(0)E_{14}-zh_n(0)(E_{21}+E_{32}+E_{43})\big\}\\
	&&\ \ \ +\mathcal{O}\left(r^4n^6\right)\left(E_{13}+E_{24}\right)+\mathcal{O}\left(r^3n^4\right),\ \ 0\leq|z|<r. \end{eqnarray*}
\er
\bp 
\label{structure} The matrix coefficients $\{K_j\}_{j=1}^{\infty}$ appearing in the asymptotic expansion \eqref{mat1} display the following structure,
\begin{equation}\label{coeffstruc}
	K_j=\begin{cases}
	\le[\begin{smallmatrix}
	0 & 0 & 0 & \ast\\
	\ast & 0 & 0 & 0\\
	0 & \ast & 0 & 0\\
	0 & 0 & \ast & 0\\
	\end{smallmatrix}\ri],&j\equiv 1\mod 4\smallskip\\
	\le[\begin{smallmatrix}
	0 & 0 & \ast & 0\\
	0 & 0 & 0 & \ast\\
	\ast& 0 & 0 & 0\\
	0 & \ast & 0 & 0\\
	\end{smallmatrix}\ri],&j\equiv 2\mod 4
	\end{cases}  \hspace{0.95cm}\textnormal{and}\hspace{0.5cm}\ K_j=\begin{cases}
	\le[\begin{smallmatrix}
	0 & \ast & 0 & 0\\
	0 & 0 & \ast & 0\\
	0 & 0 & 0 & \ast\\
	\ast & 0 & 0 & 0\\
	\end{smallmatrix}\ri],&j\equiv 3\mod 4\smallskip\\
	\le[\begin{smallmatrix}
	\ast & 0 & 0 & 0\\
	0 & \ast & 0 & 0\\
	0& 0 & \ast & 0\\
	0 &0 & 0 & \ast\\
	\end{smallmatrix}\ri],&j\equiv 4\mod 4
	\end{cases} 
\end{equation}
\ep
\begin{proof} The line of argument is almost identical to the last Proposition. Notice that $Q(z)\big(M(z)\big)^{-1}$ has no jump on $\mathbb{R}\backslash\{0\}$. Hence the coefficients in the asymptotic equality \eqref{mat1} have to be meromorphic in $z$. As we have just seen, $z^{-\frac{1}{8}\lambda_4}H(z)z^{\frac{1}{8}\lambda_4}$ is an entire function, thus the coefficients in the formal series
\begin{equation*}
	z^{-\frac{1}{8}\lambda_4}\Big[I+\sum_{j=1}^{\infty}K_j\z^{-\frac{j}{4}}\Big]z^{\frac{1}{8}\lambda_4}
\end{equation*}
can contain only integer powers of $z$. Since $\z^{-\frac{1}{4}}=3^{\frac{3}{4}}(2ne_1(z))^{-1}z^{-\frac{1}{4}}$ where $e_1(z)$ is analytic, we obtain \eqref{coeffstruc} by simply conjugating the formal series by $z^{-\frac{1}{8}\lambda_4}$ and collecting integer powers.
\end{proof}
Our goal is to achieve a matching relation between the model functions $Q(z)$ and $M(z)$ as $n\rightarrow\infty$, valid on a disk boundary $\partial D(0,r)$, compare \eqref{desire}. As can be seen from \eqref{mat1} and \eqref{Hdef} the presence of the function $H(z)$ forces us to work with a contracting radius $r=r_n$
\begin{equation}\label{shrink}
	r_n=n^{-2+\epsilon},\ \ \  \ \  0<\epsilon<\frac{1}{7}\ \ \textnormal{fixed}.
\end{equation}
Shrinking the radius in this way we obtain from \eqref{Hprop}, as $n\rightarrow\infty$ uniformly for $|z|=r_n$,
\begin{equation}\label{Hshrink}
	z^{-\frac{1}{8}\lambda_4}H(z)z^{\frac{1}{8}\lambda_4} = I+\mathcal{O}(n)E_{14}+\mathcal{O}\left(n^{\epsilon}\right)(E_{13}+E_{24})+\mathcal{O}\left(n^{-1+3\epsilon}\right)E_{14}+\mathcal{E}_n(z),
\end{equation}
with $|\mathcal{E}_n(z)|\leq c\,n^{-1+2\epsilon},c>0$. This estimate contains terms which are unbounded in $n$, but which are all analytic functions in the spectral variable $z$. Now
following Proposition \ref{structure}, we find the bound 
\beas
	z^{-\frac{1}{8}\lambda_4}\Big[I+\sum_{j=1}^{\infty}K_j\,\z^{-\frac{j}{4}}\Big]z^{\frac{1}{8}\lambda_4}&\& =I+k_1^{14}E_{14}\frac{\alpha}{nz}+(k_2^{13}E_{13}+k_2^{24}E_{24})\frac{\alpha^2}{n^2z}\\
	+(k_1^{21}E_{21}+k_1^{32}E_{32}+k_1^{43}E_{43}&\& -k_1^{14}E_{14}\beta)\frac{\alpha}{n}+(k_3^{12}E_{12}+k_3^{23}E_{23}+k_3^{34}E_{34})\frac{\alpha^3}{n^3z}+\wh {\mathcal{E}}_n(z),
\eeas
valid as $n\rightarrow\infty$ for $z\in\partial D(0,r_n)$ with $|\wh {\mathcal{E}}_n(z)|\leq c n^{-1-2\epsilon},c>0$. We used the notation $K_j=\big[k_j^{m\ell}\big]_{m,\ell=1}^4$ and
\be 
	\alpha=\frac{3^{\frac{3}{4}}}{2e_1(0)},\ \ \ \beta=\frac{e_1'(0)}{e_1(0)}.
\ee 
\br\label{struc2} Also here, we require more detail on the structure of the error term $\wh {\mathcal{E}}_n(z)$, as $n\rightarrow\infty$ uniformly for $z\in\partial D(0,r_n)$,
\be 
	\wh {\mathcal{E}}_n(z)=k_5^{14}E_{14}\frac{\alpha^5}{n^5z^2}+\big\{(k_2^{31}E_{31}+k_2^{42}E_{42})-2(k_2^{13}E_{13}+k_2^{24}E_{24})\beta\big\}\frac{\alpha^2}{n^2}+\mathcal{O}\left(n^{-2-\epsilon}\right)
\ee 
\er
Let us summarize, as $n\rightarrow\infty$ uniformly for $z\in\partial D(0,r_n)$,
\begin{eqnarray*}
	z^{-\frac{1}{8}\lambda_4}H(z)z^{\frac{1}{8}\lambda_4} &=& I+\mathcal{O}(n)+\mathcal{O}\left(n^{\epsilon}\right)+\mathcal{O}\left(n^{-1+3\epsilon}\right),\\
	z^{-\frac{1}{8}\lambda_4}\Big[I+\sum_{j=1}^{\infty}K_j\z^{-\frac{j}{4}}\Big] z^{\frac{1}{8}\lambda_4}&=& I+\mathcal{O}\left(n^{1-\epsilon}\right)+\mathcal{O}\left(n^{-\epsilon}\right)+\mathcal{O}\left(n^{-1}\right)+\mathcal{O}\left(n^{-1-\epsilon}\right).
\end{eqnarray*}
We fix $r=r_n$ as in \eqref{Hshrink} and first eliminate the unbounded terms in $z^{-\frac{1}{8}\lambda_4}H(z)z^{\frac{1}{8}\lambda_4}$ by successively redefining the left analytic multiplier $B_0(z)$.
This shall be accomplished in the {\em three steps} detailed below.\bigskip

{\it Changing $B_0(z)$ - step one}. Recall \eqref{Hzero} and move from $B_0(z)$ as in \eqref{multi1} to $B_{0,1}(z)$ given by
\begin{equation}\label{multi2}
	B_{0,1}(z)=\wh {M}(z)\left(I-h_n(0)E_{14}+\frac{z}{2}h_n^2(0)(E_{13}+E_{24})+\frac{z^3}{120}h_n^5(0)E_{14}\right)\big(\wh {M}(z)\big)^{-1}B_0(z).
\end{equation}
The parametrix $Q(z)$ defined as in \eqref{3matori} but with $B_{0,1}(z)$ instead of $B_0(z)$ still has the same analytical properties near $z=0$, however the matching \eqref{mat1} is replaced by
\begin{eqnarray}
	Q(z)\big(M(z)\big)^{-1}&=&\wh {M}(z)\Big[I+k_1^{14}E_{14}\frac{\alpha}{nz}+zh_n^2(0)E_{13}+\big(k_1^{21}E_{24}-k_1^{43}E_{13}-k_1^{14}E_{13}\big)\frac{\alpha}{n}h_n(0)\nonumber\\
	&&-zh_n(0)E_{21}+\tilde{\mathcal{E}}_n(z)\Big]\big(\wh {M}(z)\big)^{-1},\ \ \ n\rightarrow\infty,\ \ z\in\partial D(0,r_n)\label{better1}
\end{eqnarray}
where the error term $\tilde{\mathcal{E}}_n(z)$ has the following structure
\begin{eqnarray*}
	\tilde{\mathcal{E}}_n(z)&=&\mathcal{O}\left(n^{-1+3\epsilon}\right)
	\le[\begin{smallmatrix}
	0 & 0 & 0 & \ast\\
	0 & 0 & 0 & 0\\
	0 & 0 & 0 & 0\\
	0 & 0 & 0 & 0\\
	\end{smallmatrix}\ri]+\mathcal{O}\left(n^{-1+2\epsilon}\right)
	\le[\begin{smallmatrix}
	0 & \ast & 0 & \ast\\
	0 & 0 & \ast & 0\\
	0 & 0 & 0 & \ast\\
	0 & 0 & 0 & 0\\
	\end{smallmatrix}\ri]+
	\mathcal{O}\left(n^{-1+\epsilon}\right)
	 \le[\begin{smallmatrix}
	0 & \ast & 0 & \ast\\
	0 & 0 & \ast & 0\\
	0 & \ast & 0 & \ast\\
	0 & 0 & \ast & 0\\
	\end{smallmatrix}\ri]+\mathcal{O}\left(n^{-1}\right)\le[\begin{smallmatrix}
	0 & \ast & 0 & \ast\\
	\ast & 0 & \ast & 0\\
	0 & \ast & 0 & \ast\\
	0 & 0 & \ast & 0\\
	\end{smallmatrix}\ri]\\
	&& +\,\mathcal{O}\left(n^{-1-\epsilon}\right).
\end{eqnarray*}
This information is derived by directly applying Proposition \ref{Hprop} and recalling Remarks \ref{struc1} and \ref{struc2}, in principle we could compute $\tilde{\mathcal{E}}_n(z)$ explicitly. Still, estimation \eqref{better1} is not of the form \eqref{desire} since, as $n\rightarrow\infty$ uniformly for $z\in\partial D(0,r_n)$,
\be 
	k_1^{14}E_{14}\frac{\alpha}{nz}+zh_n^2(0)E_{13}+\big(k_1^{21}E_{24}-k_1^{43}E_{13}-k_1^{14}E_{13}\big)\frac{\alpha}{n}h_n(0)=\mathcal{O}\left(n^{1-\epsilon}\right)+\mathcal{O}\left(n^{\epsilon}\right)+\mathcal{O}(1).
\ee 
We now ``peel off'' the analytic terms in the latter expression by redefining the multiplier for a second time.\bigskip

{\it Changing $B_0(z)$ - step two}. Replace $B_{0,1}(z)$ by
\begin{equation}\label{multi3}
	B_{0,2}(z)=\wh {M}(z)\left(I-zh_n^2(0)E_{13}-(k_1^{21}E_{24}-k_1^{43}E_{13}-k_1^{14}E_{13})\frac{\alpha}{n}h_n(0)+zh_n(0)E_{21}\right)\big(\wh {M}(z)\big)^{-1}B_{0,1}(z).
\end{equation}
Again, the analytical properties of the parametrix $Q(z)$ with $B_{0,2}(z)$ instead of $B_0(z)$ remain unchanged, only the matching relation now reads as
\begin{equation}\label{better2}
	Q(z)\big(M(z)\big)^{-1}=\wh {M}(z)\Big[I+k_1^{14}E_{14}\frac{\alpha}{nz}+k_1^{14}E_{24}h_n(0)\frac{\alpha}{n}+\dot{\mathcal{E}}_n(z)\Big]\big(\wh {M}(z)\big)^{-1},
\end{equation}
and the error term $\dot{\mathcal{E}}_n(z)$ has to leading order the same structure as $\tilde{\mathcal{E}}_n(z)$, i.e.
\begin{equation}\label{mat6}
	\dot{\mathcal{E}}_n(z)=\mathcal{O}\left(n^{-1+3\epsilon}\right)\le[\begin{smallmatrix}
	0 & 0 & 0 & \ast\\
	0 & 0 & 0 & 0\\
	0 & 0 & 0 & 0\\
	0 & 0 & 0 & 0\\
	\end{smallmatrix}\ri]+\mathcal{O}\left(n^{-1+2\epsilon}\right)\le[\begin{smallmatrix}
	0 & \ast & 0 & \ast\\
	0 & 0 & \ast & 0\\
	0 & 0 & 0 & \ast\\
	0 & 0 & 0 & 0\\
	\end{smallmatrix}\ri]+\mathcal{O}\left(n^{-1+\epsilon}\right)\le[\begin{smallmatrix}
	0 & \ast & 0 & \ast\\
	0 & 0 & \ast & 0\\
	0 & \ast & 0 & \ast\\
	0 & 0 & \ast & 0\\
	\end{smallmatrix}\ri]+\mathcal{O}\left(n^{-1}\right)
\end{equation}
as $n\rightarrow\infty$, uniformly for $z\in\partial D(0,r_n)$. The leading growth in \eqref{better2} originates from the term $k_1^{14}\frac{\alpha}{nz}=\mathcal{O}(n^{1-\epsilon})$ which is not analytic in the disk $D(0,r_n)$, hence we cannot absorb it by another change of the analytic multiplier $B_0(z)$ -- we can only remove the constant term $k_1^{14}h_n(0)\frac{\alpha}{n}$ in this way.\bigskip

{\it Changing $B_0(z)$ - step three}. In this final step, we replace $B_{0,2}(z)$ by
\begin{equation}\label{multi4}
	B_{0,3}(z)=\wh {M}(z)\left(I-k_1^{14}E_{24}h_n(0)\frac{\alpha}{n}\right)\big(\wh {M}(z)\big)^{-1}B_{0,2}(z),
\end{equation}
and summarize our estimations in the following Proposition.
\bp\label{matprop} Let $r_n=n^{-2+\epsilon}$ with $0<\epsilon<\frac{1}{7}$ fixed. The origin parametrix $Q(z),z\in D(0,r)$ is given by \eqref{3matori} with $B_0(z)$ replaced by $B_{0,3}(z)$ as in \eqref{multi2}, \eqref{multi3} and \eqref{multi4}. Moreover, as $n\rightarrow\infty$, we have an asymptotic matching relation between the model functions $Q(z)$ and $M(z)$ of the form
\begin{equation}\label{mat7}
	Q(z)\big(M(z)\big)^{-1} = \wh {M}(0)\left[I+k_1^{14}E_{14}\frac{\alpha}{nz}+\dot{\mathcal{E}}_n(z)\right]\big(\wh {M}(0)\big)^{-1},
\end{equation}
uniformly for $z\in\partial D(0,r_n)$ where $\dot{\mathcal{E}}_n(z)$ is estimated in \eqref{mat6}.
\ep
The last Proposition completes the construction of the origin parametrix. We now briefly discuss 
\subsubsection{Parametrices near $z=a$ and $z=b$} Two remaining parametrices need to be constructed inside the disks
\be\nonumber 
	D(a,r)=\big\{z\in\mathbb{C}:\ |z-a|<r\big\},\ \ \ D(b,r)=\big\{z\in\mathbb{C}:\ |z-b|<r\big\}
\ee 
with $r>0$ sufficiently small and fixed. As for $z\in D(b,r)\cap(b,\infty)$,
\be\nonumber 
	\omega_{12}(z)=\omega_{34}(z)=-C(z-b)^{\frac{3}{2}}\left(1+\mathcal{O}(z-b)\right),\ \ C>0
\ee 
with similar expansions for $z\in\gamma_1^{\pm}\cap D(b,r)$ as well as on the jump contours near $z=a$, the relevant model functions are constructed with the help of Airy functions. These constructions are well known in the literature, see \cite{DKMVZ} for the standard Airy parametrices in the $2\times 2$ context.\footnote{The Airy parametrices of \cite{DKMVZ} were embedded in \cite{BGS2} into the $3\times 3$ situation of the Cauchy two matrix model, here we would simply embed them into the given $4\times 4$ context.} We skip the details as they are not relevant for our purposes and only list the matching relations between the endpoint parametrices $P_j(z)$ and the outer parametrix $M(z)$,
\begin{equation}\label{mat8}
	P_j(z)=\left(I+\mathcal{O}\left(n^{-1}\right)\right)M(z),\ \ n\rightarrow\infty,
\end{equation}
uniformly for $z\in\partial D(a,r)\cup \partial D(b,r)$.

\subsubsection{Ratio problem and final transformation} We introduce
\begin{equation}\label{R:1}
	R(z)=\begin{cases}
	S(z)\big(P_1(z)\big)^{-1},&|z-a|<r\\
	S(z)\big(P_2(z)\big)^{-1},&|z-b|<r\\
	S(z)\big(Q(z)\big)^{-1},&|z|<r_n\\
	S(z)\big(M(z)\big)^{-1},&|z|>r_n,\ |z-a|>r,\ |z-b|>r
	\end{cases}
\end{equation}
where $Q(z)$ is in \eqref{3matori} (with $B_0(z)$ replaced by $B_{0,3}(z)$ as in \eqref{multi4}), $P_{1,2}(z)$ as in \eqref{mat8} and $M(z)$ in Proposition \ref{propM}. 
The radius  $0<r<\frac{2}{3}$ remains fixed and $r_n=n^{-2+\epsilon}$ with $0<\epsilon<\frac{1}{7}$. This transformation leads to a ratio-RHP for $R(z)$ on a contour $\Sigma_R$ which is depicted in Figure \ref{ratioj} below.
\begin{figure}[tbh]
\begin{center}
\includegraphics[width=0.8\textwidth]{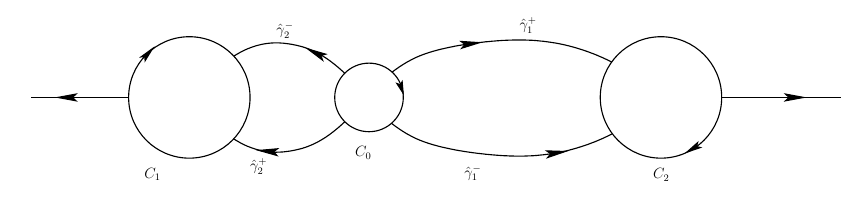}
\end{center}
\caption{Jump contour $\Sigma_R$ in the ratio problem for $R(z)$}
\label{ratioj}
\end{figure}

\begin{problem} Determine the $4\times 4$ piecewise analytic function $R(z)$ such that
\begin{itemize}
	\item $R(z)$ is analytic for $z\in\mathbb{C}\backslash\Sigma_R$ with $\Sigma_R=(-\infty,a-r)\cup(b+r,\infty)\cup C_0\cup C_1\cup C_2\cup\bigcup_{j=1}^2\wh {\gamma}_j^{\pm}$
	\item The jumps on $\Sigma_R$ are as follows
	\begin{eqnarray*}
		R_+(z)&=&R_-(z)M(z)S_{L_j}^{(\pm)}(z)\big(M(z)\big)^{-1},\ \ \ y\in\wh {\gamma}_j^{\pm},\ \ j=1,2\\
		R_+(z)&=&R_-(z)M(z)\left(\bigoplus_{j=1,3}\begin{bmatrix}
		1 & z^{a_j}e^{n\omega_{j,j+1}(z)}\\
		0 & 1\\
		\end{bmatrix}\right)\big(M(z)\big)^{-1},\ \ \ z>b+r\\
		R_+(z)&=&R_-(z)M(z)\left(1\oplus\begin{bmatrix}
		1 & (-z)^{a_2}e^{n\omega_{23}(z)}\\
		0 & 1\\
		\end{bmatrix}\oplus 1\right)\big(M(z)\big)^{-1},\ \ \ z<a-r\\
		R_+(z)&=&R_-(z)Q(z)\big(M(z)\big)^{-1},\ \ \ z\in C_0\\
		R_+(z)&=&R_-(z)P_j(z)\big(M(z)\big)^{-1},\ \ \ z\in C_j,\ \ j=1,2\\
	\end{eqnarray*}
	\item We emphasize that $R(z)$ is analytic at $z=0$, this follows from \eqref{parae} and definition \eqref{R:1}
	\item As $z\rightarrow\infty$, we have $R(z)\rightarrow I$.
\end{itemize}
\end{problem}
In order to proceed, we estimate the behavior of the latter jumps $G_R(z,n)$ as $n\rightarrow\infty$ and $z\in\Sigma_R$: on the contours of $\Sigma_R$ which extend to infinity, this is done by recalling Proposition \ref{behave}. Since $0<r_1<\frac{2}{3}$ remains fixed, we have there
\begin{equation}\label{DZ:es1}
	\|G_R(\cdot,n)-I\|_{L^2\cap L^{\infty}(b+r,\infty)}\leq d_1e^{-d_2n},\ \ \ \|G_R(\cdot,n)-I\|_{L^2\cap L^{\infty}(-\infty,a-r)}\leq d_3 e^{-d_4n},\ \ d_j>0.
\end{equation}
Next for the parts $\wh {\gamma}_j^{\pm}$ which are part of the original lens boundaries: we notice that
\be\nonumber 
	\sup_{z\in\gamma_j^{\pm}}\big|G_R(z,n)-I\big|=\sup_{z\in C_0\cap\gamma_j^{\pm}}\big|G_R(z,n)-I\big|
\ee 
and
\be\nonumber 
	\|S_{L_1}^{(\pm)}(z,n)-I\|\leq d_5|z|^{-\max\{a_1,a_3\}}e^{n\Re\,\varphi_1(z)},\ z\in\wh {\gamma}_1^{\pm};\hspace{0.5cm}\|S_{L_2}^{(\pm)}(z,n)-I\|\leq d_6|z|^{-a_2}e^{n\Re\,\varphi_2(z)},\ z\in\wh {\gamma}_2^{\pm}.
\ee  
Thus with \eqref{Adef},
\begin{equation}\label{DZ:es2}
	\|G_R(\cdot,n)-I\|_{L^2\cap L^{\infty}(\gamma_1^{\pm})}\leq d_7n^{\frac{3}{2}(1-\frac{\epsilon}{2})}e^{-d_8n^{\frac{1}{2}+\frac{\epsilon}{4}}},\ \ \ \|G_R(\cdot,n)-I\|_{L^2\cap L^{\infty}(\gamma_2^{\pm})}\leq d_9n^{\frac{3}{2}(1-\frac{\epsilon}{2})}e^{-d_{10}n^{\frac{1}{2}+\frac{\epsilon}{4}}},
\end{equation}
which ensures that, even with a shrinking disk $C_0$, the lens boundary contributions are exponentially close to the identity matrix in the limit $n\rightarrow\infty$. On the circles $C_j,j=1,2$ we obtain a power like decay from \eqref{mat8},
\begin{equation}\label{DZ:es3}
	\|G_R(\cdot,n)-I\|_{L^2\cap L^{\infty}(C_j)}\leq \frac{d_{11}}{n},\ \ n\rightarrow\infty,\ \ j=1,2.
\end{equation}
As for the corresponding estimation on $C_0$, we have already seen in \eqref{mat7}, that $G_R(z,n)=Q(z)\big(M(z)\big)^{-1},z\in C_0$ is not uniformly close to the identity matrix. We resolve this issue with another transformation: note that (with $\wh M(z)$ as defined in \eqref{outereasy})
\be\nonumber 
	F(z,n)=\left[\wh {M}(0)\left(I+k_1^{14}E_{14}\frac{\alpha}{nz}\right)\big(\wh {M}(0)\big)^{-1}\right]^{-1}=\wh {M}(0)\left(I-k_1^{14}E_{14}\frac{\alpha}{nz}\right)\big(\wh {M}(0)\big)^{-1}
\ee 
exists and
\be\nonumber 
	F(z,n)=I+\mathcal{O}\left(z^{-1}\right),\ \ z\rightarrow\infty.
\ee 
We define
\begin{equation}\label{R:2}
	X(z)=\begin{cases}
		R(z),&|z|\leq r_n\\
		R(z)F(z,n),& |z|>r_n,
	\end{cases}  \hspace{0.5cm} \textnormal{with}\ \ r_n=n^{-2+\epsilon},\ \ 0<\epsilon<\frac{1}{7}\ \ \textnormal{fixed}
\end{equation}
and obtain a RHP for $X(z)$ which is posed on the same contour $\Sigma_R$ as shown in Figure \ref{ratioj} 
\begin{problem} Determine the $4\times 4$ piecewise analytic function $X(z)$ such that
	\begin{itemize}
	\item $X(z)$ is analytic for $z\in\mathbb{C}\backslash\Sigma_R$
	\item The jumps equal
	\begin{eqnarray*}
	X_+(z)&=&X_-(z)G_R(z)F(z,n),\ \ z\in C_0\\
	X_+(z)&=&X_-(z)\big(F(z,n)\big)^{-1}G_R(z)F(z,n),\ \ z\in\Sigma_R\backslash C_0
	\end{eqnarray*}
	\item $X(z)$ is analytic at the origin
	\item As $z\rightarrow\infty$,
	\be\nonumber 
		X(z)=I+\mathcal{O}\left(z^{-1}\right),\ \ z\rightarrow\infty.
	\ee 
	\end{itemize}
\end{problem}
Since for $n\rightarrow\infty$,
\be\nonumber 
	\big(F(z,n)\big)^{\pm 1} = I+\mathcal{O}\left(n^{-1}\right),\ \ z\in C_j,\ j=1,2;\hspace{0.5cm} \big(F(z,n)\big)^{\pm 1}=I+\mathcal{O}\left(n^{1-\epsilon}\right),\ \ z\in C_0,
\ee 
we obtain
\begin{equation}\label{DZ:es4}
	\|G_X(\cdot,n)-I\|_{L^2\cap L^{\infty}(\wh {\gamma}_j^{\pm})}\leq d_{12}n^{\frac{5}{2}(1-\frac{7}{10}\epsilon)}e^{-d_{13}n^{\frac{1}{2}+\frac{\epsilon}{4}}},\ \ j=1,2
\end{equation}
as well as estimations on $C_j,j=1,2$ and $(-\infty,a-r)\cup(b+r,\infty)$ which are identical to \eqref{DZ:es3} and \eqref{DZ:es1}. For the relevant estimation on $C_0$, we recall \eqref{mat7} and in particular \eqref{mat6}. The latter expansion shows that right multiplication of $\dot{\mathcal{E}}_n(z)$ with $E_{14}$ does not affect the terms in \eqref{mat6} up to $\mathcal{O}\left(n^{-1}\right)$. But this means that we have the following estimation
\begin{equation}\label{DZ:es5}
	\|G_X(\cdot,n)-I\|_{L^2\cap L^{\infty}(C_0)}\leq\frac{d_{14}}{n^{\epsilon}},\ \ n\geq n_0,
\end{equation}
which, combined with \eqref{DZ:es1},\eqref{DZ:es3} and \eqref{DZ:es4}, guarantees the unique solvability of the $X$-RHP (cf. \cite{DZ}) for sufficiently large $n$.
\subsubsection{Iterative solution of the $X$-RHP}
The $X$-RHP is equivalent to solving the singular integral equation
\begin{equation}\label{integraleq}
	X_-(z)=I+\frac{1}{2\pi i}\int_{\Sigma_R}X_-(\lambda)\big(G_X(\lambda)-I\big)\frac{\d\lambda}{\lambda-z},\ \ z\in\Sigma_R.
\end{equation}
As we have seen in the latter subsection, there exists $n_0>0$ such that
\be\nonumber 
	\|G_X(\cdot,n)-I\|_{L^2\cap L^{\infty}(\Sigma_R)}\leq \frac{c}{n^{\epsilon}},\ \ \forall\ n\geq n_0,\ \ 0<\epsilon<\frac{1}{7}\ \ \textnormal{fixed}
\ee 
and therefore \eqref{integraleq} can be solved uniquely in $L^2(\Sigma_R)$ via iteration. The solution satisfies
\be\nonumber 
	\|X_-(\cdot,n)-I\|\leq \frac{c}{n^{\epsilon}},\ \ n\geq n_0
\ee 
and we have
\begin{equation}\label{itesti}
	X(z)=\mathcal{O}\left(\frac{n^{-\epsilon}}{1+|z|}\right),\ \ n\geq n_0,\ \ z\in\mathbb{C}\backslash\Sigma_R.
\end{equation}
The latter estimation completes the asymptotical analysis of the initial RHP 
\ref{ChainRHP} for $p=3$ and the choice \eqref{CLaguerre}.

\subsection{Proof of Conjecture \ref{conjpchain} for the Cauchy-Laguerre three matrix model}
Following our general discussion in Section \ref{sec1}, we need to analyze nine correlation kernels, compare \eqref{eynardmeh} and \eqref{rep}. We scale $x$ and $y$ as
\begin{equation}\label{scaleker}
	x=\frac{27}{16}\frac{\xi}{n^4},\ \ \ \ \ y=\frac{27}{16}\frac{\eta}{n^4},\ \ \xi,\eta>0 ,
\end{equation}
and are now interested in the $n\to \infty$ behavior of $\mathbb{K}_{j\ell}(x,y)$  given in \eqref{MK}, Theorem \ref{theo2}. We need to unravel the sequence of transformations
\be\nonumber 
	\Gamma(z;n=N)\mapsto Y(z)\mapsto S(z)\mapsto R(z)\mapsto X(z)
\ee 
to solve the initial $\Gamma$-RHP.  Through the first transformation \eqref{g:0},
\begin{eqnarray}
	\mathbb{K}_{j\ell}(x,y)&=&\frac{(-1)^{\ell-1}}{(-2\pi i)^{j-\ell+1}}e^{-\frac{1}{2}U_j(x)-\frac{1}{2}U_{\ell}(y)}e^{\frac{n}{4}(\mathfrak{l}_{j+1}-\mathfrak{l}_{\ell})}e^{n(\mathfrak{g}_+^{(\ell)}(y(-1)^{\ell-1})-\mathfrak{g}_+^{(j+1)}(x(-1)^{j+1}))}\nonumber\\
	&&\times\left[\frac{Y_+^{-1}(w)Y_+(z)}{w-z}\right]_{j+1,\ell}\Bigg|_{w=x(-1)^{j+1},\ z=y(-1)^{\ell-1}}\nonumber\\
	&=&\frac{(-1)^{\ell-1}}{(-2\pi i)^{j-\ell+1}}x^{\frac{1}{2}a_j}y^{\frac{1}{2}a_{\ell}}\exp\left[n\int_0^{w}y_{j+1,+}(\lambda)\d\lambda-n\int_0^{z}y_{\ell,+}(\lambda)\d\lambda\right]\label{scsi:1}\\
	&&\times\,\left[\frac{Y_+^{-1}(w)Y_+(z)}{w-z}\right]_{j+1,\ell}\Bigg|_{w=x(-1)^{j+1},\ z=y(-1)^{\ell-1}}\nonumber.
\end{eqnarray}
To obtain \eqref{scsi:1}, one uses the explicit expressions for the $\mathfrak{g}^{(k)}(z)$ functions. With the help of the transformation sequence $Y(z)\mapsto S(z)\mapsto R(z)\mapsto X(z)$, we have for $z\in\mathbb{R}$ with $|z|=\mathcal{O}(n^{-4})$,
\be\label{inter:1} 
	Y_+(z) = X(z)B_{0,3}(z)\mathbb G^{(+)}(\z(z))\left(\frac{2n}{3^{\frac{3}{4}}}e_1(z)\right)^{-A}\begin{cases}
	\exp\left[4\z_+^{\frac{1}{4}}(z)\Omega+\frac{n}{2}3^{-\frac{1}{4}}z_+^{\frac{3}{4}}e_2(z)\tilde{\Omega}\right],& z>0\smallskip\\
	\exp\left[4\z_+^{\frac{1}{4}}(z)\tilde{\Omega}+\frac{n}{2}3^{-\frac{1}{4}}z_+^{\frac{3}{4}}e_2(z)\Omega\right],&z<0
	\end{cases} 
\ee 
as the effect of the opening of lenses transformation $Y(z)\mapsto S(z)$ is compensated in the definition of the origin parametrix $Q(z)$, more precisely through \eqref{GRHdef}, the conjugation with $(\cdots)^{-A}$ and the piecewise defined exponential factors in the last equality. Also, we chose to approach $z\in\mathbb{R}$ from the $(+)$ side, as this choice was immaterial, compare Theorem \ref{theo2}. Thus for $|z|,|w|=\mathcal{O}(n^{-4})$,
\begin{align}
	&\left[Y^{-1}_+(w)Y_+(z)\right]_{j+1,\ell}=\left(\frac{2n}{3^{\frac{3}{4}}}e_1(w)\right)^{A_{j+1}}\left(\frac{2n}{3^{\frac{3}{4}}}e_1(z)\right)^{-A_{\ell}}\begin{cases}
	e^{-4\z_+^{\frac{1}{4}}(w)\Omega_{j+1}-\frac{n}{2}3^{-\frac{1}{4}}w_+^{\frac{3}{4}}e_2(w)\tilde{\Omega}_{j+1}},&w>0\\
	e^{-4\z_+^{\frac{1}{4}}(w)\tilde{\Omega}_{j+1}-\frac{n}{2}3^{-\frac{1}{4}}w_+^{\frac{3}{4}}e_2(w)\Omega_{j+1}},&w<0
	\end{cases}
	\label{scsi:2}\\
	&\times\begin{cases}
	e^{4\z_+^{\frac{1}{4}}(z)\Omega_{\ell}+\frac{n}{2}3^{-\frac{1}{4}}z_+^{\frac{3}{4}}e_2(z)\tilde{\Omega}_{\ell}},& z>0\\
	e^{4\z_+^{\frac{1}{4}}(z)\tilde{\Omega}_{\ell}+\frac{n}{2}3^{-\frac{1}{4}}z_+^{\frac{3}{4}}e_2(z)\Omega_{\ell}},&z<0
	\end{cases}  \,\times
	\Big[\big( \mathbb G^{(+)}(\z(w))\big)^{-1}B_{0,3}^{-1}(w)X^{-1}(w)X(z)B_{0,3}(z)\mathbb G^{(+)}(\z(z))\Big]_{j+1,\ell}\nonumber
\end{align}
where we use the notation $\Omega=[\Omega_j\delta_{jk}]_{j,k=1}^4$ and similarly $\tilde{\Omega}=[\tilde{\Omega}_j\delta_{jk}]_{j,k=1}^4$. Now we check that for $w>0$ and $w=\mathcal{O}(n^{-4})$,
\begin{eqnarray*}
	\int_0^wy_{1+}(\lambda)\,\d\lambda&=&-\frac{8}{3^{\frac{3}{4}}}e^{-i\frac{\pi}{4}}w^{\frac{1}{4}}e_1(w)-\frac{3^{-\frac{1}{4}}}{2}e^{i\frac{\pi}{4}}w^{\frac{3}{4}}e_2(w)\\
	\int_0^wy_{2+}(\lambda)\,\d\lambda&=&-\frac{8}{3^{\frac{3}{4}}}e^{i\frac{\pi}{4}}w^{\frac{1}{4}}e_1(w)-\frac{3^{-\frac{1}{4}}}{2}e^{-i\frac{\pi}{4}}w^{\frac{3}{4}}e_2(w)
\end{eqnarray*}
as well as for $w<0$ and $w=\mathcal{O}(n^{-4})$,
\begin{eqnarray*}
	\int_0^wy_{1+}(\lambda)\,\d\lambda&=&-\frac{8}{3^{\frac{3}{4}}}|w|^{\frac{1}{4}}e_1(w)+\frac{3^{-\frac{1}{4}}}{2}|w|^{\frac{3}{4}}e_2(w)\\
	\int_0^wy_{2+}(\lambda)\,\d\lambda&=&\frac{8}{3^{\frac{3}{4}}}e^{i\frac{\pi}{2}}|w|^{\frac{1}{4}}e_1(w)+\frac{3^{-\frac{1}{4}}}{2}e^{i\frac{\pi}{2}}|w|^{\frac{3}{4}}e_2(w).
\end{eqnarray*}
Combining the latter in \eqref{scsi:2} with \eqref{scsi:1},
\begin{eqnarray}
	\mathbb{K}_{j\ell}(x,y)&=&\frac{(-1)^{\ell-1}}{(-2\pi i)^{j-\ell+1}}x^{\frac{1}{2}a_j}y^{\frac{1}{2}a_{\ell}}\left(\frac{2n}{3^{\frac{3}{4}}}e_1(w)\right)^{A_{j+1}}\left(\frac{2n}{3^{\frac{3}{4}}}e_1(z)\right)^{-A_{\ell}}\frac{1}{x(-1)^{j+1}-y(-1)^{\ell-1}}\label{scsi:3}\\
	&&\times\,\Big[\big(\mathbb G^{(+)}(\z(w))\big)^{-1}B_{0,3}^{-1}(w)X^{-1}(w)X(z)B_{0,3}(z)\mathbb G^{(+)}(\z(z))\Big]_{j+1,\ell}\Bigg|_{w=x(-1)^{j+1},\ z=y(-1)^{\ell-1}}\nonumber
\end{eqnarray}
valid for $x,y=\mathcal{O}(n^{-4})$. For the remaining matrix use \eqref{itesti} and recall the definitions of the analytic multipliers $B_{0,k}(z)$, thus for $w=x(-1)^{j+1}$ and $z=y(-1)^{\ell-1}$
\beas
	 \big(\mathbb G^{(+)}(\z(w))\big)^{-1}B_{0,3}^{-1}(w)X^{-1}(w)X(z)B_{0,3}(z) \mathbb G^{(+)} (\z(z))=\big( \mathbb G^{(+)} (&\&\xi(-1)^{j+1})\big)^{-1} \mathbb G^{(+)} (\eta(-1)^{\ell-1})\\
	 &\& +\mathcal{O}\left(\frac{\xi(-1)^{j+1}-\eta(-1)^{\ell-1}}{n^{\epsilon+\frac{5}{4}}}\right).
\eeas
It is important to observe that in the last equality the choice of the limiting values $(\pm)$ would lead to different results as we are not choosing specific entries of the matrix product $( \mathbb G^{(\pm)} (w))^{-1} \mathbb G^{(\pm)} (z)$. This is however irrelevant for our purposes since \eqref{scsi:3} selects concrete entries.\smallskip

Notice now that all explicit $n$ dependent terms in the right hand side of \eqref{scsi:3} are taken to the exponent
\be 
	\kappa_{j\ell}=-\frac{1}{2}(p+1)(a_j+a_{\ell})+A_{j+1}-A_{\ell},\ \ 1\leq j,\ell\leq p,
\ee 
in \eqref{scsi:3} with the special choice $p=3$. In order to complete the proof of Theorem \ref{thmmain1} for this special choice as well as to state the general conjecture \ref{conjpchain}, we require the following Lemma
\bl
\label{lemmaeta}
Let $\{A_j\}_{j=1}^{p+1}$ be solutions of the linear system $A_{j+1}-A_j=(p+1)a_j$ which add up to zero. Then 
\be
\kappa=\big[\kappa_{j\ell}\big]_{j,\ell=1}^p:\hspace{0.5cm}\kappa_{j\ell} = -\frac{1}{2}(p+1)(a_j + a_\ell) + A_{j+1} - A_\ell\ ,\ \ \ 1 \leq \ell, j \leq p
\ee
is a skew--symmetric $p\times p$ matrix and 
\be
\label{etaconstants}
\kappa_{j\ell}  = \varpi_j - \varpi_\ell, \ \ \textnormal{with}\ \ \ \varpi_j   = (p+1) \le(a_{1j} - \frac {a_j}2\ri).
\ee
\el
\begin{proof}  If $j=\ell$ it is immediately seen that $\kappa_{jj}=0$. Assume now $\ell<j$, then 
\begin{eqnarray*}
	\kappa_{j\ell} &=& -\frac {p+1}2 (a_j + a_\ell) + A_{j+1} - A_\ell=
-\frac {p+1}2 (a_j + a_\ell) + (p+1) \sum_{k=\ell} ^{j} a_k\nonumber\\
&=&(p+1)\sum_{k=\ell}^{j-1}a_k+\frac{1}{2}(p+1)(a_j-a_{\ell})
 =(p+1) \le(a_{1j} - \frac {a_j}2 \ri)  -(p+1) \le(a_{1\ell} - \frac {a_\ell}2\ri) = \varpi_j - \varpi_\ell.\\
	\kappa_{\ell j}&=&-\frac{p+1}{2}(a_{\ell}+a_j)-(A_j-A_{\ell+1})=-\frac{p+1}{2}(a_{\ell}+a_j)-(p+1)\sum_{k=\ell+1}^{j-1}a_k\nonumber\\
	&=&-(p+1)\sum_{k=\ell}^{j-1}a_k-\frac{1}{2}(p+1)(a_j-a_{\ell})=-\kappa_{j\ell}\nonumber
\end{eqnarray*}
which implies the stated skew-symmetry.
\end{proof}
Up to this point we have thus proven
\bt\label{theorig} For any $1\leq j,\ell\leq p$, $p=3$ with $c_0=\frac{27}{16}$,
\begin{equation}\label{scalelimit}
	 \lim_{n\rightarrow\infty} \frac{c_0}{n^{p+1}}n^{\eta_{\ell}-\eta_j}\mathbb{K}_{j\ell}\left(\frac{c_0}{n^{p+1}}\xi,\frac{c_0}{n^{p+1}}\eta\right)=\frac{(-1)^{\ell-1}c_0^{\frac{1}{p+1}(\varpi_{\ell}-\varpi_j)}}{(-2\pi i)^{j-\ell+1}}\xi^{\frac{1}{2}a_j}\eta^{\frac{1}{2}a_{\ell}}\left[\frac{\mathbb G^{-1}(w) \mathbb G(z)}{w-z}\right]_{j+1,\ell}\Bigg|_{\substack{w=\xi(-1)^{j+1}\\ z=\eta(-1)^{\ell-1}}}
\end{equation}
where the choice of limiting values $(\pm)$ in the matrix entries upon evaluation at $w=\xi(-1)^{j+1}$ and $z=\eta(-1)^{\ell-1}$ is immaterial 
and the stated convergence is uniform for $\xi,\eta$ chosen from compact subsets of the half line $(0,\infty)\subset\mathbb{R}$.\et
\begin{proof}
We only need to address the independence of choice of the limiting values and here our argument already appeared (implicitly) in the computations which lead to Theorem \ref{theo2}. Also the same logic applies to the general $p\in\mathbb{Z}_{\geq 2}$ bare parametrix $ \mathbb G(\z)$ which is constructed in the next section. Note that (compare Theorem \ref{originparmsol} below, in particular \eqref{Gpm}, or also \eqref{GRHdef0})
\begin{eqnarray*}
	 \mathbb G_+(\z)&=& \mathbb G_-(\z)\left(1\oplus \begin{bmatrix}
	1 & (-\z)^{a_2}\\
	0 & 1\\
	\end{bmatrix}\oplus 1\right),\ \ \z<0\\
	 \mathbb G_+(\z)&=& \mathbb G_-(\z)\left(\begin{bmatrix}
	1 & \z^{a_1}\\
	0 & 1\\
	\end{bmatrix}\oplus\begin{bmatrix}
	1 & \z^{a_3}\\
	0 & 1\\
	\end{bmatrix}\right),\ \ \z>0
\end{eqnarray*}
which shows the same (block) jump structure as the original $\Gamma$-RHP. Qualitatively it tells us that the first column of $ \mathbb G(\z)$ is an entire function and subsequently all even numbered columns are analytic in $\mathbb{C}\backslash[0,\infty)$ while all odd numbered ones are analytic in $\mathbb{C}\backslash(-\infty,0]$. For $( \mathbb G(\z))^{-1}$ the situation is reversed, there the last row is entire and subsequently all even numbered rows are analytic in $\mathbb{C}\backslash(-\infty,0]$ and all odd numbered in $\mathbb{C}\backslash[0,\infty)$. But since the entries under consideration are as follows
\beas
	j,\ell\equiv 1\mod 2:&\hspace{0.2cm} \left[\big( \mathbb G (w)\big)^{-1} \mathbb G(z)\big)\right]_{j+1,\ell}\bigg|_{ {w=\xi>0\atop  z=\eta>0}}\\
	j\equiv 1,\,\, \ell\equiv  0\mod 2:&\hspace{0.2cm} \left[\big( \mathbb G (w)\big)^{-1} \mathbb G (z)\big)\right]_{j+1,\ell}\bigg|_{ {w=\xi>0\atop  z=-\eta<0}}\\
	j\equiv 0,\,\, \ell\equiv  1\mod 2:&\hspace{0.2cm} \left[\big( \mathbb G (w)\big)^{-1} \mathbb G (z)\big)\right]_{j+1,\ell}\bigg|_{{w=-\xi<0\atop  z=\eta>0}}\\
	j,\ell\equiv 0\mod 2:&\hspace{0.2cm}\left[\big( \mathbb G (w)\big)^{-1} \mathbb G (z)\big)\right]_{j+1,\ell}\bigg|_{{w=-\xi<0\atop z=-\eta<0}}
\eeas
it is now evident that the choice of limiting values in the matrix entries upon evaluation is  immaterial.
\end{proof}
The latter Theorem proves that all local scaling limits of the correlation kernels in the given Cauchy-Laguerre three matrix chain are determined by specific entries of $\mathbb G^{-1} (w) \mathbb G (z)$, with $ \mathbb G (\z)$ being constructed out of Meijer G-functions, compare Corollary \ref{bareprop}. We expect that for general $p\in\mathbb{Z}_{\geq 2}$ similar identities as \eqref{scalelimit} hold, compare Conjecture \ref{conjpchain}, that is the limits of the correlation functions $\mathbb{K}_{j\ell}(x,y)$ to be proportional to the ratio  
\be\nonumber 
	\left[\frac{\mathbb G^{-1}(w) \mathbb G(z)}{w-z}\right]_{j+1,\ell}\bigg|_{w=\xi(-1)^{j+1},\ z=\eta(-1)^{\ell-1}}.
\ee 
For $w,z\in\mathbb{C}\backslash\mathbb{R}$ the explicit computation of  $\mathbb G^{-1}(w) \mathbb G (z)$ is achieved in the following section.
\subsubsection{General origin parametrix}\label{originparm}
The analog of the RHP for the bare parametrix $G^{(p)} (\z)$ in the general $p\geq 2$ chain can be evinced by repeating the steps that we have taken for $p=3$. 
\begin{problem}[Bare Meijer-G parametrix for $p$-chain] \label{pcMGRHP}
 Let $G^{(p)}(\z)$ be a $(p+1)\times (p+1)$ piecewise analytic matrix function analytic in $\C$ minus the rays 
 $\mathfrak r_{0} =  \R_+$,  $\mathfrak r_{5} = - \R_+$ $\mathfrak r_{1,2} =  {\rm e}^{\pm i\frac \pi 4} \R_+$\ , $\mathfrak r_{3,4} ={\rm e}^{\pm i\frac {3\pi} 4} \R_+$ which are all oriented from the origin towards $\z=\infty$. 
With
 \be 
	\lambda_{p+1}=\textnormal{diag}\left[p, p-2,p-4, \ldots, -p\right],\ \ \ A=\textnormal{diag}\left[A_1,\ldots,A_{p+1}\right]\ ,
\ee 
where $A_{j+1}-A_j=(p+1)a_j,1\leq j\leq p$ such that $\sum_{j=1}^{p+1}A_j=0$, the jumps on the $6$ rays $\mathfrak r_j$ equal
\begin{equation*}
 	 G^{(p)}_+(\z) =  G^{(p)}_-(\z) J_\ell\ \  \textnormal{for}\ \z\in \mathfrak r_\ell,\ \ \ell=0,\dots, 5.
\end{equation*}
As $\z\rightarrow 0$, we have a singular behavior as in \eqref{Gammasing} and \eqref{zerobeh} approaching the origin from the top and bottom sectors. Furthermore, the asymptotic behavior at infinity in the half planes is given by:
 \begin{equation}\label{bareexpinf}
	G^{(p)} (\z) = \z^{-\frac{\lambda_{p+1}}{2(p+1)}}U^{\pm}\left(I+\mathcal{O}\left(\z^{-\frac{1}{p+1}}\right)\right)\z^{\frac{A}{p+1}}
	\exp\left[-(p+1)\z^{\frac{1}{p+1}}\,\Omega_{\pm} \right],\ \ \z\in \mathbb H^{\pm}.
\end{equation}
Here the constants $U^{\pm}$ and $\Omega_\pm$ as well as the jump matrices take the following forms depending on the parity of $p$.\par
For $p \equiv 1  \mod 2$ we have,
\begin{eqnarray*}
	J_{1,2}&=& \bigoplus_{k=0}^{\frac{1}{2}(p-1)}\begin{bmatrix}
	1 & 0\\
	\z^{-a_{2k+1}} & 1\\
	\end{bmatrix},\hspace{0.5cm}
	J_{3,4}=\left(1\oplus\bigoplus_{k=0}^{\frac{1}{2}(p-3)}\begin{bmatrix}
	1 & 0\\
	\z^{-a_{2k+2}}e^{\pm i\pi a_{2k+2}} & 1\\
	\end{bmatrix}\oplus 1\right),\\
	J_0 &=&\bigoplus_{k=0}^{\frac{1}{2}(p-1)}\begin{bmatrix}
	0 & \z^{a_{2k+1}}\\
	-\z^{-a_{2k+1}} & 0\\
	\end{bmatrix},\ \ 
	J_5 =\left(1\oplus\bigoplus_{k=0}^{\frac{1}{2}(p-3)}\begin{bmatrix}
	0 & (-\z)^{a_{2k+2}}\\
	-(-\z)^{-a_{2k+2}} & 0\\
	\end{bmatrix}\oplus 1\right);\\
	U^+&=&
	\left[(-1)^{k+j-1}\omega^{(-1)^k(p-2\lfloor\frac{k-1}{2}\rfloor)}\omega^{(-1)^{k-1}\frac{2}{p}(p-2\lfloor\frac{k-1}{2}\rfloor)(j-1)}\right]_{j,k=1}^{p+1}
	\bigoplus_{k=1}^{\frac{1}{2}(p+1)}\omega^{\frac{2}{p}(\sum_{j=2k}^{p+1}A_j)\sigma_3},\\
	U^-&=& U^+ \left(\bigoplus_{k=1}^{\frac{1}{2}(p+1)}(-i\sigma_2)\right)
\ ,\qquad	\Omega_\pm =\bigoplus_{k=0}^{\frac{1}{2}(p-1)}\omega^{\pm \frac{2}{p}(p-2k)\sigma_3}
\ ,\qquad 
\omega\equiv e^{i\frac{\pi}{2}\frac {p}{p+1}}.\bigskip
\end{eqnarray*}

On the other hand, for $p\equiv 0\mod 2$:
\begin{eqnarray*}
	J_{1,2} &=&\left(\bigoplus_{k=0}^{\frac{1}{2}(p-2)}\begin{bmatrix}
	1 & 0\\
	\z^{-a_{2k+1}} & 1\\
	\end{bmatrix}\oplus 1\right),\ \ J_{3,4} = \left(1\oplus\bigoplus_{k=0}^{\frac{1}{2}(p-2)}\begin{bmatrix}
	1 & 0\\
	\z^{-a_{2k+2}}e^{\pm i\pi a_{2k+2}} & 1\\
	\end{bmatrix}\right),\\
	J_0&=&\left(\bigoplus_{k=0}^{\frac{1}{2}(p-2)}\begin{bmatrix}
	0 & \z^{a_{2k+1}}\\
	-\z^{-a_{2k+1}} & 0\\
	\end{bmatrix}\oplus 1\right),\ \ 
	J_5 = \left(1\oplus\bigoplus_{k=0}^{\frac{1}{2}(p-2)}\begin{bmatrix}
	0 & (-\z)^{a_{2k+2}}\\
	-(-\z)^{-a_{2k+2}} & 0\\
	\end{bmatrix}\right).\\
	U^+&=&
\left[(-1)^{j-1}\omega^{(-1)^k(p-2\lfloor\frac{k-1}{2}\rfloor)}\omega^{(-1)^{k-1}\frac{2}{p}(p-2\lfloor\frac{k-1}{2}\rfloor)(j-1)}\right]_{j,k=1}^{p+1}	\left(\bigoplus_{k=1}^{\frac{p}{2}}\omega^{\frac{2}{p}(\sum_{j=2k}^{p+1}A_j)\sigma_3}\oplus 1\right)\\
	U^-&=&U^+\left(\bigoplus_{k=1}^{\frac{p}{2}}\,\,(-i\sigma_2)\oplus 1\right) \ ,\qquad \Omega_\pm =\bigoplus_{k=0}^{\frac{1}{2}(p-2)}\omega^{\pm \frac{2}{p}(p-2k)\sigma_3}\oplus 1.
\end{eqnarray*}
\end{problem}
\bt[Solution of the RHP \ref{pcMGRHP}]\label{originparmsol}
Let $\sigma_j = (j+1)\mod 2$ and 
\bea\label{com:1} 
	g_j^{(\pm)}(\z)=\frac{c_j}{2\pi i}\int_{L}\frac{\prod_{\ell=1}^{j}\Gamma(s+a_{\ell,j-1})}{\prod_{\ell=j}^p\Gamma(1+a_{j\ell}-s)}e^{\pm i\pi s\sigma_j}\z^{-s}\,\d s,\ \ \ 1\leq j\leq p+1\, \ \ \z\in \C\backslash(-\infty,0].
\\
\label{cjs}
	c_j = (2\pi i)^{p+1-j} \sqrt{\frac{p+1}{(2\pi)^p}},
\eea
and the contour of integration $L$ leaves all possible singularities of the integrands in \eqref{com:1} to the left. Let
\begin{equation*}
	\mathbb G^{(\pm)}(\z)=\left[\left(\Delta_{\z}-a_{1,k-1}\right)^{j-1}g_k^{(\pm)}(\z)\right]_{j,k=1}^{p+1} ,\ \ \z\in\mathbb{C}\backslash(-\infty,0],\hspace{0.5cm} \Delta_\z := \z \frac {\d}{\d \z}.
\end{equation*}
and assemble 
$\ds
	\mathbb{G}(\z)=\begin{cases}
		\mathbb{G}^{(+)}(\z),&\z\in\mathbb{H}^+\\
		\mathbb{G}^{(-)}(\z),&\z\in\mathbb{H}^-
		\end{cases}.
$
With this, the solution $G^{(p)}(\z)$ to the bare RHP \ref{pcMGRHP} is given by
\begin{equation}\label{genp:1}
	G^{(p)}(\z)=\begin{cases}
	\mathbb{G}(\z),& \textnormal{arg}\,\z\in(-\frac{3\pi}{4},-\frac{\pi}{4})\cup(\frac{\pi}{4},\frac{3\pi}{4})\\
	\mathbb G(\z)\bigg(1\oplus\bigoplus_{k=0}^{\frac{1}{2}(p-3)}\begin{bmatrix}
	1 & 0\\
	\z^{-a_{2k+2}}e^{i\pi a_{2k+2}} & 1\\
	\end{bmatrix}\oplus 1\bigg),&\textnormal{arg}\,\z\in(\frac{3\pi}{4},\pi)\\
	\mathbb G(\z)\bigoplus_{k=0}^{\frac{1}{2}(p-1)}\begin{bmatrix}
	1 & 0\\
	-\z^{-a_{2k+1}} & 1\\
	\end{bmatrix},&\textnormal{arg}\,\z\in(0,\frac{\pi}{4})\\
	\mathbb G(\z)\bigg(1\oplus\bigoplus_{k=0}^{\frac{1}{2}(p-3)}\begin{bmatrix}
	1 & 0\\
	-\z^{-a_{2k+2}}e^{-i\pi a_{2k+2}} & 1\\
	\end{bmatrix}\oplus 1\bigg),&\textnormal{arg}\,\z\in(-\pi,-\frac{3\pi}{4})\\
	\mathbb G(\z)\bigoplus_{k=0}^{\frac{1}{2}(p-1)}\begin{bmatrix}
	1 & 0\\
	\z^{-a_{2k+1}} & 1\\
	\end{bmatrix},&\textnormal{arg}\,\z\in(-\frac{\pi}{4},0)
	\end{cases}
\end{equation}
in case $p\equiv 1\mod 2$, and for even $p\equiv 0\mod 2$ by
\begin{equation}\label{genp:2}
	G^{(p)}(\z)=\begin{cases}
	\mathbb G(\z)&\textnormal{arg}\,\z\in(-\frac{3\pi}{4},-\frac{\pi}{4})\cup(\frac{\pi}{4},\frac{3\pi}{4})\\
	\mathbb G(\z) \bigg( 1\oplus\bigoplus_{k=0}^{\frac{1}{2}(p-2)}\begin{bmatrix}
	1 & 0\\
	\z^{-a_{2k+2}}e^{i\pi a_{2k+2}} & 1\\
	\end{bmatrix}\bigg),& \textnormal{arg}\,\z\in(\frac{3\pi}{4},\pi)\\
	\mathbb G(\z) \bigg( \bigoplus_{k=0}^{\frac{1}{2}(p-2)}\begin{bmatrix}
	1 & 0\\
	-\z^{-a_{2k+1}} & 1\\
	\end{bmatrix}\oplus 1\bigg),& \textnormal{arg}\,\z\in(0,\frac{\pi}{4})\\
	\mathbb G(\z) \bigg( 1\oplus\bigoplus_{k=0}^{\frac{1}{2}(p-2)}\begin{bmatrix}
	1 & 0\\
	-\z^{-a_{2k+2}}e^{-i\pi a_{2k+2}} & 1\\
	\end{bmatrix}\bigg),& \textnormal{arg}\,\z\in(-\pi,-\frac{3\pi}{4})\\
	\mathbb G(\z) \bigg( \bigoplus_{k=0}^{\frac{1}{2}(p-2)}\begin{bmatrix}
	1 & 0\\
	\z^{-a_{2k+1}} & 1\\
	\end{bmatrix}\oplus 1\bigg),& \textnormal{arg}\,\z\in(-\frac{\pi}{4},0)
	\end{cases}
\end{equation}
\et
We will split the proof of Theorem \ref{originparmsol} in several parts, starting with the jump conditions and the singular behavior at the origin $\z=0$.
\bl\label{lemmamonodromy} The function $g_1^{(\pm)}(\z),\z\in\mathbb{C}$ is an entire function, whereas $\{g_j^{(\pm)}(\z)\}_{j=2}^{p+1}$ are defined and analytic for $\z\in\mathbb{C}\backslash(-\infty,0]$. In particular, for $2\leq j\leq p+1$, we have the monodromy relations
\begin{equation}\label{gen:0}
	g_j^{(+)}\left(\z e^{2\pi i}\right)-g_j^{(+)}(\z) = -\z^{a_{j-1}}e^{i\pi a_{j-1}\sigma_{j-1}}g_{j-1}^{(+)}\left(\z e^{2\pi i\sigma_{j-1}}\right),
\end{equation}
valid on the entire universal covering of the punctured plane. Also, the behavior of $g_{\ell+1}^{(\pm)}(\z)$ at $\z=0$ for $1\leq \ell\leq p$ is the same as the behavior of the iterated Cauchy transforms $\mathcal{C}_{\ell+1}$ given in \eqref{zerobeh}.
\el
\begin{proof} The singularities in the integrand of $g_1^{(\pm)}(\z)$ are solely located at $\z=-n,n\in\mathbb{Z}_{\geq 0}$. Thus retracting the contour $L$ to $-\infty$ we pick up a residue at each nonpositive integer point equal to
\begin{equation*}
	\res_{s=-n}\Gamma(s)=\frac{(-1)^n}{n!}.
\end{equation*}
Since the remainder of the integral tends to zero by the properties of the Gamma function, we get
\begin{equation*}
	g_1^{(\pm)}(\z)=c_1\sum_{k=0}^{\infty}\frac{(-1)^k}{\prod_{\ell=1}^p\Gamma(1+a_{1\ell}+k)}\frac{\z^k}{k!},\ \ \z\in\mathbb{C},
\end{equation*}
which implies that $g_1^{(\pm)}(\z)$ is entire. The same argument applied to the remaining $\{g_j^{(\pm)}(\z)\}_{j=2}^{p+1}$ shows directly that they are analytic in $\mathbb{C}$ with a cut along the negative real axis. Suppose now that $2\leq j\leq p+1$ and start with
\begin{equation}\label{gen:1}
	g_j^{(+)}\left(\z e^{2\pi i}\right)-g_j^{(+)}(\z) = \frac{c_j}{2\pi i}\int_L\frac{\prod_{\ell=1}^j\Gamma(s+a_{\ell,j-1})}{\prod_{\ell=j}^p\Gamma(1+a_{j\ell}-s)}e^{i\pi s\sigma_j}\z^{-s}\left(e^{-2\pi is}-1\right)\,\d s.
\end{equation}
Since
\be\nonumber 
	e^{-2\pi is}-1=-e^{-i\pi s}\frac{2\pi i}{\Gamma(s)\Gamma(1-s)},\hspace{0.5cm} \prod_{\ell=1}^j\Gamma(s+a_{\ell,j-1})=\Gamma(s)\prod_{\ell=1}^{j-1}\Gamma(s+a_{\ell,j-1}),
\ee 
we can change the variable of integration in \eqref{gen:1} as $s=u-a_{j-1,j-1}\equiv u-a_{j-1}$, and are lead to
\beas
	g_j^{(+)}\left(\z e^{2\pi i}\right)-g_j^{(+)}(\z) =-\frac{c_{j-1}}{2\pi i}\int\limits_{L+a_{j-1}}\frac{\prod_{\ell=1}^{j-1}\Gamma(u+a_{\ell,j-1}-a_{j-1})}{\prod_{\ell=j}^p\Gamma(1+a_{j\ell}+a_{j-1}-u)}\frac{e^{i\pi u\sigma_j}}{\Gamma(1+a_{j-1}-u)}\,e^{-i\pi u}\z^{-u}\,\d u\\
	\times\,\z^{a_{j-1}}e^{i\pi a_{j-1}\sigma_{j-1}}\left(\frac{2\pi i\,c_j}{c_{j-1}}\right)\\
	=-\z^{a_{j-1}}e^{i\pi a_{j-1}\sigma_{j-1}}\frac{c_{j-1}}{2\pi i}\int\limits_{L+a_{j-1}}\frac{\prod_{\ell=1}^{j-1}\Gamma(u+a_{\ell,j-2})}{\prod_{\ell=j-1}^p\Gamma(1+a_{j\ell}-u)}e^{i\pi u\sigma_{j-1}}\left(\z e^{2\pi i\sigma_{j-1}}\right)^{-u}\,\d u\\
	=-\z^{a_{j-1}}e^{i\pi a_{j-1}\sigma_{j-1}}g_{j-1}^{(+)}\left(\z e^{2\pi i\sigma_{j-1}}\right).
\eeas
In the last equality we used that there are no singularities of the integrand between $L+a_{j-1}$ and $L$ since $a_{j-1}>-1$. As for the singular behavior at $\z=0$, we simply use analyticity of $g_1^{(\pm)}(\z)$ and apply the monodromy relations iteratively. This combined with the Plemelj-Sokhotskii formula leads to a behavior as in \eqref{zerobeh}.
\end{proof}
We are now ready to derive the jump behavior of $G^{(p)}(\z)$ as stated in Theorem \ref{originparmsol}
\begin{proof}[Proof of Theorem \ref{originparmsol} - jump and singular behavior] The matrix $\mathbb G(\z)$ is analytic in the upper/lower half plane and thus the jumps on the four rays $\mathfrak r_{1,2,3,4}$ follow at once from the definition of $G^{(p)}(\z)$. 
Now it follows from $\sigma_j  \equiv (j+1)\mod 2$ that  for odd $j$ the functions $g_{j}^{(\pm)}(\z)$ coincide. For even $j =2k$ we have instead that 
\begin{equation*}
	g_{2k}^{(+)}(\z e^{2\pi i})=g_{2k}^{(-)}(\z),\ \ \z\in\mathbb{C}\backslash(-\infty,0]
\end{equation*} 
and thus with Lemma \ref{lemmamonodromy}
\be\nonumber
	g_{2k}^{(+)}(\z)  = g_{2k}^{(-)}(\z) +\z^{a_{2k-1}} g_{2k-1}^{(-)}( \z),\ \ \ \z\in\mathbb{C}\backslash(-\infty,0].
\ee
Hence, the functions $\mathbb G^{(\pm)}(\z)$ are related by 
\be 
\label{Gpm}
	\mathbb G^{(+)} (\z)=
	\mathbb G^{(-)} (\z)\begin{cases}
	\bigoplus_{k=0}^{\frac{1}{2}(p-1)}\begin{bmatrix}
	1 & \z^{a_{2k+1}}\\
	0 & 1\\
	\end{bmatrix},&p\equiv 1\mod 2\smallskip\\
	\bigoplus_{k=0}^{\frac{1}{2}(p-2)}\begin{bmatrix}
	1 & \z^{a_{2k+1}}\\
	0 & 1\\
	\end{bmatrix}\oplus 1,&p\equiv 0\mod 2\\
	\end{cases}
\ee 
From this, the remaining jumps on the real line, i.e. on $\mathfrak r_{0,5}$, follow by matrix multiplication applying the Definitions \eqref{genp:1}, \eqref{genp:2} and using that $\z_+^{\gamma}=\z_-^{\gamma}$ for $\z>0$ as well as $\z^{\gamma}_+=\z^{\gamma}_-e^{-2\pi i\gamma}$ for $\z<0$. As for the singular behavior near $\z=0$, this is dictated by the result of Lemma \ref{lemmamonodromy} and the Definitions \eqref{genp:1}, \eqref{genp:2}.
\end{proof}
We move on to the asymptotics at $\z=\infty$. Since
\begin{equation*}
	g_{p+1}^{(+)}(\z)=c_{p+1}G_{0,p+1}^{\,p+1,0}\!\left(\le.{--\atop a_{1p},a_{2p},a_{3p},\ldots,a_{pp},a_{p+1,p}}\; \right| \,e^{-i\pi\sigma_{p+1}}\z \right),
\end{equation*}
we get from \cite{B,F} that, as $\z\rightarrow\infty$ with $|\textnormal{arg}\,\z|<\pi(p+1)$,
\begin{equation}\label{Fieldsexp}
	g_{p+1}^{(+)}(\z)=\z^{-\frac{p}{2(p+1)}+\frac{1}{p+1}(\sum_{1}^pa_{jp})}\omega^{\sigma_{p+1}}\omega^{-\frac{2}{p}(\sum_{1}^pa_{jp})\sigma_{p+1}}\exp\left[-(p+1)\omega^{-\frac{2}{p}\sigma_{p+1}}\z^{\frac{1}{p+1}}\right]\left(1+\mathcal{O}\left(\z^{-\frac{1}{p+1}}\right)\right),
\end{equation}
Here we put
\begin{equation*}
	\omega=\omega_p=e^{i\frac{\pi}{2}\frac{p}{p+1}},
\end{equation*}
and all subsequent expansions of $\{g^{(+)}_j(\z)\}_{j=1}^p$ at $\z=\infty$ can now be derived from \eqref{Fieldsexp} by substituting into \eqref{gen:0}. We summarize
\bl
\label{genasy} Let $\epsilon>0$ be fixed. As $\z\rightarrow\infty$,
\be\nonumber 
	g_{2k}^{(+)}(\z)=\z^{-\frac{p}{2(p+1)}}\omega^{2+p-2k}\,\z^{\frac{A_{2k}}{p+1}}\omega^{-\frac{2}{p}\sum_{2k}^{p+1}\!A_j}\exp\left[-(p+1)\omega^{-\frac{2}{p}(2+p-2k)}\z^{\frac{1}{p+1}}\right]\left(1+\mathcal{O}\left(\z^{-\frac{1}{p+1}}\right)\right)
\ee 
uniformly for $\textnormal{arg}\,\z\in(-\pi,\pi-\epsilon]$ and any
\be\nonumber 
	k=\begin{cases}
		1,\ldots,\frac{1}{2}(p-1),&p\equiv 1\mod 2\\
		1,\ldots,\frac{1}{2}p,& p\equiv 0\mod 2.
		\end{cases}  
\ee 
Secondly with $\mathbb{H}^{\pm}=\{\z\in\mathbb{C}:\ \textnormal{sgn}(\Im\,\z)=\pm 1\}$, as $\z\rightarrow\infty$
\be\nonumber 
	g_{2k+1}^{(+)}(\z)=\begin{cases}
	\z^{-\frac{p}{2(p+1)}}\omega^{p-2k}\z^{\frac{A_{2k+1}}{p+1}}\omega^{-\frac{2}{p}\sum_{2k+2}^{p+1}\!A_j}\exp\left[-(p+1)\omega^{-\frac{2}{p}(p-2k)}\right]\left(1+\mathcal{O}\left(\z^{-\frac{1}{p+1}}\right)\right),&\!\!\!\!\!\z\in\mathbb{H}^-\\
	(-1)^p\z^{-\frac{p}{2(p+1)}}\omega^{-(p-2k)}\z^{\frac{A_{2k+1}}{p+1}}\omega^{\frac{2}{p}\sum_{2k+2}^{p+1}\!A_j}\exp\left[-(p+1)\omega^{\frac{2}{p}(p-2k)}\z^{\frac{1}{p+1}}\right]\left(1+\mathcal{O}\left(\z^{\frac{1}{p+1}}\right)\right)\!,&\!\!\!\!\!\z\in\mathbb{H}^+
	\end{cases}  
\ee 
uniformly for $\textnormal{arg}\,\z\in(-\pi,-\epsilon]$ in the lower half-plane, for $\textnormal{arg}\,\z\in[\epsilon,\pi)$ in the upper half-plane and any
$k = 0,1, \dots, \le\lfloor \frac {p-1}2 \ri\rfloor$
\el
In addition 
\bc Let $\epsilon>0$ be fixed, then as $\z\rightarrow\infty$, uniformly for $\textnormal{arg}\,\z\in[-\pi+\epsilon,0)$
\begin{align}
	g_{2k}^{(+)}&(\z)-\z^{a_{2k-1}}g_{2k-1}^{(+)}(\z)=g_{2k}^{(+)}\left(\z e^{2\pi i}\right)=g_{2k}^{(-)}(\z)\label{genp:6}\\
	&=(-1)^{p-1}\z^{-\frac{p}{2(p+1)}}\omega^{-(2+p-2k)}\z^{\frac{A_{2k}}{p+1}}\omega^{\frac{2}{p}\sum_{2k}^{p+1}A_j}\exp\left[-(p+1)\omega^{\frac{2}{p}(2+p-2k)}\z^{\frac{1}{p+1}}\right]\left(1+\mathcal{O}\left(\z^{-\frac{1}{p+1}}\right)\right)\nonumber
\end{align}
for any  $k = 1,\ldots, \le \lfloor \frac {p+1}2\ri\rfloor$.
\ec
We can now complete the proof of Theorem \ref{originparmsol}

\begin{proof}[Proof of Theorem \ref{originparmsol} - asymptotic behavior at $\z=\infty$] The sectorial asymptotics of $\mathbb G^{(\pm)}(\z)$ follow from Lemma \ref{genasy} and careful algebra. The jump-matrices do not affect the sectorial asymptotic because by construction of $\mathbb G^{(p)}(\z)$, the asymptotics  of $ \mathbb G^{(\pm)}(\z)$ and $G^{(p)}(\z)$ are the same as $\z\rightarrow\infty$. This follows from the Definitions \eqref{genp:1} and \eqref{genp:2} in the sectors $\textnormal{arg}\,\z\in(-\frac{3\pi}{4},-\frac{\pi}{4})\cup(\frac{\pi}{4},\frac{3\pi}{4})$ and  estimations of the form (here only  for $p\equiv 1\mod 2$) 
\begin{align*}
	\z^{\frac{A}{p+1}}e^{-(p+1)\Omega\z^{\frac{1}{p+1}}}\left(\bigoplus_{k=0}^{\frac{1}{2}(p-1)}
	\begin{bmatrix}
	1 & 0\\
	-\z^{-a_{2k+1}} & 1\\
	\end{bmatrix}\right)e^{(p+1)\Omega\z^{\frac{1}{p+1}}}\z^{-\frac{A}{p+1}} &=I+\mathcal{O}\left(\z^{-\infty}\right),\ \ \textnormal{arg}\,\z\in\left(0,\frac{\pi}{4}\right)\\
	\z^{\frac{A}{p+1}}e^{-(p+1)\Omega\z^{\frac{1}{p+1}}}\left(1\oplus\bigoplus_{k=0}^{\frac{1}{2}(p-3)}
	\begin{bmatrix}
	1 & 0\\
	\z^{-a_{2k+2}}e^{i\pi a_{2k+2}} &1\\
	\end{bmatrix}\oplus 1\right)e^{(p+1)\Omega\z^{\frac{1}{p+1}}}\z^{-\frac{A}{p+1}} &=I+\mathcal{O}\left(\z^{-\infty}\right),\ \ \textnormal{arg}\,\z\left(\frac{3\pi}{4},\pi\right)
\end{align*}
as $\z\rightarrow\infty$ with similar ones in the sectors in the lower half-plane. 
\end{proof}
\subsubsection{Computation of the right hand side in \eqref{scalelimit} for general $p\geq 2$} Our next goal is to express the entries under consideration in the matrix product $\mathbb{G}^{-1}(w)\mathbb{G}(z)$ as double contour integrals. To this end it is convenient to pass from the functions $g_j^{(\pm)}(\z)$  and $\mathbb G^{(\pm)}(\z)$ to the functions $\{f_j^{(\pm)}(\z)\}_{j=1}^{p+1}$, $\mathbb F^{(\pm)}(\z)$ defined through
\bea
	f_j^{(\pm)}(\z)&\&=\z^{-a_{1,j-1}}g_j^{(\pm)}(\z),\ \ \z\in\mathbb{C}\backslash(-\infty,0],\hspace{1cm} j=1,\ldots,p+1,\\
	\mathbb F^{(\pm)} (\z) &\&= \bigg[\Delta_\z^{j-1} f_k^{(\pm)}(\z) \bigg]_{j,k=1}^{p+1}  = \mathbb G^{(\pm)}(\z) \z^D
\qquad
\label{defD}
	D:=\textnormal{diag}\left[0, a_1, a_{12}, a_{13}, \dots, a_{1p}\right].
\eea
Note in particular that all functions $f_j^{(\pm)}(\z)$ admit a contour integral representation, with $\z\in\C\backslash(-\infty,0]$,
\begin{equation}\label{gen:2}
	f_j^{(\pm)}(\z)=\frac{c_j}{2\pi i}\int_{L_j}F_j^{(\pm)}(s)\z^{-s}\,\d s,\hspace{0.5cm}F_j^{(\pm)}(s)=\frac{\prod_{\ell=0}^{j-1}\Gamma(s-a_{1\ell})}{\prod_{\ell=j}^p\Gamma(1+a_{1\ell}-s)}e^{\pm i\pi(s-a_{1,j-1})\sigma_j}.
\end{equation}
We also define for convenience the following functions 
\be 
\label{Fhatpm}
	\hat{f}_j^{\,(\pm)}(\z)=\frac{\hat{c}_j}{2\pi i}\int_{\hat{L}_j}\hat{F}_j^{(\pm)}(s)\z^{-s}\,\d s,\ \hspace{0.5cm}\hat{F}_j^{(\pm)}(s)=\frac{\prod_{\ell=j-1}^p\Gamma(s+a_{1\ell})}{\prod_{\ell=0}^{j-2}\Gamma(1-s-a_{1\ell})}e^{\pm i\pi(s+a_{1,j-1})\sigma_{j-1}},
\ee 
and analogously as before,
\be\nonumber
 \widehat{\mathbb F}^{(\pm)}(\z)\equiv[\Delta_{\z}^{j-1}\hat{f}_k^{\,(\pm)}(\z)]_{j,k=1}^{p+1},\ \ \ \z\in\mathbb{C}\backslash(-\infty,0].
\ee 
The normalization constants $\hat{c}_j$ are defined through $c_j$ in \eqref{cjs} as
\begin{equation}\label{correctnorm}
	\hat{c}_{j+1}=-2\pi i\hat{c}_j.
\end{equation}

The goal of this section is to prove 
\bt
\label{thmConco} For $w,z\in\mathbb{R}$,
\begin{equation}\label{summary1}
	\bigg[\mathbb G^{-1}(w)\mathbb G(z)\bigg]_{jk} 
 =c_j\hat{c}_k w^{-a_{1,j-1}} z^{a_{1,\ell-1}}\int_{L}\int_{\wh{L}}F_k^{(\pm)} (u)\hat{F}_j^{(\pm)} (-v)\frac{K(u)-K(v)}{u-v}\,w^{v}z^{-u}\,\frac{\d v\,\d u}{(2\pi i)^2},
\end{equation}
where the signs $(\pm)$ are chosen according to whether the corresponding variable belongs to $\mathbb H^\pm$. Also, the multi-valued functions $\z^{\gamma}$ have to be evaluated with principal branches and the integration contours are chosen as in Definition \ref{defGjl}.
\et

We split the proof of the latter Theorem into several steps

\bp The functions $\{f_j^{(\pm)}(\z)\}_{j=1}^{p+1}$  and  $\{\hat{f}_j^{\,(\pm)}(\z)\}_{j=1}^{p+1}$ defined for $\z\in\mathbb{C}\backslash(-\infty,0]$ are linearly independent solutions of the classically adjoint  differential equations
\bea\nonumber
\prod_{\ell=0}^p\left(\Delta_{\z}+a_{1\ell}\right)f(\z)=-\z f(\z)\ ,\qquad
\prod_{\ell=0}^p\left(\Delta_{\z}-a_{1\ell}\right)\hat{f}(\z)=-\z \hat{f}(\z),\hspace{0.5cm}\Delta_{\z}=\z\frac{\d}{\d\z}
\eea
which follows from the functional relation of the kernel functions
\be 
\label{funcrel}
	F_j^{(\pm)}(s+1)= F_j^{(\pm)}(s)K(s),\
\qquad  \hat{F}_j^{(\pm)}(s+1)=\hat{F}_j^{(\pm)}(s)K(-s),\ \hspace{0.5cm}K(s)=(-1)^p\prod_{\ell=0}^p(s-a_{1\ell}).
\ee 
\ep
\begin{proof}
The functional relations \eqref{funcrel} follow simply from the standard relation $\Gamma(1 + s) = s\Gamma(s)$.
The stated differential equations are then derived by differentiation in \eqref{gen:2}, \eqref{Fhatpm} and application of the latter functional relations for the integrands.
\end{proof}

\bd[Bilinear Concomitant, see \cite{I}]
\label{bilconco}
For $\z\in\mathbb{C}\backslash(-\infty,0]$, introduce the bilinear form,
\begin{equation}\label{gen:3}
	\mathcal{B}(f_j,\hat{f}_k)(\z)=\frac{c_j\hat{c}_k}{(2\pi i)^2}\int_{L_j}\int_{\hat{L}_k}\frac{F_j(u)\hat{F}_k(v)}{u+v}\Big[K(u)-K(-v)\Big]\z^{-u-v}\,\d v\,\d u,\ \ \ \ 1\leq j,k\leq p+1
\end{equation}
or written equivalently without double integrals,
\begin{equation}\label{comm}
	\mathcal{B}(f_j,\hat{f}_k)(\z)=\left[\hat{f}_k(\z),\Delta_{\z}\hat{f}_k(\z),\Delta_{\z}^2\hat{f}_k(\z),\ldots,\Delta_{\z}^{p}\hat{f}_k(\z)\right]\mathcal{K}\left[f_j(\z),\Delta_{\z}f_j(\z),\Delta_{\z}^2f_j(\z),\ldots,\Delta_{\z}^{p}f_j(\z)\right]^T
\end{equation}
where
\be\nonumber 
	\mathcal{K}=\left[(-1)^{p+k-1}\frac{K^{(j+k-1)}(0)}{(j+k-1)!}\right]_{j,k=1}^{p+1}.
\ee 
Here, $f_j(\z)$ or $\hat{f}_k(\z)$ can be replaced by any function of the collection $\{f_j^{(\pm)}(\z)\}$ or $\{\hat{f}^{(\pm)}(\z)\}$.
\ed

\bp
The bilinear form in Definition \ref{bilconco} is piecewise constant in $\z$.
\ep
\begin{proof}
From the functional equations of $F_j(s)$ and $\hat{F}_j(s)$ (here $F_j(s)$ can represent any of the $F_j^{(\pm)}(s)$, similarly for $\hat{F}_j(s)$),
\begin{eqnarray*}
	\frac{\d}{\d\z}\mathcal{B}(f_j,\hat{f}_k)(\z)=-\frac{c_j\hat{c}_k}{(2\pi i)^2}\int_{L}\int_{\wh{L}}F_j(u)\hat{F}_k(v)\Big[K(u)-K(-v)\Big]\z^{-u-v-1}\,\d v\,\d u\\
	=-\frac{c_j \hat{c}_k}{(2\pi i)^2}\int_{L+1}F_j(u)\z^{-u}\,\d u\,\int_{\wh{L}}\hat{F}_k(v)\z^{-v}\,\d v
	+\frac{c_j\hat{c}_k}{(2\pi i)^2}\int_{L}F_j(u)\z^{-u}\,\d u
	\int_{\wh{L}+1}\hat{F}_k(v)\z^{-v}\,\d v\\
	=-f_j(\z)\hat{f}_k(\z)+f_j(\z)\hat{f}_k(\z)\equiv 0,\ \ \z\in\mathbb{C}\backslash(-\infty,0]
\end{eqnarray*}
where we used Cauchy Theorem in the last equality. 
\end{proof}
The particular choice of the expressions \eqref{gen:2}, \eqref{Fhatpm} is explained by the following Proposition.
\bp 
For independent choices of signs $(\pm)$, we have 
\begin{equation}\label{resicomp}
	\mathcal{B}(f_j^{(\pm)},\hat{f}_k^{\,(\pm)})(\z)\equiv \delta_{jk},\ \ j,k=1,\ldots,p+1.
\end{equation}
\ep
\begin{proof} 
The proof is technically simpler if we impose the {\it non-resonance} condition
\begin{equation}\label{gen:4}
	a_{k\ell}=\sum_{j=k}^{\ell}a_j\notin\mathbb{Z},\ \ \ 1\leq k\leq\ell\leq p.
\end{equation}
This condition can then be lifted a posteriori since the result is independent of the $a_j$'s.
As $\mathcal{B}(f_j,\hat{f}_k)(\z)$ is defined through a double contour integral we shall apply residue theorem to retract first the contour $\hat{L}_k$ to $-\infty$. This procedure amounts to picking up the residues of the inner integrand which by assumption \eqref{gen:4} are all originating from {\it simple poles} of the expression $\hat{F}_k(-v)$. Let 
$ 
	\mathcal{P}=\left\{a_{11},a_{12},\ldots,a_{1p}\right\}$: note that our assumption \eqref{gen:4} implies  $\mathcal{P}\cap\mathbb{Z} = \emptyset.$
Then the poles of $F_j(u)$ are in general located on the lattice $(\mathcal{P}\cup\{0\})-\mathbb{N}$ whereas the poles of $\hat{F}_k(-v)$ are in general centered at $(\mathcal{P}\cup\{0\})+\mathbb{N}$. Retracting the contours as indicated, we create certain double series  of the form
\begin{equation}\label{gen:5}
	\mathcal{B}(f_j^{(\pm)},\hat{f}_k^{(\pm)})(\z)=\sum_{m\in(\mathcal{P}\cup\{0\})-\mathbb{N}}\,\,\,\sum_{\ell\in(\mathcal{P}\cup\{0\})+\mathbb{N}}R_{m,\ell;j,k}^{(\pm)}\,\,\z^{\ell-m}
\end{equation}
with coefficients $R_{m,\ell;j,k}^{(\pm)}$ determined through residue evaluations. The so obtained series defines an analytic function in $\mathbb{C}\backslash(-\infty,0]$. Now we know that $\mathcal{B}(f_j^{(\pm)},\hat{f}_k^{\,(\pm)})(\z)$ is $\z$-independent and  hence the computation of \eqref{gen:5} only requires from us to inspect those coefficients $R_{m,\ell;j,k}^{(\pm)}$ which can give a contribution to the $\mathcal{O}\left(\z^0\right)$ terms in \eqref{gen:5}. Also, as \eqref{gen:4} is in place, we only have to compute the residues of the integrand at the elements of the finite set $\mathcal{P}$. Concretely we obtain
\bea
	\mathcal{B}(f_j^{(\pm)},\hat{f}_k^{(\pm)})(\z)=\frac{c_j\hat{c}_k}{(2\pi i)^2}\int_{L}\int_{\wh{L}}\frac{\prod_{\ell=0}^{j-1}\Gamma(u-a_{1\ell})}{\prod_{\ell=j}^p\Gamma(1+a_{1\ell}-u)}\frac{\prod_{m=k-1}^p\Gamma(-v+a_{1m})}{\prod_{m=0}^{k-2}\Gamma(1+v-a_{1m})}e^{\pm i\pi(u-a_{1,j-1})\sigma_j}\nonumber\\
	\times\,e^{\pm i\pi(-v+a_{1,k-1})\sigma_{k-1}}\z^{-u+v}\frac{K(u)-K(v)}{u-v}\,\d v\,\d u\nonumber\\
	\equiv c_j\hat{c}_k\sum_{\ell=0}^{j-1}\sum_{m=k-1}^p\frac{\prod_{{n=0\atop  n\neq \ell}}^{j-1}\Gamma(a_{1\ell}-a_{1n})}{\prod_{n=j}^p\Gamma(1+a_{1n}-a_{1\ell})}\frac{\prod_{{n=k-1\atop  n\neq m}}^{p}\Gamma(-a_{1m}+a_{1n})}{\prod_{n=0}^{k-2}\Gamma(1+a_{1m}-a_{1n})}\frac{K(a_{1\ell})-K(a_{1m})}{a_{1\ell}-a_{1m}}\label{gen:6}\\
	\ \ \ \times e^{\pm i\pi(a_{1\ell}-a_{1,j-1})\sigma_j}e^{\pm i\pi(-a_{1m}+a_{1,k-1})\sigma_{k-1}}\z^{a_{1m}-a_{1\ell}}.\nonumber
\eea
Since by construction
\be 
	\frac{K(a_{1\ell})-K(a_{1m})}{a_{1\ell}-a_{1m}}=\delta_{\ell m}K'(a_{1\ell}),
\ee 
we see from \eqref{gen:6} that $\mathcal{B}(f_j^{(\pm)},\hat{f}_k^{(\pm)})(\z)\equiv 0$ for $j<k$ in the corresponding half-planes. For $j=k$,
\begin{eqnarray*}
	\mathcal{B}(f_j^{(\pm)},\hat{f}_j^{(\pm)})(\z)&\equiv& c_j\hat{c}_j\,\frac{\prod_{n=0}^{j-2}\Gamma(a_{1,j-1}-a_{1n})}{\prod_{n=j}^p\Gamma(1+a_{1n}-a_{1,j-1})}\frac{\prod_{n=j}^p\Gamma(-a_{1,j-1}+a_{1n})}{\prod_{n=0}^{j-2}\Gamma(1+a_{1,j-1}-a_{1n})}K'(a_{1,j-1})\\
=c_j\hat{c}_j\,&\&\left(\prod_{n=0}^{j-2}\frac{1}{a_{1,j-1}-a_{1n}}\right)\left(\prod_{n=j}^p\frac{1}{a_{1n}-a_{1,j-1}}\right)K'(a_{1,j-1})	=c_j\hat{c}_j\,e^{i\pi\sigma_j}=1,
\end{eqnarray*}
where we used the normalization \eqref{correctnorm}. Thus $\mathcal{B}(f_j^{(\pm)},\hat{f}_j^{(\pm)})(\z)\equiv 1$ for all $j=1,\ldots,p+1$ in the half-planes. It remains to consider the situation when $j>k$,
\begin{align}
	\mathcal{B}&(f_j^{(\pm)},\hat{f}_k^{(\pm)})(\z)=c_j\hat{c}_k\!\!\!\!\sum_{m=k-1}^{j-1}\frac{\prod_{{n=0\atop  n\neq m}}^{j-1}\Gamma(a_{1m}-a_{1n})}{\prod_{n=j}^p\Gamma(1+a_{1n}-a_{1m})}\frac{\prod_{{n=k-1\atop  n\neq m}}^p\Gamma(-a_{1m}+a_{1n})}{\prod_{n=0}^{k-2}\Gamma(1+a_{1m}-a_{1n})}\,K'(a_{1m})\nonumber\\
	&\,\,\times e^{\pm i\pi(a_{1m}-a_{1,j-1})\sigma_j}e^{\pm i\pi(-a_{1m}+a_{1,k-1})\sigma_{k-1}}\nonumber\\
	&=c_j\hat{c}_k\sum_{m=k-1}^{j-1}\prod_{{n=k-1\atop n\neq m}}^{j-1}\Gamma(a_{1m}-a_{1n})\Gamma(a_{1n}-a_{1m})\frac{K'(a_{1m})e^{\pm i\pi(a_{1m}-a_{1,j-1})\sigma_j}e^{\pm i\pi(-a_{1m}+a_{1,k-1})\sigma_{k-1}}}{\prod_{n=j}^p(a_{1n}-a_{1m})\prod_{n=0}^{k-2}(a_{1m}-a_{1n})}\nonumber\\
	&=c_j\hat{c}_k\sum_{m=k-1}^{j-1}\prod_{{n=k-1\atop  n\neq m}}^{j-1}\frac{\pi}{\sin\pi(a_{1m}-a_{1n})}\frac{K'(a_{1m})}{\prod_{{n=0\atop n\neq m}}^p(a_{1n}-a_{1m})}\,e^{i\pi\sigma_k}e^{\pm i\pi(a_{1m}-a_{1,j-1})\sigma_j}e^{\pm i\pi(-a_{1m}+a_{1,k-1})\sigma_{k-1}}\nonumber\\
	&=\frac{c_j}{c_k}e^{\pm i\pi(a_{1,k-1}\sigma_{k-1}-a_{1,j-1}\sigma_j)}\sum_{m=k-1}^{j-1}e^{\pm i\pi a_{1m}(\sigma_j-\sigma_{k-1})}\prod_{{n=k-1\atop  n\neq m}}^{j-1}\frac{\pi}{\sin\pi(a_{1m}-a_{1n})}. \label{lattersum}
\end{align}
The last sum vanishes identically: to see that, we consider  the meromorphic functions
\be \nonumber
	\varphi_{jk}^{(\pm)}(z)=e^{\pm i\pi z(\sigma_j-\sigma_{k-1})}\prod_{n=k-1}^{j-1}\frac{\pi}{\sin\pi(z-a_{1n})},\ \ j>k,
\ee 
which are periodic $\varphi_{jk}^{(\pm)}(z+1)=\varphi_{jk}^{(\pm)}(z)$. 
In this latter expression we can assume without loss of generality that $a_{1n}\in[0,1)$ for all $k-1\leq n\leq j-1$. 
Let  $B_{R,\epsilon}$ be the rectangular box with sides
\be\nonumber 
 	\big\{1+\epsilon+it,\,t\in[-R,R]\big\}\cup\big\{t+iR,\,t\in[\epsilon,1+\epsilon]\big\}\cup\big\{\epsilon+it,\,t\in[-R,R]\big\}\cup\big\{-iR+t,\,t\in[\epsilon,1+\epsilon]\big\}.
\ee 
Then  we can always find $\epsilon\in\mathbb{R}$ such that
\begin{eqnarray*}
	0=\frac{1}{2\pi i}\left[\int_{\epsilon+i\infty}^{\epsilon-i\infty}+\int_{1+\epsilon-i\infty}^{1+\epsilon+i\infty}\right]\varphi_{jk}^{(\pm)}(z)\,\d z&=&\lim_{R\rightarrow\infty}\frac{1}{2\pi i}\oint_{\partial B_{R,\epsilon}}\varphi_{jk}^{(\pm)}(z)\,\d z,
\end{eqnarray*}
and the latter integrals yields the sum of residues inside, which equals exactly  the sum in \eqref{lattersum}.
\end{proof}
The last result is now put into use in the following way: with
\begin{equation*}
	\mathcal{B}(\z)=\big(\wh{\mathbb{F}}(\z)\big)^T\mathcal{K}\mathbb{G}(\z)\z^{-D},\hspace{0.5cm} \textnormal{where}\hspace{0.5cm} \wh{\mathbb{F}}(\z)=\begin{cases}
		\mathbb{F}^{(+)}(\z),& \z\in\mathbb{H}^+\\
		\mathbb{F}^{(-)}(\z),& \z\in\mathbb{H}^-
		\end{cases}
\end{equation*}
we get from \eqref{comm} that
\begin{equation*}
	\big[\mathcal{B}(\z)\big]_{jk}=\begin{cases}
	\mathcal{B}(f_k^{(+)},\hat{f}_j^{\,(+)})(\z),&\,\,\,\,0<\textnormal{arg}\,\z<\pi\\
	\mathcal{B}(f_k^{(-)},\hat{f}_j^{\,(-)})(\z), &-\pi<\textnormal{arg}\,\z<0
	\end{cases}
\end{equation*}
and thus \eqref{resicomp} shows that $\mathcal{B}(\z)\equiv I$ in the separate half-planes. In other words, we have computed the matrix inverse
\begin{equation}\label{inv:1}
	\big(\mathbb{G}(\z)\big)^{-1}=\z^{-D}\big(\wh{\mathbb{F}}(\z)\big)^T\mathcal{K},\ \ \z\in\mathbb{C}\backslash\mathbb{R}.
\end{equation}
\br A direct computation in fact shows that $\z^{-D}\big(\wh{\mathbb{F}}(\z)\big)^T$ has the same jumps on the real line as $(\mathbb{G}(\z))^{-1}$. For this we would notice that
\begin{equation*}
	\z^{-D}(\wh{\mathbb{F}}^{(+)}(\z))^T=\left[(\Delta_{\z}+a_{1,j-1})^{k-1}\hat{g}_j^{\,(+)}(\z)\right]_{j,k=1}^{p+1}
\end{equation*}
where we introduced the ``dual functions" $\{\hat{g}_j^{\,(+)}(\z)\}$ to $\{g_j^{\,(+)}(\z)\}$, namely
\begin{equation*}
	\hat{g}_j^{\,(+)}(\z)=\frac{\hat{c}_j}{2\pi i}\int_{\hat{L}_j}\frac{\prod_{\ell=j-1}^p\Gamma(s+a_{j\ell})}{\prod_{\ell=1}^{j-1}\Gamma(1-s+a_{\ell,j-1})}e^{i\pi s\sigma_{j-1}}\z^{-s}\,\d s,\ \ \ \z\in\mathbb{C}\backslash(-\infty,0].
\end{equation*}
Here, $\hat{g}_{p+1}^{\,(+)}(\z)$ is an entire function whereas $\{\hat{g}_j^{\,(+)}(\z)\}_{j=1}^p$ are defined and analytic for $\z\in\mathbb{C}\backslash(-\infty,0]$, also (compare \eqref{gen:0}) we have a monodromy relation
\begin{equation}\label{dualmono}
	\hat{g}_j^{\,(+)}\left(\z e^{2\pi i}\right)-\hat{g}_j^{\,(+)}(\z)=\z^{a_j}e^{i\pi a_j\sigma_j}\hat{g}_{j+1}^{\,(+)}\left(\z e^{2\pi i\sigma_j}\right),\ \ 1\leq j\leq p
\end{equation}
valid on the entire universal covering of the punctured plane. Then one checks with \eqref{dualmono} that the jumps of $\z^{-D}(\wh{\mathbb{F}}(\z))^T$ are indeed identical to the ones of $(\mathbb{G}(\z))^{-1}$.
\er
In order to complete the proof of Theorem \ref{thmConco}, we use \eqref{inv:1}, thus for $w,z\in\mathbb{C}\backslash\mathbb{R}$
\begin{equation*}
	\mathbb{G}^{-1}(w)\mathbb{G}(z)=w^{-D}\big(\wh{\mathbb{F}}(w)\big)^T\mathcal{K}\,\mathbb{F}(z)z^D.
\end{equation*}
This motivates the following generalization of \eqref{gen:3}
\bd[Generalized Bilinear Concomitant] For $w,z\in\mathbb{C}\backslash(-\infty,0]$, let
\begin{equation}\label{gen:9}
	\bar{\mathcal{B}}(f_j,\hat{f}_k)(w,z)=\frac{c_j\hat{c}_k}{(2\pi i)^2}\int_{L}\int_{\wh{L}}F_j(u)\hat{F}_k(-v)\frac{K(u)-K(v)}{u-v}\,w^{v}z^{-u}\,\d v\,\d u,\ \ w,z\in\mathbb{C}\backslash(-\infty,0]
\end{equation}
where $f_j, \hat f_j$ stand for any of $f_j^{(\pm)}$, $\hat f_j^{(\pm)}$, with integration contours chosen as in the definition of $\mathcal{B}(f_j,\hat{f}_k)(\z)$, compare \eqref{gen:3}. Equivalently, without any contour integrals,
\begin{equation}\label{idea2}
	\bar{\mathcal{B}}(f_j,\hat{f}_k)(w,z)=\left[\hat{f}_k(w),\Delta_{w}\hat{f}_k(w),\Delta_{w}^2\hat{f}_k(w),\ldots,\Delta_{w}^{p}\hat{f}_k(w)\right]\mathcal{K}\left[f_j(z),\Delta_{z}f_j(z),\Delta_{z}^2f_j(z),\ldots,\Delta_{z}^{p}f_j(z)\right]^T.
\end{equation}
\ed

\begin{proof}[Proof of Theorem \ref{thmConco}] Let
\begin{equation*}
	\bar{\mathcal{B}}(w,z)=\big(\wh{\mathbb{F}}(w)\big)^T\mathcal{K}\,\mathbb{F}(z),\ \ \ w\in\mathbb{C}\backslash\mathbb{R}
\end{equation*}
and observe from \eqref{idea2}, that
\begin{equation}\label{idea3}
	\big[\bar{\mathcal{B}}(w,z)\big]_{jk}=\begin{cases}
		\bar{\mathcal{B}}(f_k^{(+)},\hat{f}_j^{(\pm)}),& w\in\mathbb{H}^{\pm},\ \ z\in\mathbb{H}^+\\
		\bar{\mathcal{B}}(f_k^{(-)},\hat{f}_j^{(\pm)}),& w\in\mathbb{H}^{\pm},\ \ z\in\mathbb{H}^-.
		\end{cases}
\end{equation}
Since
\begin{equation*}
	\mathbb{G}^{-1}(w)\mathbb{G}(z)=w^{-D}\bar{\mathcal{B}}(w,z)z^D
\end{equation*}
we have thus proven Theorem \ref{thmConco}.
\end{proof}
An alternative formulation of the matrix \eqref{idea2}, which is also used in Definition \ref{defGjl} is given below
\bp\label{further} For any $1\leq j,k\leq p+1$, with $\mathcal{P}_0=\mathcal{P}\cup\{0\}=\{0,a_{11},a_{12},\ldots,a_{1p}\}$,
\bea\nonumber
	\frac{\bar{\mathcal{B}}(f_j,\hat{f}_k)(w,z)}{w-z}=\frac{c_j\hat{c}_k}{(2\pi i)^2}\int_{L}\int_{\wh{L}}F_j^{(\pm)}(u)\hat{F}_k^{(\pm)}(-v)\frac{w^{v}z^{-u}}{1-u+v}\,\d v\,\d u\\
	-c_j\hat{c}_k\sum_{s\in\mathcal{P}_0}\res_{v=s}F_j(v+1)\hat{F}_k(-v)\frac{w^vz^{-v}}{w-z}\nonumber
\eea
Here the integrations around $L, \wh{L}$ are taken in the indicated order and thus mean the evaluation of the residues in the $v$ variable first at the poles of $\wh F_k(-v)$ followed by evaluation of the residues in $u$ at the poles of $F_j$.
\ep
\begin{proof}
Start from \eqref{gen:9} and first use the functional equations for $F_j(u)$ and $\hat{F}_k(-v)$, i.e.
\begin{align*}
	\bar{\mathcal{B}}(f_j,\hat{f}_k)(w,z)&=\frac{c_j\hat{c}_k}{(2\pi i)^2}\int_{L}\int_{\wh{L}}F_j(u+1)\hat{F}_k(-v)\frac{w^{v}z^{-u}}{u-v}\,\d v\,\d u\\
	&-\frac{c_j\hat{c}_k}{(2\pi i)^2}\int_{L}\int_{\wh{L}}F_j(u)\hat{F}_k(1-v)\frac{w^{v}z^{-u}}{u-v}\,\d v\,\d u \equiv I_1-I_2
\end{align*}
where each $I_j$ is now dependent on the order of integration. By the residue theorem, with $\mathcal{P}_0=\mathcal{P}\cup\{0\}$,
\begin{align}
	I_2&=\frac{c_j\hat{c}_k}{2\pi i}\sum_{s\in\mathcal{P}_0+1+\mathbb{N}}\res_{v=s}\hat{F}_k(1-v)\int_{L}F_j(u)\frac{w^{v}z^{-u}}{u-v}\,\d u-\frac{c_j\hat{c}_k}{2\pi i}\int_{L\cap\textnormal{Int}(\wh{L})}F_j(u)\hat{F}_k(1-u)w^{u}z^{-u}\,\d u\nonumber\\
	&=c_j\hat{c}_k\!\!\!\!\!\sum_{s\in\mathcal{P}_0+1+\mathbb{N}}\sum_{t\in\mathcal{P}_0-\mathbb{N}}\res_{u=t}\res_{v=s}\hat{F}_k(1-v)F_j(u)\frac{w^{v}z^{-u}}{u-v}+c_j\hat{c}_k\!\!\!\!\!\!\!\!\!\!\!\!\!\!\!\sum_{s\in(\mathcal{P}_0+1+\mathbb{N})\cap\textnormal{Int}(L)}\!\!\!\!\!\!\!\!\!\!\!\!\!\!\!\!\res_{v=s}\hat{F}_k(1-v)F_j(v)w^{v}z^{-v}\nonumber\\
	&\ \ \ \ -\frac{c_j\hat{c}_k}{2\pi i}\int_{L\cap\textnormal{Int}(\wh{L})}F_j(u)\hat{F}_k(1-u)w^{u}z^{-u}\,\d u.\label{gen:11}
\end{align}
Notice that from the functional relations we have 
$F_j(v)\hat{F}_k(1-v)=F_j(v+1)\hat{F}_k(-v),$
and thus,
\begin{align*}
	s\in(\mathcal{P}_0+1+\mathbb{N})\cap\textnormal{Int}(L):\ \ \ \ \res_{v=s}\hat{F}_k(1-v)F_j(v)w^vz^{-v}&=\frac{1}{2\pi i}\int_{\partial D(s,\epsilon)}F_j(v)\hat{F}_k(1-v)w^vz^{-v}\,\d v\\
	&=\res_{v=s}F_j(v+1)\hat{F}_k(-v)w^vz^{-v}.
\end{align*}
Back to \eqref{gen:11} with the help of the functional relations once more,
\begin{align*}
	I_2&=c_j\hat{c}_k\!\!\!\!\!\sum_{s\in\mathcal{P}_0+1+\mathbb{N}}\sum_{t\in\mathcal{P}-\mathbb{N}}\res_{u=t}\res_{v=s}\hat{F}_k(1-v)F_j(u)\frac{w^{v}z^{-u}}{u-v}+c_j\hat{c}_k\!\!\!\!\!\!\!\!\!\!\!\!\!\!\!\sum_{s\in(\mathcal{P}_0+1+\mathbb{N})\cap\textnormal{Int}(L)}\!\!\!\!\!\!\!\!\!\!\!\!\!\!\!\!\res_{v=s}F_j(v+1)\hat{F}_k(-v)w^{v}z^{-v}\\
	&\ \ \ \ -\frac{c_j\hat{c}_k}{2\pi i}\int_{L\cap\textnormal{Int}(\wh{L})}F_j(u+1)\hat{F}_k(-u)w^{u}z^{-u}\,\d u\\
	&=c_j\hat{c}_k\!\!\!\!\!\sum_{s\in\mathcal{P}_0+\mathbb{N}}\sum_{t\in\mathcal{P}_0-\mathbb{N}}\res_{u=t}\res_{v=s}\hat{F}_k(-v)F_j(u)\frac{w^{v+1}z^{-u}}{u-v-1}+c_j\hat{c}_k\!\!\!\!\!\!\!\!\!\!\!\!\!\!\!\sum_{s\in(\mathcal{P}_0+1+\mathbb{N})\cap\textnormal{Int}(L)}\!\!\!\!\!\!\!\!\!\!\!\!\!\!\!\!\res_{v=s}\hat{F}_k(-v)F_j(v+1)w^{v}z^{-v}\\
	&\ \ \ \ -\frac{c_j\hat{c}_k}{2\pi i}\int_{L\cap\textnormal{Int}(\wh{L})}F_j(u+1)\hat{F}_k(-u)w^{u}z^{-u}\,\d u.
\end{align*}
Now move on to $I_1$, by similar reasoning,
\begin{align*}
	I_1	&=c_j\hat{c}_k\!\!\!\sum_{s\in\mathcal{P}_0+\mathbb{N}}\sum_{t\in\mathcal{P}_0-\mathbb{N}}\res_{u=t}\res_{v=s}\hat{F}_k(-v)F_j(u)\frac{w^{v}z^{-u+1}}{u-v-1}+c_j\hat{c}_k\!\!\!\!\!\!\!\!\!\!\!\!\!\!\!\sum_{s\in(\mathcal{P}_0+\mathbb{N})\cap\textnormal{Int}(L)}\!\!\!\!\!\!\!\!\!\!\!\!\res_{v=s}\hat{F}_k(-v)F_j(v+1)w^{v}z^{-v}\\
	&-\frac{c_j\hat{c}_k}{2\pi i}\int_{L\cap\textnormal{Int}(\wh{L})}F_j(u+1)\hat{F}_k(-u)w^{u}z^{-u}\,\d u.
\end{align*}
and subtracting, we have proven the Proposition.
\end{proof}

In order to obtain the expression of the kernels in Definition \ref{defGjl} and also completely prove Theorem \ref{thmmain1}, we need to 
express explicitly the right side in Theorem \ref{theorig}, that is we have to compute
\begin{equation}\label{finformula}
	\mathcal{C}_{j\ell}(\xi,\eta)=\frac{(-1)^{\ell-1}}{(-2\pi i)^{j-\ell+1}}c_0^{\frac{\varpi_{\ell}-\varpi_j}{p+1}}\xi^{\frac{1}{2}a_j}\eta^{\frac{1}{2}a_{\ell}}\left[\frac{\mathbb G^{-1}(w)\mathbb G(z)}{w-z}\right]_{j+1,\ell}\Bigg|_{\substack{w=\xi(-)^{j+1}\\  z=\eta(-)^{\ell-1}}}
\end{equation}
where $j,\ell=1,\ldots,p$ and $c_0,\xi,\eta>0$ with  $\{\varpi_k\}$ as in \eqref{etaconstants}. For this, we need to use Theorem \ref{thmConco}, the explicit formul\ae\ for $F_j(u), \hat F_j(v)$ \eqref{gen:2}, \eqref{Fhatpm} combined with \eqref{idea3}, the expressions for $c_j, \hat{c}_j$ in \eqref{cjs}, \eqref{correctnorm} and then simplify so as to obtain the expression in Conjecture \ref{conjpchain}.
\subsubsection{The one-matrix ``chain''}
\label{onechain}
We show here that for $p=1$ the Meijer-G field is nothing but the ordinary Bessel random point field \cite{BGS3}. We make use of
\be 
	B_{\nu}(\z)=\frac{1}{2\pi i}\int_{\gamma}\frac{\Gamma(u)}{\Gamma(1+\nu-u)}\z^{-u}\d u\equiv \z^{-\frac{\nu}{2}}J_{\nu}(2\sqrt{\z})
\ee 
with the Bessel function $J_{\nu}(\cdot)$ of first kind. Thus, using \eqref{Gpsecond}, we have 
\beas
\mathcal G^{(1)}_{11}(\xi,\eta)&\&=
\int\hspace{-4pt}\int \frac{\Gamma(u)}{\Gamma(1+a_1-u)}\frac{\Gamma(-v+a_1)}{\Gamma(1+v)}\frac{\xi^v\eta^{-u}}{1-u+v}\frac{\d v\,\d u}{(2\pi i)^2}
=\\
&&= \int_0^1B_{a_1}(t\eta)B_{a_1}(t\xi)t^{a_1}\,\d t = 4K_{\textnormal{Bess},a_1}(4\xi,4\eta)
\eeas
where we used the expression of the Bessel kernel as given in \cite{BGS3}, formul$\ae$ $(4.26)$ and $(4.27)$. 

\subsubsection{Comparison with \cite{BGS3}, two matrix chain}
\label{comparep2}
In \cite{BGS3} (Theorem 2.2) the chain $p=2$ was studied; we can compare those results with our situation. The four kernels defining the Meijer-G field were introduced in \cite{BGS3} as 
\bea
	\mathcal{G}_{00}(\z,\xi) &\& =\frac{1}{(2\pi i)^2}\int\hspace{-4pt}\int _{\gamma^2}\frac{\Gamma(u+a)}{\Gamma(1-u)\Gamma(1+b-u)}\frac{\Gamma(v+b)}{\Gamma(1-v)\Gamma(1+a-v)}\frac{\z^{-u}\xi^{-v}}{1-u-v}\,\d v\,\d u,
	\nonumber
	\\
	\mathcal{G}_{01}(\z,\xi) &\& =\frac{1}{(2\pi i)^2}\int\hspace{-4pt}\int _{\gamma^2}\frac{\Gamma(u+a)}{\Gamma(1-u)\Gamma(1+b-u)}\frac{\Gamma(v)\Gamma(v+b)}{\Gamma(1+a-v)}\frac{\z^{-u}\xi^{-v}}{1-u-v}\,\d v\,\d u,
	\nonumber 
	\\
	\mathcal{G}_{10}(\z,\xi) &\& =\frac{1}{(2\pi i)^2}\int\hspace{-4pt}\int _{\gamma^2}\frac{\Gamma(u)\Gamma(u+a)}{\Gamma(1+b-u)}\frac{\Gamma(v+b)}{\Gamma(1-v)\Gamma(1+a-v)}\frac{\z^{-u}\xi^{-v}}{1-u-v}\,\d v\,\d u,
	\nonumber\\
	\mathcal{G}_{11}(\z,\xi) &\& =\frac{1}{(2\pi i)^2}\int\hspace{-4pt}\int _{\gamma^2}\frac{\Gamma(u)\Gamma(u+a)}{\Gamma(1+b-u)}\frac{\Gamma(v)\Gamma(v+b)}{\Gamma(1+a-v)}\frac{\z^{-u}\xi^{-v}}{1-u-v}\,\d v\,\d u-\frac{1}{\z+\xi},
	\label{Gfield2}
\eea
The indexing of the four kernels follows a different convention and thus we need to compare 
\be\nonumber
\mathcal G_{00} \leftrightarrow \mathcal G^{(2)}_{12}\ ,\ \ 
\mathcal G_{01} \leftrightarrow \mathcal G^{(2)}_{11}\ ,\ \ 
\mathcal G_{10} \leftrightarrow \mathcal G^{(2)}_{22}\ ,\ \ 
\mathcal G_{11} \leftrightarrow \mathcal G^{(2)}_{21}\ . 
\ee
It is then a simple verification that 
\bea
\mathcal G_{00}(\xi,\eta) = \le(\frac \xi \eta\ri)^{a} \mathcal G^{(2)}_{12} (\eta,\xi;\{a,b\})\ ,\ \ 
\mathcal G_{01}(\xi,\eta)=\le(\frac \xi \eta\ri)^{a}  \mathcal G^{(2)}_{11} (\eta,\xi;\{a,b\})\\
\mathcal G_{10}(\xi,\eta)=\le(\frac \xi \eta\ri)^{a}  \mathcal G^{(2)}_{22}(\eta,\xi;\{a,b\})\ ,\ \ 
\mathcal G_{11}(\xi,\eta) =\le(\frac \xi \eta\ri)^{a}  \mathcal G^{(2)}_{21}(\eta,\xi;\{a,b\})\ . 
\eea
This implies the equivalence of the determinantal point fields.
\subsubsection{Comparison with \cite{BZ}, singular values of products of Ginibre matrices}
\label{comparep3}
In \cite{BZ}, Theorem $5.3.$, Kuijllaars and Zhang obtained the following limiting kernel in the cause of a local scaling analysis, 
\begin{equation*}
	K_{\nu}^M(x,y)=\int_0^1G_{0,M+1}^{\,1,0} \!\left( \le.{ -- \atop  -\nu_0,-\nu_1 \dots, -\nu_M } \; \right| \, tx \right)G_{0,M+1}^{\,M,0} \!\left( \le.{ -- \atop  \nu_1, \dots,\nu_M,\nu_0 } \; \right| \, ty \right)\d t
\end{equation*}
where $\nu_j=N_j-N_0\in\mathbb{Z}_{\geq 0}$ and $M\in\mathbb{Z}_{\geq 1}$. Recalling \eqref{MG:1}, we have equivalently
\begin{equation*}
	K_{\nu}^M(x,y)=\frac{1}{(2\pi i)^2}\iint_{\gamma^2}\frac{\Gamma(u)}{\prod_{s=1}^M\Gamma(1+\nu_s-u)}\frac{\prod_{s=0}^M\Gamma(\nu_s+v)}{\Gamma(1-v)}\frac{y^{-v}x^{-u}}{1-u-v}\,\d v\,\d u
\end{equation*}
and thus with \eqref{Gpsecond}
\begin{equation*}
	K_{\nu}^M(x,y)=\mathcal{G}_{11}^{(M)}\big(y,x;\{\nu_1,\nu_2-\nu_1,\nu_3-\nu_2,\ldots,\nu_M-\nu_{M-1}\}\big).
\end{equation*}
Here we observe that in 
\cite{BZ} and in \cite{AIK} only the correlation kernel of one product  was considered; thus we can only compare it to one (the $(1,1)$ specifically) of the kernels we obtain. 
It is possible to speculate that   if one could construct the joint correlation functions for the singular values of all  the intermediate products of Ginibre matrices in \cite{BZ, AIK}, then also the remaining kernels $\mathcal G_{ij}^{(M)}$ would match. This would reinforce the universal character of these new kernels.
%
\subsection{Limiting random point fields and chain separation}
\label{chainsep}
We now provide the verification of Theorem \ref{chainsepn}.
In the study of these limits, we use Stirling's approximation for the Gamma functions 
\begin{equation*}
	\Gamma(z+\delta)=\left(\frac{z}{e}\right)^zz^{\delta}(2\pi z)^{\frac{1}{2}}\left(1+\mathcal{O}\left(z^{-1}\right)\right),\ \ z\rightarrow\infty,\ |\textnormal{arg}\, z|<\pi-\epsilon
\ \Rightarrow\ 
\frac{\Gamma(z+\delta)}{\Gamma(z+\rho)} = z^{\delta-\rho}\left(1+\mathcal{O}\left(z^{-1}\right)\right).
\end{equation*}

\begin{proof}[Proof of Theorem \ref{chainsepn}]
For the purposes of this proof we introduce the notation 
\bea\nonumber
{\bf \Gamma}_p(u,v;\{\vec a\})\equiv 
\frac{\prod_{s=0}^{\ell-1}\Gamma(u-a_{1s})}{\prod_{s=\ell}^p\Gamma(1+a_{1s}-u)}
\frac{\prod_{s=j}^p\Gamma(a_{1s}-v)}{\prod_{s=0}^{j-1}\Gamma(1+v-a_{1s})}
\ ,\qquad
\nabla K(u,v)\equiv \frac {K(u)- K(v)}{u-v}
\eea
and $K=K(u)$ as in \eqref{Kdef}.
The expression $\nabla K(u,v)$ obeys the Leibniz rule
\be\nonumber
\nabla (K_1K_2)(u,v) =
 K_1(u) \nabla K_2(u,v)  + K_2(v) \nabla K_1(u,v)
\ee 
Now we shall write 
\begin{align*}
	K_p(u;\vec a\,) &= (-1)^p \prod_{s=0}^{p} (u - a_{1s}) = (-1)^{q-1} \prod_{s=0}^{q-1} (u - a_{1s})  (-1)^{p-q-1}  \prod_{s=q}^{p} (u - a_{1s}) = \\
	&\equiv K_{q-1}(u;\{a_1,\dots, a_{q-1}\}) K_{p-q}(u -a_{1q};\{ a_{q+1},\dots, a_p\})=K_{q-1}(u) K_{p-q}(u -a_{1q})
\end{align*}
where in the last writing the parametric dependence on the $a_j$'s is  understood. Note that $K_{q-1}(u), K_{p-q}(u)$ are independent of $a_q$. 
We analyze the integrand in \eqref{scalelimit3}
\bea
{\bf \Gamma}_p(u,v;\{\vec a\})\nabla K(u,v) = 
\frac{\prod_{s=0}^{\ell-1}\Gamma(u-a_{1s})}{\prod_{s=\ell}^p\Gamma(1+a_{1s}-u)}
\frac{\prod_{s=j}^p\Gamma(a_{1s}-v)}{\prod_{s=0}^{j-1}\Gamma(1-a_{1s}+v)} \nabla K(u,v)
\eea
and need to consider $9$ types of situations, depending on the positioning of the indices: $j,\ell$ less, equal or greater than $q$.
The large parameter in these computations is $a_q = \Lambda$.\bigskip

{\it Case: $ j,\ell <q$.} We look at the asymptotic behavior of the integrand for the kernels in this block under the two scalings; the computation requires to consider the following steps
\begin{align}
	{\bf \Gamma}_p(u,v;\{\vec a\})\nabla K(u,v)&=\overbrace{\frac{\prod_{s=0}^{\ell-1}\Gamma(u-a_{1s})  \prod_{s=j}^{q-1}\Gamma(a_{1s}-v)}
	{\prod_{s=\ell}^{q-1}\Gamma(1+a_{1s}-u) \prod_{s=0}^{j-1}\Gamma(1-a_{1s}+v)}}^{{\bf \Gamma}_{q-1}(u,v;\{a_1,\dots, a_{q-1}\})}\frac{\prod_{s=q}^{p}\Gamma(a_{1s}-v)}
	{\prod_{s=q}^{p}\Gamma(1+a_{1s}-u) }\nonumber\\
	&\,\times \le( K_{p-q}(u-a_{1q}) \nabla K_{q-1} (u,v)+ K_{q-1}(v) \nabla K_{p-q} (u-a_{1q},v-a_{1q})\ri)\nonumber\\
	&={\bf \Gamma}_{q-1}(u,v;\{a_1,\dots, a_{q-1}\})\nabla K_{q-1} (u,v){\Lambda}^{p-q+1} {\Lambda}^{(p-q+1)(u-v-1)}\big(1 + \mathcal O(\Lambda^{-1})\big)\nonumber\\
	&={\bf \Gamma}_{q-1}(u,v;\{a_1,\dots, a_{q-1}\})\nabla K_{q-1} (u,v)  {\Lambda}^{(p-q+1)(u-v)}\big(1 + \mathcal O(\Lambda^{-1})\big)\label{511}
\end{align}
If we plug \eqref{511} into the formula for the kernels we find 
\begin{align*}
	\Lambda^{p-q+1} &\mathcal G^{(p)}_{j\ell}(\Lambda^{p-q+1}\xi,\Lambda^{p-q+1} \eta; \{a_1,\dots, a_p\}) =  \frac{1}{(-1)^{\ell} \eta - (-1)^{j} \xi}\\
	&\times\, \frac{1}{(2\pi i)^2}\iint{\bf \Gamma}_{q-1}(u,v;\{a_1,\dots, a_{q-1}\})\nabla K_{q-1} (u,v)  {\Lambda}^{(p-q+1)(u-v)} \big(1 + \mathcal O(\Lambda^{-1})\big)\\
 &\times\, \le(\Lambda^{p-q+1}\xi\ri)^{v}\le(\Lambda^{p-q+1}\eta\ri)^{-u}\,\d v\,\d u  = \mathcal  G^{(q)}_{j\ell}(\xi,\eta;\{a_1,\dots, a_{q-1}\})\big(1 + \mathcal O(\Lambda^{-1})\big).
\end{align*}
In the other scaling we need to show that the latter block of kernels tends to zero; to this end we also need the behavior of the integrand ${\bf \Gamma}_p(u,v;\{\vec a\})\nabla K(u,v)$ with the shift $u = u' + a_{1q}, \ v = v'+a_{1q}$. In the computation below we use Euler's reflection formula
\begin{align}
	&\frac{\prod_{s=0}^{\ell-1}\Gamma(u'+a_{1q} -a_{1s})}{\prod_{s=\ell}^{q-1}\Gamma(1+a_{1s}-a_{1q}-u')}\frac{\prod_{s=j}^{q-1}\Gamma(a_{1s}-v'-a_{1q})}{\prod_{s=q}^p
	\Gamma(1+a_{q+1,s}-u')}\frac{\prod_{s=q}^p\Gamma(a_{q+1,s}-v')}{\prod_{s=0}^{j-1}\Gamma(1+v'+a_{1q} -a_{1s})}\nabla K(u'+a_{1q},v'+a_{1q} )\nonumber\\
	&=\prod_{s=q}^p\frac{\Gamma(a_{q+1,s}-v')}{\Gamma(1+a_{q+1,s}-u')}\prod_{s=0}^{q-1}\frac{\Gamma(a_{1q}+u'-a_{1s})}{\Gamma(1+v'+a_{1q}-a_{1s})}
	\frac{\prod_{s=\ell}^{q-1}\pi^{-1}\sin\pi(a_{1q}+u'-a_{1s})}{\prod_{s=j}^{q-1}(-\pi)^{-1}\sin\pi(a_{1q}+v'-a_{1s})}\mathcal{O}\left(\Lambda^q\right)\nonumber\\
	&=\prod_{s=q}^p\frac{\Gamma(a_{q+1,s}-v')}{\Gamma(1+a_{q+1,s}-u')}\frac{\prod_{s=\ell}^{q-1}\pi^{-1}\sin\pi(a_{1q}+u'-a_{1s})}{\prod_{s=j}^{q-1}(-\pi)^{-1}\sin\pi(a_{1q}+v'-a_{1s})}\Lambda^{q(u'-v')}\left(\mathcal{O}(1)+\mathcal{O}\left(\Lambda^{-1}\right)\right).\label{514}
\end{align}
Substituting \eqref{514} into the formula for the kernels, we obtain
\begin{align*}
	\Lambda^q&\mathcal G^{(p)}_{j\ell}(\Lambda^q\xi,\Lambda^q\eta; \{a_1,\dots, a_p\}) =\frac{1}{(-1)^{\ell}\eta-(-1)^j\xi}\frac{1}{(2\pi i)^2}\iint\prod_{s=q}^p\frac{\Gamma(a_{q+1,s}-v')}{\Gamma(1+a_{q+1,s}-u')}\\
	&\times\,\frac{\prod_{s=\ell}^{q-1}\pi^{-1}\sin\pi(a_{1q}+u'-a_{1s})}{\prod_{s=j}^{q-1}(-\pi)^{-1}\sin\pi(a_{1q}+v'-a_{1s})}\Lambda^{q(u'-v')}\left(\Lambda^q\xi\right)^{v'+a_{1q}}\left(\Lambda^q\eta\right)^{-u'-a_{1q}}\,\d v'\,\d u'\,\mathcal{O}(1)\left(1+\mathcal{O}\left(\Lambda^{-1}\right)\right)\\
	&=\frac{(\xi/\eta)^{a_{1q}}}{(-1)^{\ell}\eta-(-1)^j\xi}\frac{1}{(2\pi i)^2}\iint\prod_{s=q}^p\frac{\Gamma(a_{q+1,s}-v')}{\Gamma(1+a_{q+1,s}-u')}\frac{\prod_{s=\ell}^{q-1}\pi^{-1}\sin\pi(a_{1q}+u'-a_{1s})}{\prod_{s=j}^{q-1}(-\pi)^{-1}\sin\pi(a_{1q}+v'-a_{1s})}\d v'\,\d u'\,\mathcal{O}(1)
\end{align*}
In principle, at this point, one expects an expression that contributes to order $\mathcal O(1)$ in $\Lambda$; but notice that the integrand is entire in the integration variable $u'$ and thus a simple argument using Cauchy theorem shows that it vanishes. Thus the leading contribution must come from the next order in $\Lambda$, namely, $\mathcal{O}(\Lambda^{-1})$.\bigskip

{\it Case: $ j,\ell >q$.} This is entirely analogous to the above and left to the reader.\bigskip

{\it Case: $ j<q<\ell$.} We proceed following the same logic as before.
\begin{align}
	&\frac{\prod_{s=0}^{q-1}\Gamma(u-a_{1s})\prod_{s=j}^{q-1}\Gamma(a_{1s}-v)}{\prod_{s=0}^{j-1}\Gamma(1+v-a_{1s})}\prod_{s=q}^{\ell-1}\frac{\pi}{\sin\pi(u-a_{1s})}\prod_{s=q}^p\frac{\Gamma(a_{1s}-v)}{\Gamma(1+a_{1s}-u)}\nabla K(u,v)\nonumber\\
	&=\frac{\prod_{s=0}^{q-1}\Gamma(u-a_{1s})\prod_{s=j}^{q-1}\Gamma(a_{1s}-v)}{\prod_{s=0}^{j-1}\Gamma(1+v-a_{1s})}\prod_{s=q}^{\ell-1}\frac{\pi}{\sin\pi(u-a_{1s})}\Lambda^{(p-q+1)(u-v-1)}\left(1+\mathcal{O}\left(\Lambda^{-1}\right)\right)\mathcal{O}\left(\Lambda^{p-q+1}\right)\nonumber\\
	&=\frac{\prod_{s=0}^{q-1}\Gamma(u-a_{1s})\prod_{s=j}^{q-1}\Gamma(a_{1s}-v)}{\prod_{s=0}^{j-1}\Gamma(1+v-a_{1s})}\prod_{s=q}^{\ell-1}\frac{\pi}{\sin\pi(u-a_{1s})}\Lambda^{(p-q+1)(u-v)}\mathcal{O}(1).\label{515}
\end{align}	
Substituting \eqref{515} into the formula for the kernels thus yields
\begin{align*}
	\Lambda^{p-q+1}&\mathcal G^{(p)}_{j\ell}(\Lambda^{p-q+1}\xi,\Lambda^{p-q+1}\eta; \{a_1,\dots, a_p\})=\frac{1}{(-1)^{\ell}-(-1)^j\xi}\iint\frac{\prod_{s=0}^{q-1}\Gamma(u-a_{1s})\prod_{s=j}^{q-1}\Gamma(a_{1s}-v)}{(2\pi i)^2\prod_{s=0}^{j-1}\Gamma(1+v-a_{1s})}\\
	&\times\,\prod_{s=q}^{\ell-1}\frac{\pi}{\sin\pi(u-a_{1s})}\Lambda^{(p-q+1)(u-v)}\left(\Lambda^{p-q+1}\xi\right)^v\left(\Lambda^{p-q+1}\eta\right)^{-u}\,\mathcal{O}(1)\d v\,\d u\,\\
	=&\frac{1}{(-1)^{\ell}-(-1)^j\xi}\iint\frac{\prod_{s=0}^{q-1}\Gamma(u-a_{1s})\prod_{s=j}^{q-1}\Gamma(a_{1s}-v)}{(2\pi i)^2\prod_{s=0}^{j-1}\Gamma(1+v-a_{1s})}\prod_{s=q}^{\ell-1}\frac{\pi}{\sin\pi(u-a_{1s})}\xi^v\eta^{-u}\,\mathcal{O}(1)\d v\d u\,
	=\mathcal{O}(1)
\end{align*}
For the other scaling we use again a shift of $u,v$, thus obtaining an estimate of $\mathcal O(1)$. Details are omitted.

{\it Case: $\ell < q<j$.} The computation proceeds similarly to the previous case; this time we obtain a  leading order term $\mathcal O(1)$ in the integrand that is entire  in one of the two variables and thus vanishes by Cauchy's theorem. Hence we get a leading order term of order $\mathcal O(\Lambda^{-1})$.

{\it Remaining cases.} They are all handled along the same lines; the verification is left to the reader because there is really no further surprise in the computation.
\end{proof}

\end{document}